\documentclass{article}

\usepackage{arxiv}
\usepackage[utf8]{inputenc} 
\usepackage[T1]{fontenc}    
\usepackage{hyperref}       
\usepackage{url}            
\usepackage{booktabs}       
\usepackage{nicefrac}       
\usepackage{microtype}      
\usepackage{lipsum}
\usepackage{graphicx}

\usepackage{amsmath,amsfonts}
\usepackage{algorithmic}
\usepackage{xcolor}
\usepackage{textcomp}
\usepackage{subfig}

\usepackage{lscape}
\usepackage{array,multirow}
\usepackage{longtable}
\usepackage{titletoc}

\usepackage{pifont}

\DeclareUnicodeCharacter{21E8}{\weirdarrow}
\newcommand*\rot{\rotatebox{90}}
\newcommand*\OK{\ding{51}}

\title{Graph models for Cybersecurity - A Survey}

\author{
 Jasmin Wachter \\ \\
  University of Klagenfurt\\
  Universitätsstraße 65-67\\
  9020 Klagenfurt, Austria\\ \\
Northwestern University\\
1800 Sherman Ave\\
Evanston, Illinois, USA, 60201 \\
  \texttt{jasmin.wachter@aau.at} \\
}

\begin{document}
\maketitle

\begin{abstract}
  Graph models are helpful means of analyzing computer networks as well as complex system architectures for security. In this paper we evaluate the current state of research for \emph{representing} and \emph{analysing} cyber-attack using graph models, i.e. attack graph (AG) formalisms. We propose a taxonomy on attack graph formalisms, based on 70 models, which we analysed with respect to their \textit{graph semantic}, involved \textit{agents} and \textit{analysis features}. Additionally, we adress which formalisms allow for \textit{automatic attack graph generation} from raw or processes data inputs. Our taxonomy is especially designed to help users and applied researchers identify a suitable AG model for their needs. A summary of the individual AG formalisms is provided as supplementary material. 
\end{abstract}


\begin{figure}[htb]
\centering
 \subfloat[Hyper-alert type graph \cite{ning2004building}] {\includegraphics[width=.49\textwidth]{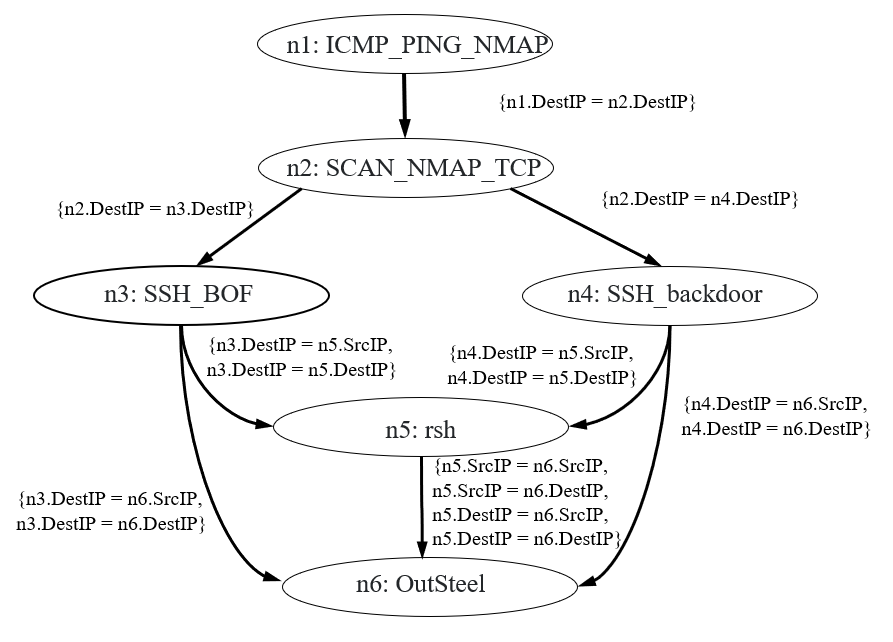}}
 \subfloat[Security Goal Indicator Tree \cite{peine2008security}] {\includegraphics[width=.34\textwidth]{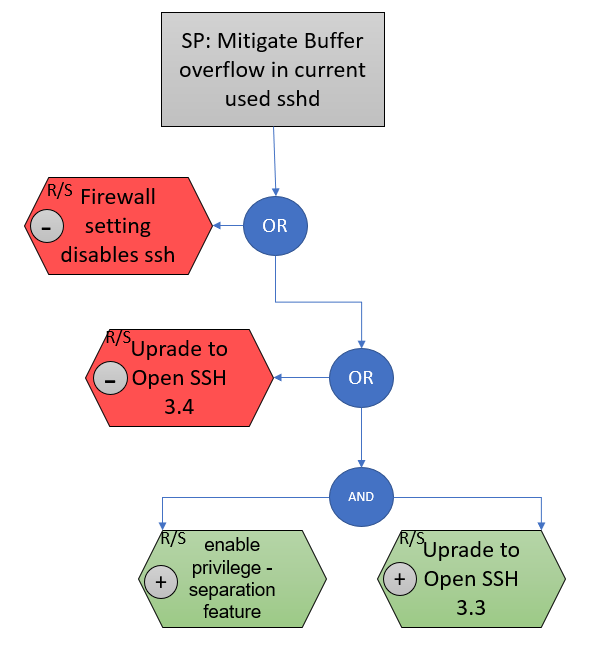}}
  \caption{Illustration: two different graph models for security for one and the same use case}\label{fig:iigraph}
\end{figure}

\section{Introduction}
\label{sec:introduction}
Graph-based cyber-attack models, often referred to as attack graphs (AG), are employed to model computer networks with respect to cyber-attacks. These models abstract information about the cyber-"battlefield", such as network components, their connectivity, information on vulnerabilities, exploits or mitigations, in terms of nodes and edges, resulting in attack (path) graphs or, more general, graph cyber-attack models. 

Ever since their introduction in the 1990s (cf. \cite{weiss1991system, dacier1994privilege, phillips1998graph}) and their popularisation in the early 2000s \cite{schneier1999attack, ammann2002scalable}, a large number of graph-based cyber-attack models and analysis frameworks were (re)developed to analyze computer networks as well as complex system architectures for security. These formalisms model attacker engagement (cf., e.g., \cite{ammann2002scalable, mcqueen2006quantitative,ning2002analyzing}), defensive actions \cite{ardi2006towards,byers2006modeling} or attacker-defender interaction \cite{kordy2014attackdef,edge2007use,baca2010prioritizing}. 

Graph formalisms not only offer an intuitive and system-oriented view on potential attack-steps, vulnerabilities, mitigations and combinations there-of, but they allow for a formal in-context analysis of security scenarios. Due to the large number of formal semantics and analysis methods, it is hard to identify the most suitable representation for one's purpose. Our survey facilitates this issue: we evaluate the current state of research for representing graph models for cybersecurity, and categorize them according to their common underlying graph structure and the semantic of the components. Additionally, we review their offensive/defensive character as well as their analysis features, and summarize the results in a taxonomy. This taxonomy is especially designed to help users and applied researchers identify a suitable graph model for their needs.


\subsection{Objectives and Contribution}
\begin{enumerate}
    \item We review, categorize and summarize the different theoretical approaches of \textit{knowledge representation} for graph models for cyber-attacks with respect to their \textit{graph representation/semantics}. 
    \item In addition, we investigate the \textit{agents involved} in the model (i.e. whether the model serves as an attacker, defender, or attacker-defender model), whether the formalism allows for \textit{automatic attack graph generation} and which network \textit{analysis features} the formalism offers. 
    \item We present a \textit{taxonomy} to classify attack graph frameworks based on the aforementioned components, which facilitates a user in choosing a suitable attack graph model based on their requirements.
      \item We present an overview of existing graph based (cyber-) attack models to prevent the reinvention of existing features. This overview is provided as supplementary material. It especially includes a running example (\emph{the malicious insider}), which is applied to each attack graph type. Thus, we facilitate the reader in understanding the individual attack graph types, and compare and contrast their features in the same context.
\end{enumerate}


\subsection{Methodology}
We performed a systematic review of literature, including related work and past surveys. Then, we categorized 70 attack graph representations with respect to knowledge representations and summarized the resulting categories in our taxonomy. The attack graph analysis goal category-system was identified though inductive categorisation of the corresponding literature item introducing the attack graph formalism. 
Note that we only included attack graph types with a designated name (e.g. "evidence graph") in our research. Models with the generic name "attack graph" were not considered to avoid ambiguity. 
\subsubsection{Main Aspect: Attack Graph Representation and Structure}
We categorise existing attack graph according to their graph representation, especially the semantics of the nodes and edges and identified several attack graph \textit{meta-types}, i.e. common meta-structures present in various attack graph types. Our resulting taxonomy with respect to these AG meta-types helps the reader identify a suitable model and find the answer to the question \textit{HOW} to model the system under study with respect to cyber-attacks. In addition, we categorize the formalisms according to the involved agents and attack graph analysis features, and indicate if the formalism supports automatic model generation from data. 
\subsubsection{Additional Aspects: Agents, Analysis Features and Automatic Graph Generation}
\subsubsection*{Agents: Offense/Defense/Interaction Cyber-attack Models}
AG models may have an attack, defense or interactive character, depending on the agent we want to model. Offensive models concentrate on vulnerabilities, exploits and attacker actions. Defensive formalisms model the defender domain, including preventive aspects, reactive and network hardening components as well as cause elimination. Attacker-defender interaction models focus on the interplay of offensive and defensive aspects. 
The analysis of this feature find the answer to the question \textit{WHOM} we want to model - the attacker or the defender (or both). 
\subsubsection*{Attack Graph Analysis Features}
This find the answer to the question \textit{WHAT ELSE} we can \textit{infer} from the model. We identified \textit{Simulation}, \textit{Security Quantification} and \textit{Intrusion Detection} as frequent inference goals of AG analysis and therefore summarize, which of these forms of analysis are supported by the according formalisms. 
\subsubsection*{Automatic Generation} For each attack graph formalism, we indicate whether the formalism allows for (semi-)automatic attack graph generation from raw or processed data or not. This information helps practitioners choose the right formalism for their needs. 

\subsection{Article Structure}
First, in Section \ref{sec:prelim} we define the most important terms in the context of graph cyber-attack models and attack graphs (cf. \ref{sec:defsandgloss}). Next, in Section \ref{sec:evolution_of_ag} we illustrate the early development of graph-based modelling for cybersecurity. This section serves as a soft introduction to the formalisms and make the reader familiar with the context. Last but not least, this an overview on related surveys and their scope is provided in \ref{sec:related}. 

Section \ref{sec:AG-meta} is dedicated to the first aspect of our taxonomy, \textit{Graph Representation and Structure}. First, in Section \ref{sec:agmetainlit} we review how existing attack graph types were classified with respect to their graphical representation and edge-node semantics in existing literature. In \ref{sec:tax} we introduce our own categorisation of attack graph knowledge representation, and introduce a hierarchic system of attack graph meta-types, categorizing the common features of attack graph representations we identified in our study. The aim of this section is to describe, compare and contrast attack graph meta-types and their particularities. 
Section \ref{sec:aspects} is dedicated to the additional criteria our taxonomy, i,e, \textit{Agents} (cf. Section \ref{sec:aspect2}), \textit{Attack Graph Analysis Features} (cf. Section \ref{sec:aspect3}) as well as \textit{Automatic Generation} (cf. Section \ref{sec:aspect4}). We close the survey with a discussion of some open questions and selected future directions in our conclusion, cf. Section \ref{sec:conclusion}.

An overview of the 70 attack graph types under study is given in the supplementary material (cf. Section \ref{sec:knowledge_rep}). This material may serve as a tutorial on attack graph types, and it includes a running example (\emph{the malicious insider}). This section shall help the user or researcher identify suitable AG formalism on a more fine granular level, providing more details on the individual attack graph types. 

\section{Preliminaries and Terminology} \label{sec:prelim}
\subsection{Definitons and Glossary} \label{sec:defsandgloss}
In this Section, we introduce and clarify the terminology used in this paper. Is serves as a glossary or reference guide for the most important terms related to graph cyber-attack models formalisms and provide a common language.

\paragraph{Agent} An agent is an entity capable of perceiving an environment and acting upon it to change its state. In our cyber-security context, the environment we consider is the network or asset under study. We differentiate between malicious agents, or \textit{attackers}, that try to change the system state to the worse, and benevolent agents, aka. \textit{defenders} trying to maintain the system state or improve it. 
\paragraph{Attacker} A malicious agent in the cyber-security domain is usually referred to as \textit{attacker} or \textit{adversary} (especially in the cryptography domain). We will use these terms, as well as the terms \textit{malicious individual, intruder, offender} interchangeably. 
\paragraph{Defender} A benevolent agent in the cyber-security domain is usually referred to as \textit{Defender}. We will treat the termsm \textit{defender, network administrator, system administrator} as well as \textit{security officer} as synonymes. 

\paragraph{Graphs, Nodes} Informally, graph models are mathematical models frequently used to model network or grid structures. The aim is to model and analyse the pairwise relations between the objects under study. Formally, a \textit{graph} is a set consisting of nodes (aka. points, vertices) that are connected by edges. Thus, nodes are the basic components from which the graph is constructed.  

\paragraph{Edges, directed edges, (un)directed graph} connect two nodes and model the pairwise relation between them. Formally, an egde (link, connection) between two nodes is a pair consisting of the two nodes it connects. Edges may be \textit{directed}, i.e. one of the endpoints of the edges marks the startnode while the other endpoint is the endnode. A \textit{directed graph} is thus a graph with directed edges, while in \textit{undirected graphs} start- and endnode of each edge are interchangeable. Graphically, directed edges are represented by arrows, while undirected edges are represented as lines. Edges may be labelled. In case the edge-label is a numerical, we speak of weighted edges or weights. 

\paragraph{Directed Acyclic Graphs, Trees} 
We call a graph \textit{acyclic}, if it does not connect any cycles or loops. Special classes of acyclic graphs are \textit{directed acyclic graphs} (DAG) or \textit{trees} - acyclic graphs, where there exists exacly one path from each node to every other node. 

\subsection{Related Surveys} \label{sec:related}
Numerous literature surveys on attack graphs have been conducted to analyze the state of the art of graph models for security in a certain sector. In this section, a selection of past literature surveys on this matter is presented in chronological order. Cf. table \ref{tab:past_surveys} for an aggregation of research items presented below. 

\begin{table} \centering
    \begin{tabular}{llllllll|cc}
        & & \multicolumn{7}{c}{Scope of Survey} \\[2ex]
        & & \rot{AG Representation} & \rot{AG Generation} & \rot{Analysis of Security (\& IDS) } & \rot{Complexity} 
        & \rot{Practical Aspects (\& Usability)} & \rot{Year 20xx} & \rot{\# Articles} \\
    \cmidrule{2-9}
        & Lipmann \& Ingols \cite{lippmann2005annotated}           & (\OK) & \OK & (\OK)  & \OK &  &  '05 & 29  \\
        & Kordy et al. \cite{kordy2010foundations}               & \OK & & (\OK) & & & '10 & 30  \\
        & Alhomidi \& Reed \cite{alhomidi2012attack}           & \OK &  &  &  &  &  '12 & 34 \\
        & Aslanyan et al. \cite{aslanyan2013comparative}  & \OK &  &  &  &  & '12 & 34 \\
        & Kordy et al. \cite{kordy2014dag} & \OK & & \OK & & & '14 & 329  \\
        & Shandilya et al. \cite{shandilya2014use}  & (\OK) & \OK & \OK & & &'14 & 30 \\
        & Mell \& Harang \cite{mell2015minimizing} & (\OK) & &  & \OK &   \OK & '15& 30 \\
        & Barik et al. \cite{barik2016attack}  & (\OK) & \OK & \OK & & &  '16 & 35 \\
        & Kaynar \cite{kaynar2016taxonomy}  & (\OK) & \OK & \OK & \OK & & '16 & 90 \\
        \rot{\rlap{~Research Item}}& Haque et al. \cite{haque2017evolutionary}  & \OK & & \OK & \OK & & '17 & 32  \\
        & Noel \cite{noel2018review}  & (\OK) & \OK & \OK &  & \OK& '18 & 97 \\
        & Zeng et al. \cite{zeng2019survey}   &  &  & \OK &  &  & '19 & 54 \\
        & Lallie et al. \cite{LALLIE2020100219}  & \OK &  &  &  &  \OK & '20 & 433 \\
        \cmidrule[1pt]{2-9}
         & Wachter et al.  & \OK & (\OK) & \OK &  & \OK & '23 & 179 \\
    \end{tabular}
    \caption{Aggregation of survey papers on attack graphs} \label{tab:past_surveys}
\end{table}
Lipmann and Ingols \cite{lippmann2005annotated} largely focus on research papers describing approaches to generate attack graphs as well as how these frameworks scale to larger networks. What is interesting about their review is the identification of limitations of past attack-graph research that need to be addressed in the future, i.e. scalability; obtaining information needed to build attack graphs and automated approaches thereof; determination of reachability (in terms of computational complexity as well as model validity); generating recommendations for network defense from attack graphs. Despite their work being from 2005, the issues raised in this past work remain partly unresolved until the time of writing this article.

Kordy et al. \cite{kordy2010foundations} focus on the semantics on attack–defense trees only and present analysis methods and use cases for them. In \cite{alhomidi2012attack}, Alhomidi and Reed (2012) conducted a literature survey that focuses on attack graph representations in terms of its nodes and edges interpretations. In \cite{aslanyan2013comparative} Aslanyan et. al (2013) summarize attack graph representations, while Shandilya et al. (2014) focus on attack graphs for network security \cite{shandilya2014use}. Their classification is based on methodologies and technologies for attack graph analysis. They consider methodologies for formal models, representation of system parameters in the model, graph analyses and property formulation as well as violation detection and response formulation. With respect to technologies, they review automated graph generation tools, technologies for graph analyses and violation detection as well as visualisation and response recommendation/implementation tools. 

Kordy et al. \cite{kordy2014dag} present an extensive comparative survey on DAG-based attack and defense models, including a taxonomy of the described formalisms. Mell and Harang's analysis \cite{mell2015minimizing} (2015) adresses the size-complexity and usability issues, such as human readability of various attack graph representations. Barik et al.'s survey focuses on attack graph generation and analysis techniques \cite{barik2016attack}, extending \cite{lippmann2005annotated} survey until 2016.
 
Kaynar \cite{kaynar2016taxonomy} also focuses on attack graph generation and analysis for security. His 2016 classification is based on the individual phases of attack graph generation and analysis. The survey commence by addressing attack graph generation and reachability analysis, followed by attack modelling and attack template determination. Last but not least, the prescription of attack graph structure, attack graph determination and pruning techniques are elaborated. Haque et al. \cite{haque2017evolutionary} present a survey of papers related to alert correlation, attack graph generation and attack tree reduction. Their 2017 paper highlights the convenience and computational efficiency of attack trees in comparison to more general attack graphs. 

Noel \cite{noel2018review} (2018) review AG representations and analyses to "help guide practitioners towards matching their use cases with existing technical approaches." The consider the phase of security operations (prevention, detection, and reaction), the operational layer (network infrastructure, security posture, cyberspace threats, mission dependencies). Besides, the architectural aspects of AGs, their automatic construction and their performance are discussed. 

Zeng et al. \cite{zeng2019survey} (2019) provide a comparative survey of attack graph analysis methods based on graph theory, Bayesian networks, Markov models, optimization as well as uncertainty analysis. Lallie et al. (2020) analysed the visual syntax of 180 attack graph frameworks\cite{LALLIE2020100219}, i.e. the used terminology, shapes and colours. They stress that unlike fault trees and Petri nets, the visual syntax of attack graphs is not standardised and call for a standardized representation. 

\subsubsection{Identified Gap in Related Work: AG Semantics and Meta-Types}
Most attack graph literature surveys concentrate on one single - mostly technical - aspect of attack graph research (cf. Table \ref{tab:past_surveys}). The main topics covered are (automatic) generation, analysis for security, visualization, usability, complexity as well as intrusion detection. 

Regarding graph representations and semantics, however, the quality of the surveys vary. As can be seen in table \ref{tab:past_surveys}, most works only treat attack graph representations superficially, while the main contribution of the paper is dedicated to other topics. The section on attack graph representations often constitutes of short summaries of selected pieces of related work. The majority of survey cover about 30 papers - an extensive overview or a categorization thereof is usually omitted. Thus, taxonomies with respect to the knowledge representations are rare. There are, however, quite a few exceptions. 

First of all, the work of Kordy et al., \cite{kordy2014dag} presents and overview of graphical attack models, that can be modelled as a DAG or tree. Their taxonomy of attack graph formalism is based on 13 criteria, and incorporates information on its purpose or tool availability.

Lallie et al. \cite{LALLIE2020100219} focus on the visual syntax of attack graph models, i.e. the forms, colours and notions of the elements used to build attack graph. Despite covering many attack graph formalisms, a description of the individual formalisms is skipped. They implicitly categorized attack graphs in their tables - the methodology of doing so, however, is not described and thus not comprehensible. 

None of the surveys, however, taxonomize the semantics of attack graph representations, i.e. \textit{their common meta-structure}. The analysis in the aforementioned past works is rather based on the description of individual AG components, not on their interplay or common meta-structure. We aim to bridge this gap by analysing the knowledge representation of 70 attack graph models. We describe, compare and contrast the representations, form categories and put them into context with existing research. In addition, we provide information on AG agents and attack graph analysis features, as well as automatisation of AG generation. Overall, our analysis facilitates the user in finding the right attack graph formalism and representation. 

\section{Attack Graph Meta-Types} \label{sec:AG-meta}
In this Section, we will present several, what we call, attack graph meta-types, i.e. common semantics we identified over various attack graph types. Throughout the whole text, we will refer to common structures over various attack graph types as meta-types. First, we review common structural components were found in the related work. We discovered that in prior works, attack-graphs were mainly classified according to their graph structure (tree vs. general graph), statefulness, node- and edge semantics. After reviewing them, we present the result of our inductive categorization of the individual attack graph formalisms into meta-types in Section \ref{sec:knowledge_rep}. 

\subsection{Attack Graph Meta-Types identified in past surveys} \label{sec:agmetainlit}
\subsubsection{Tree models vs. general graph models}
Lallie \cite{LALLIE2020100219} differentiate between attack graphs and attack trees. "The research shows a clear divide between papers that focus on attack graphs and those that focus on attack trees. In other words, authors either write about attack graphs or attack trees." \cite{LALLIE2020100219} Kordy et al's differentiation between static and sequential modes gives a justification for this divide.

\begin{figure*}[htb]
\centering
   \subfloat[State Enumeration-Type Attack Graph\label{fig:state_enumeration_ag}]{%
    \includegraphics[width=.3\textwidth]{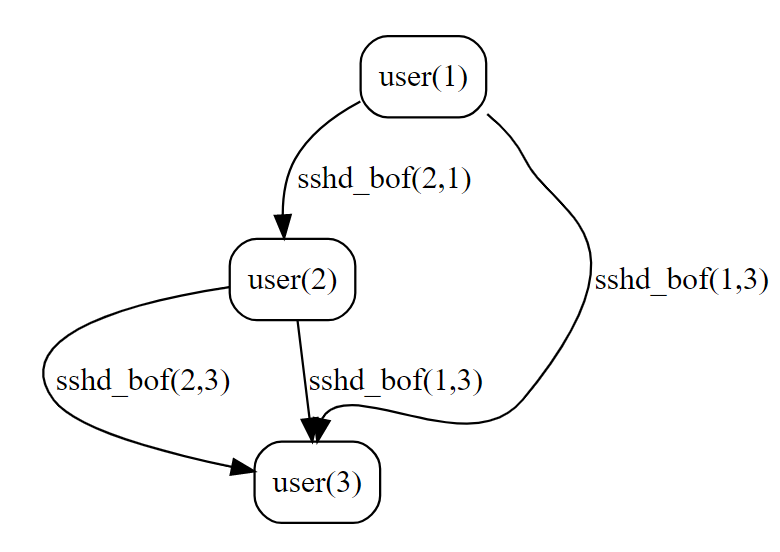}} \qquad
     \subfloat[Dependency-Type Attack Graph Type \label{fig:condition}]
  {%
    \includegraphics[width=.3\textwidth]{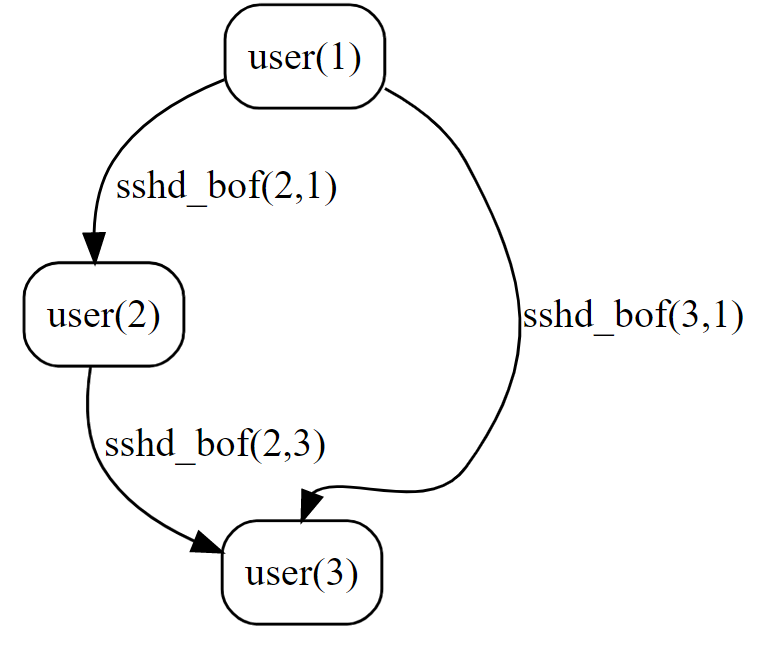}}\\
    \subfloat[Asset-oriented Attack Graph Type \label{fig:asset_ag}]
  {%
\includegraphics[width=.3\textwidth]{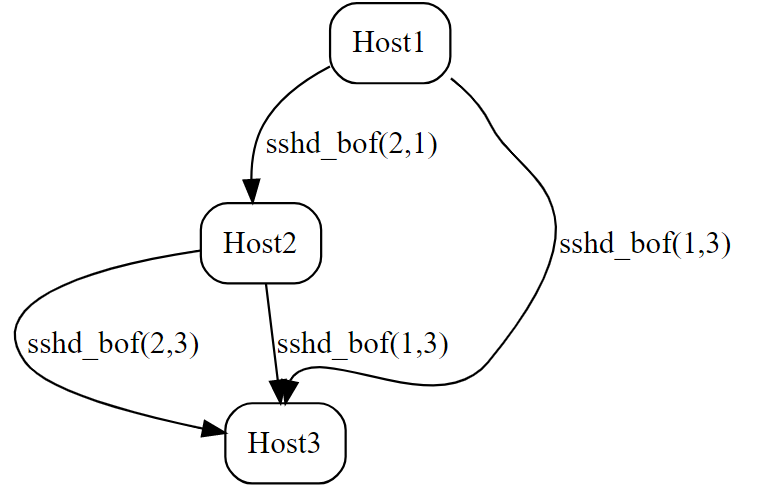}} \qquad
   \subfloat[Exploit-oriented Attack Graph Type \label{fig:exploit}]{
  \includegraphics[width=.25\textwidth]{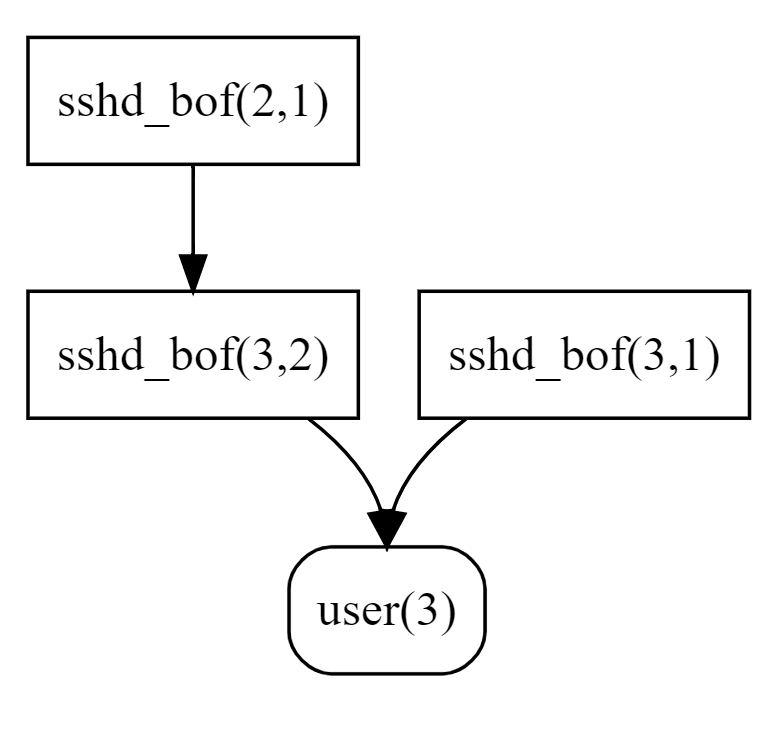} } \qquad 
  \subfloat[Condition-oriented Attack Graph Type \label{fig:condo}]{
  \includegraphics[width=.25\textwidth]{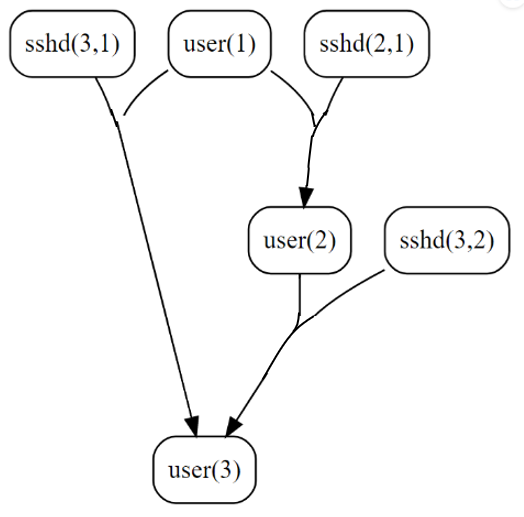} }
   \caption{Illustration of attack graph Meta-Types from related surveys. All attack graph metatypes are illustrated for the running example (\ref{fig:running_ex}) of a computer network with three hosts connected by ssh. Starting from user level privilege on the first machine, the adversary seeks to gain user access on host 3.}
\end{figure*}

\subsubsection{Stateful models vs. stateless models}
Kordy et al \cite{kordy2014dag} differentiate between static and sequential models. The static models mainly involve designated tree formalisms while the sequential models include mainly general graph and tree models. Limiting the analysis to the designated attack graphs and excluding trees,\footnote{Prior literature, such as \cite{noel2004managing,sawilla2007googling, aslanyan2013comparative} refer to dependency or state enumeration attack graphs as two important classes of attack graph representations. Note that, throughout this work, we added the suffix \textit{-type} to emphasise that we refer to a collection of attack graph models, i.e. Meta-types, rather than an individual models.}  it was argued in early works (e.g. \cite{noel2004managing, aslanyan2013comparative, sawilla2007googling}) that there exist two main structural representations of attack graphs: \textit{State Enumeration(-Type) Attack Graphs}, stateful models, where each node represents an entire network state and edges encode state transitions; and  \textit{Dependency(-Type) Attack Graphs}, on the other hand, are stateless or static models, where "a vertex does not represent the entire state of a system but rather a system condition (often in some form of logical sentence). The arcs in these graphs represent the causality relations between the system conditions"\cite{sawilla2007googling}. The causality is usually given in a requires-implies semantic for exploits. Thus, the divide between stateful and stateless models has been addressed in prior literature.


\subsubsection{Node- and Edge Semantics}
Other works, e.g. Noel and Jajodia distinguish two meta-types of attack graphs according to the semantic of the nodes and edges as exploits and conditions). Among others, Mell and Harang \cite{mell2015minimizing}, Aslanyan et. al  \cite{aslanyan2013comparative} refer to these types in their surveys. 
 
 In their first sub-type, the \emph{condition oriented attack graph type}, the nodes represent conditions and the edges mark exploits leveraging attacker privilege. 
 
 In the second type, \emph{exploit-oriented attack graph type}, the nodes represent exploits and the edges represent pre- and postcondition dependency edges. For consistency, the exploit dependency representation may include (empty) initial and goal condition nodes. Note that, when modelling attack sequences, the exploit-orientation type is similar to the \emph{vulnerability graph type} of attack graph defined in \cite{LALLIE2020100219}. One might argue that vulnerabilities and exploits are not the same. When  steps of exploits are however, associated with specific vulnerabilities, exploit and vulnerability chains are somehow isomorphic. Therefore, on a high level, we will consider exploits and vulnerabilities as one and the same construct. 
 
  \subsubsection{Attack vs. Asset view}
  While the aforementioned attack graph meta-types consider the atoms of an attack, i.e. exploits and system states as the subject of analysis, a large amount of attack graphs operate on a host- or asset level. Thus, in the early related work analysis phase, we subsumed as asset-oriented attack graph types as an important meta-type. In \textit{Asset-oriented attack graph types}, each node represents an asset in the network and transitions indicate vulnerabilities, exploits or exploit chains to reach this goal. The idea of such representations is give an overview of the assets or hosts that may be compromized rather than individual exploits. Host-access graph \cite{ammann2005host} or host-based access graph \cite{malthotra2008host} are examples of such graphs having hosts as nodes, rather than states or conditions.


%

\subsubsection{Exploit oriented Meta-Types}
 \textit{Exploit-oriented Attack Graphs}, have the nodes represent exploits and the edged represent pre- and postconditions of exploits. According to Noel and Jajodia \cite{noel2004managing}, an exploitation-oriented attack graph facilitates the visualization of the individual attack steps.

\subsection{Attack Graph Meta-Types identified in this Study} \label{sec:tax}
The aforementioned categories were identified when reviewing the related work, i.e. past surveys. Thus, they are potentially not up to date, which is why we analysed 70 individual attack graph types \ref{sec:knowledge_rep} and clustered them according to their common structural features. 

 On a high level, their common features were analysed with respect to the semantics of the nodes and edges. On a lower level, their uses or particularities are highlighted. As a result, from the analysis of the individual attack graph models, the following category system in figure \ref{fig:agcategories} was identified by the author:
\begin{figure}[htb]
\centering
    \includegraphics[width=.75\textwidth]{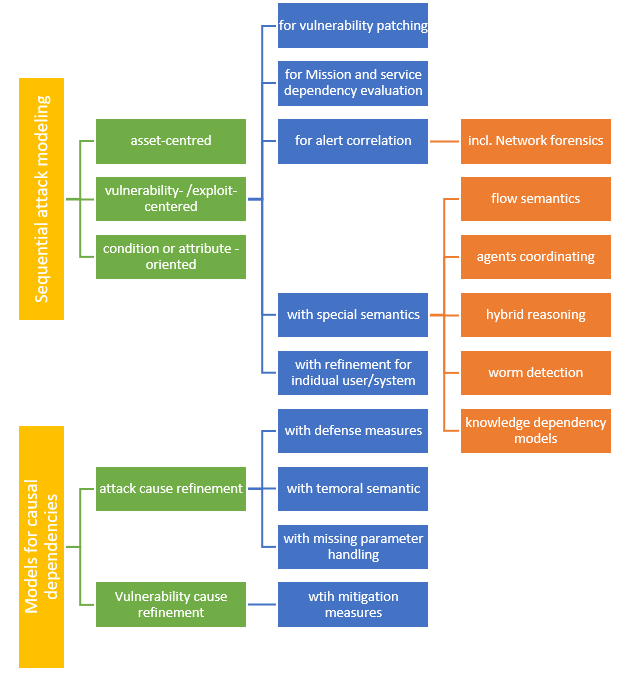}
   \caption{Knowledge representation categories in the attack graph model: the following hierarchical categories were discovered in analysing the node/edge semantics of the 70 formalism.}\label{fig:agcategories}
\end{figure}

\subsubsection{The Category System for Attack Graph Representations} 
Out taxonomy is hierarchical and consists of several levels (cf. figure \ref{fig:agcategories}). We will proceed from top-down and start with the highest levels and continue with the according subcategories. The sub-subcategories of our taxonomy are addressed in Section \ref{sec:knowledge_rep}.
\paragraph{Static vs. Sequential Models}

While in prior literature statefulness served as a distinctive feature, we rather distinguish between attack graph types for \textit{sequential attack modelling} and \textit{causal dependency models}. In sequential models, similar to \cite{kordy2014dag}, edges may represent sequential and causal dependencies, wheres edges in causal dependency models solely specify causal relationships between the nodes. We identified this distinction between static and sequential models as the top-level category. \textit{Sequential Models} are summarized in Table \ref{fig:appendix1:detailstax1}, whereas \textit{causal dependency models} are summarized in Table \ref{fig:appendix1:detailstax2}.

\paragraph{Sequential Models $\to$ Node Semantics}
On the second-level, sequential models were categorized according to the semantic of their nodes. We differentiate between \textit{asset-oriented} (or -centred), \textit{vulnerability-/exploit-oriented} and \textit{condition-/attribute-oriented} formalisms, similar to the definitions in the prior section. \textit{asset-oriented} attack graphs enable a system-component view and help identify vulnerable targets (e.g. for network segmentation). Vulnerability or exploit centred attack graphs, on the other hand, come handy for vulnerability prioritisation in the patch management process, mission and service dependency, or alert correlation for intrusion detection systems and network forensics. Condition- or attribute-oriented attack graphs help identify the individual conditions (e.g. configurations, connectivities) that might lead to compromise. 
The remaining level of sequential model is categorized according to their primary use and domain semantics, cf. \ref{fig:agcategories} for the detailed categories.

\paragraph{Causal Models $\to$ Attack vs. Vulnerability Cause refinement}
On the second level, causal models were categorized according to the refinement they provide. \textit{Attack cause refinement models} describe the causal relationship between the events or elements causing an attack, whereas \textit{vulnerability cause refinement models} describe the causal relationship between the events or elements causing a vulnerability, which may be exploited. If the requirement of the user is to prevent certain vulnerabilities from happening (e.g. in the integration of some secure software development process) the latter become handy, whereas attack cause refinement models are handy in a threat modeling context. 
On the third level, we differentiate between models with defense or mitigation measures, (partial) temporal semantic between the events or models dealing with missing parameter handling; cf. Section \ref{fig:agcategories}. We highlight that some cause refinement models with \textit{defense/mitigation measures} offer methodologies and processes for countermeasure identification and representation. 

\begin{table}[htb]
\centering
\begin{small}
\begin{tabular}{llll}
\hline
\multicolumn{4}{c}{\textbf{Sequential Graph Models for Cybersecurity}}                                                                                                              \\ \hline
\textbf{Asset-}  & Host-Compromized                                                                     &                                 &                          \\ \cline{2-4} 
\textbf{Oriented AG} & Host-Access Graph                                                                    &                                 &                          \\ \cline{2-4} 
                            & Host-Based AG                                                                        &                                 &                          \\ \cline{2-4} 
                            & Host-Centric AG                                                                      &                                 &                          \\ \cline{2-4} 
                            & Cloud-Level AG                                                                       &                                 &                          \\ \cline{2-4} 
                            & Attack Scenario Graph                                                                &                                 &                          \\ \hline
\textbf{Condition-}       & Attribute AG                                                                         &                                 &                          \\ \cline{2-4} 
\textbf{or Attribute-} & Logical AG                                                                           &                                 &                          \\ \cline{2-4} 
\textbf{Oriented} & Scenario Attacks                                                                     &                                 &                          \\ \cline{2-4} 
                            & \begin{tabular}[c]{@{}l@{}}Exploit Dependency Graph \\ by Amman et al.\end{tabular}  &                                 &                          \\ \cline{2-4} 
                            & \begin{tabular}[c]{@{}l@{}}(Exploit) Dependency Graph \\ by Noel et al.\end{tabular} &                                 &                          \\ \cline{2-4} 
                            & Multiple Prerequisites Graph                                                         &                                 &                          \\ \hline
\textbf{Vulnerability-}   & Intention-Centric Approach                                                           &                                 &                          \\ \cline{2-4} 
\textbf{or Exploit-}   & Goal-Oriented AG by Liu (2015)                                                       &                                 &                          \\ \cline{2-4} 
 \textbf{Oriented}    & \begin{tabular}[c]{@{}l@{}}Vulnerability Graph, \\ Scenario Graph\end{tabular}       &                                 &                          \\ \cline{2-4} 
                            & Probabilistic AG                                                                     &                                 &                          \\ \cline{2-4} 
                            & Full AG                                                                              &                                 &                          \\ \cline{2-4} 
                            & Predictive AG                                                                        &                                 &                          \\ \cline{2-4} 
                            & Node-Predictive AG                                                                   &                                 &                          \\ \cline{2-4} 
                            & Stage-Attribute AG                                                                   &                                 &                          \\ \cline{2-4} 
                            & Compromize Graph                                                                     &                                 &                          \\ \cline{2-4} 
                            & \textbf{for Vulnerability Patching}                                                  & Vulnerability Dependency Graph  &                          \\ \cline{2-4} 
                            & \textbf{for Mission and}                                                             & Impact Dependency Graph         &                          \\ \cline{3-4} 
                            & Service Dependency                                                                   & Mission Dependency Graph        &                          \\ \cline{3-4} 
                            &                                                                                      & Mission Impact Graph            &                          \\ \cline{3-4} 
                            &                                                                                      & Generalized Dependency Graph    &                          \\ \cline{2-4} 
                            & \textbf{for Alert Correlation}                                                       & Goal-Oriented AG by Nanda       &                          \\ \cline{3-4} 
                            &                                                                                      & Goal-Oriented AG by Liu (2010)  &                          \\ \cline{3-4} 
                            &                                                                                      & Probabilistic Temproal AG       &                          \\ \cline{3-4} 
                            &                                                                                      & Hyper-Alert Correlation Graph   &                          \\ \cline{3-4} 
                            &                                                                                      & Alert Dependency Graph          &                          \\ \cline{3-4} 
                            &                                                                                      & \textbf{for Network Forensics}  & Evidence Graph           \\ \cline{2-4} 
                            & \textbf{with Special Semantics}                                                      & \textbf{Flow Semantics}         & Insecurity Flow          \\ \cline{4-4} 
                            &                                                                                      &                                 & Risk Flow AG             \\ \cline{3-4} 
                            &                                                                                      & \textbf{Coordination Semantics} & Coordinated AG           \\ \cline{3-4} 
                            &                                                                                      & \textbf{Hybrid Reasoning}       & Hybrid AG                \\ \cline{3-4} 
                            &                                                                                      & \textbf{Worm Detection}         & Activity Graph           \\ \cline{3-4} 
                            &                                                                                      & \textbf{Knowledge} & Knowledge Graph          \\ \cline{4-4} 
                            &                                                                                      & \textbf{Dependencies}                                & Knowledge Bayesian \\ 
                            &                                                                                      &                                 &  Graph \\ \cline{2-4}
                            & \textbf{with Refinement}                                                             & Exploitation Graph              &                          \\ \cline{3-4} 
                            & \textbf{for User/System}                                                             & Personalized AG                 &                          \\ \cline{2-4} 
                            &                                                                                      &                                 &                          \\
                            &                                                                                      &                                 &                          \\
                            &                                                                                      &                                 &                         
\end{tabular}
\end{small}
\caption{Sequential Graph Models: Individual attack graph types and their meta-types} \label{fig:appendix1:detailstax1}
\end{table}

\begin{table}[h]
\centering
\begin{small}
\begin{tabular}{llll}
\hline
\multicolumn{4}{c}{\textbf{Causal Dependency Graph Models for Cybersecurity}}                                                                                    \\ \hline
\textbf{Vulnerability Cause} & Vulnerability Cause Graph                &                                                                               &  \\ \cline{2-4} 
\textbf{Refinement}          & \textbf{with Security}                   & Security Goal Indicator Tree                                                  &  \\ \cline{3-4} 
\textbf{}                    & \textbf{Indicators/Actions}              & Security Activity Graph                                                       &  \\ \cline{3-4} 
                             &                                          & Security Goal Model                                                           &  \\ \hline
\textbf{Attack Refinement}   & Attack Trees                             &                                                                               &  \\ \cline{2-4} 
\textbf{}                    & Anti-Goal Graphs                         &                                                                               &  \\ \cline{2-4} 
\textbf{}                    & Probabilistic Attack Trees               &                                                                               &  \\ \cline{2-4} 
\textbf{}                    & Attack Trees with Costs                  &                                                                               &  \\ \cline{2-4} 
                             & Multi-Parameter Attack Tree              &                                                                               &  \\ \cline{2-4} 
                             & Cyber Threat Tree                        &                                                                               &  \\ \cline{2-4} 
                             & Vulnerability Attack Tree                &                                                                               &  \\ \cline{2-4} 
                             & Intrusion Graph                          &                                                                               &  \\ \cline{2-4} 
                             & Intrusion DAG                            &                                                                               &  \\ \cline{2-4} 
                             & Bayesian AG                              &                                                                               &  \\ \cline{2-4} 
                             & \textbf{with Defense Measures}           & Defense Tree                                                                  &  \\ \cline{3-4} 
                             &                                          & Protection Tree                                                               &  \\ \cline{3-4} 
                             &                                          & Attack-Response Tree                                                          &  \\ \cline{3-4} 
                             &                                          & Attack Countermeasure Tree                                                    &  \\ \cline{3-4} 
                             &                                          & Improved Attack-Defense Tree                                                  &  \\ \cline{3-4} 
                             &                                          & Bayesian Defense Graph                                                        &  \\ \cline{3-4} 
                             &                                          & Countermeasure Graph                                                          &  \\ \cline{2-4} 
                             & \textbf{with Temporal Semantic}          & OWA Tree                                                                      &  \\ \cline{3-4} 
                             &                                          & Enhanced Attack Tree                                                          &  \\ \cline{3-4} 
                             &                                          & \begin{tabular}[c]{@{}l@{}}Improved-, \\ Sand Attack Tree\end{tabular}        &  \\ \cline{3-4} 
                             &                                          & \begin{tabular}[c]{@{}l@{}}Attack Tree with \\ Temporal Order\end{tabular}    &  \\ \cline{3-4} 
                             &                                          & \begin{tabular}[c]{@{}l@{}}Timed Probabilistic Sematics \\ of AT\end{tabular} &  \\ \cline{2-4} 
                             & \textbf{with Missing Parameter Handling} & Decorated Attack Tree                                                         &  \\ \cline{2-4} 
\end{tabular} \end{small}   
\caption{Models for Causal dependencies: Individual attack graph types and their corresponding meta-types. }\label{fig:appendix1:detailstax2}
\end{table}
\subsubsection{Summary and Statistics}
\begin{figure*}[htb]
\centering
    \includegraphics[width=.9\textwidth]{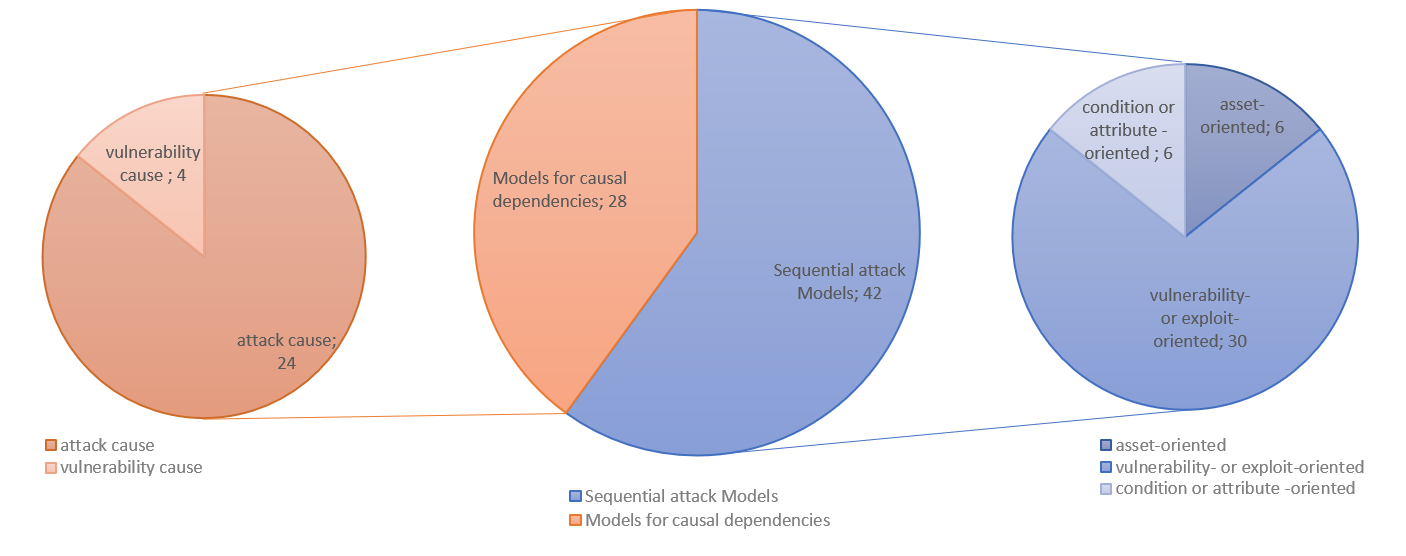}
   \caption{Knowledge representation categories in the attack graph model: the following hierarchical categoriy system was discovered.}\label{fig:statcat}
\end{figure*}

42 of 70 attack graph types are sequential models, whereas 28 model causal dependencies (cf. Figure \ref{fig:statcat}). Among these 28, 24 formalisms model the causes of an attack, whereas 4 of the 28 formalisms model the cause of a vulnerability, i.e. the reasons why the vulnerability occurred, i.e. how it was (not) prevented. Of the 42 sequential models, 30 are vulnerability- or exploit oriented, 6 are asset oriented and 6 are attribute- or condition-oriented. 
The correspondence of the individual attack graph types and their meta-types can be found in the in Tables \ref{fig:appendix1:detailstax1}(Sequential attack model types) and \ref{fig:appendix1:detailstax2} (Models for causal dependencies). 
Overall our qualitative analysis of attack graph meta-types goes extends the categories identified in the related work and orders them in a hierarchical system. The category \emph{modelling of causal dependencies} can be compared to the dependency-Meta-type. For sequential models, we still encounter asset-, vulnerability- or exploit oriented as well as condition-oriented Meta-types as the prevalent forms of knowledge representations. Furthermore, the analysis confirms that the state enumeration-type is replaced by other, more efficient forms of knowledge representation. Overall, the results are well aligned to the related work, which confirms our category system. The hierarchical character of our taxonomy, however, facilitates the user in finding the adequate attack graph representation more than isolated categories, and helps them match their requirements to the individual formalism. 

\section{Agents, Attack Analysis Features and Automatic Graph Generation} \label{sec:aspects}
As mentioned before, we also categorize the formalisms according to the involved agents and attack graph analysis features. Furthermore, whether a specific formalism supports automatic generation is adressed in this survey.

\subsection{Agents: attacker, defender and interaction modelling}\label{sec:aspect2}
Graph cyber-attack models may be modelled from different perspectives: from the attacker's perspective, the defender's perspective or their interplay. Attacker models include the components of the system under study to launch a successful attack. The attacker perspective thus concentrates on the weaknesses of the system, including vulnerabilities, exploits and attacker actions, or conditions that need to be met in order to succeed. Defense models, on the other hand, concentrate on preventive aspects, network hardening components as well as intrusion detection. Attacker-defender interaction models focus on the interplay of offensive and defensive aspects, and include both attacker and defender perspectives. We classified the 70 attack graph types under study according to the agent's view on the system under study. We distinguish the classes \textit{attacker models (A), defender models (D)} and \textit{attacker-defender models (AD)} and summarized the results in Figures \ref{tabbyy:summary} and \ref{tabbyyy:summary} in the last column, \textit{Agent}. 

\subsection{Attack Graph Analysis Features} \label{sec:aspect3}
 We analyze what kind of attack graph analysis features the formalism under study supports. In our analysis of the 70 formalisms, we identified \textit{Simulation}, \textit{Security Quantification} and \textit{Intrusion Detection} as the most frequent inference goals of AG analysis. Therefore, we summarize, which of these forms of analysis are supported by the according formalisms, to help the user pick a model that supports the according feature. The categories are defined below and the results of our analysis are summarized in the table in Figures \ref{tabbyy:summary} and \ref{tabbyyy:summary}. 

\subsubsection{Simulation} The AG formalism can be used to generate (most-likely) attack paths, predict or simulate attacks on network, or likely system behaviour e.g. from probabilities or other parametersprovided. Whether or not a certain AG formalism supports simulation is specified in the first column \textit{Simul.} in Figures \ref{tabbyyy:summary} and \ref{tabbyy:summary}. 
\subsubsection{Intrusion Detection} The AG formalism can be used to detect attacks on the network and correlate empirical data (i.e. intrusion alerts) with attack paths or goals. Formalisms featuring intrusion detection analysis are ticked \OK in the second column \textit{IDS} in Tables \ref{tabbyyy:summary} and \ref{tabbyy:summary}. 
\subsubsection{Security Quantification} The AG formalism incorporates a quantification of the security of the network, e.g. the criticality of network components, most likely attack targets, etc. This information can be employed for network hardening or optimal network design and placement of defensive resources. The third column \textit{Quant.} in Tables \ref{tabbyyy:summary} and \ref{tabbyy:summary} is designated to this feature.

\subsection{Automatic Graph Model Generation from Data}\label{sec:aspect4}
We identify which attack graph formalism under study support automatic generation of the graph model from raw or processed data inputs (e.g. network and vulnerability scanning information). The last column \textit{Gener.} in Tables \ref{tabbyyy:summary} and \ref{tabbyy:summary} is designated to this feature.

\begin{table}
\centering
\begin{tabular}{l|c|c|c|c|c|}
\hline
\multicolumn{1}{|l|}{\textbf{Attack Graph Type}}                                                                                                                 & \multicolumn{1}{l|}{Simul.}                               & \multicolumn{1}{l|}{IDS}                      &  \multicolumn{1}{l|}{Quant.}     &  \multicolumn{1}{l|}{Agent}  &  \multicolumn{1}{l|}{Gener.} \\ \hline \hline
\multicolumn{1}{|l|}{\textit{Intention-centric approach} \cite{gorodetski2002attacks} }                                                                          & \multicolumn{1}{l|}{\OK}                               & \multicolumn{1}{l|}{ }                  & \multicolumn{1}{l|}{}                                                                                                                  & \multicolumn{1}{l|}{AD}                                                                                       & \multicolumn{1}{l|}{ } \\ \hline
\multicolumn{1}{|l|}{\textit{Anti-Goal Graph} \cite{van2003system} }                                                                                             & \multicolumn{1}{l|}{ }                               & \multicolumn{1}{l|}{ }                  & \multicolumn{1}{l|}{ }                                                                                                                  & \multicolumn{1}{l|}{A}                                                                                       & \multicolumn{1}{l|}{ } \\ \hline
\multicolumn{1}{|l|}{\textit{Security Goal Indicator Tree }\cite{peine2008security}}                                                                             & \multicolumn{1}{l|}{ }                               & \multicolumn{1}{l|}{ }                  & \multicolumn{1}{l|}{ }                                                                                                                  & \multicolumn{1}{l|}{D}                                                                                       & \multicolumn{1}{l|}{ } \\ \hline
\multicolumn{1}{|l|}{\textit{Goal Oriented AG} \cite{nanda2007highly}}                                                                                           & \multicolumn{1}{l|}{\OK}                               & \multicolumn{1}{l|}{ \OK}                  & \multicolumn{1}{l|}{ }                                                                                                                  & \multicolumn{1}{l|}{A}                                                                                       & \multicolumn{1}{l|}{ } \\ \hline
\multicolumn{1}{|l|}{\textit{Goal Oriented AG} \cite{liu2010goal}}                                                                                               & \multicolumn{1}{l|}{\OK}                               & \multicolumn{1}{l|}{\OK}                  & \multicolumn{1}{l|}{\OK}                                                                                                                  & \multicolumn{1}{l|}{A}                                                                                       & \multicolumn{1}{l|}{ } \\ \hline
\multicolumn{1}{|l|}{\textit{Goal Oriented AG} \cite{liu2015approach}}                                                                                           & \multicolumn{1}{l|}{ }                               & \multicolumn{1}{l|}{ }                  & \multicolumn{1}{l|}{\OK}                                                                                                                  & \multicolumn{1}{l|}{A}                                                                                       & \multicolumn{1}{l|}{ } \\ \hline
\multicolumn{1}{|l|}{\textit{Threat Logic Tree \cite{weiss1991system}; Attack Tree }\cite{schneier1999attack}}                                                   & \multicolumn{1}{l|}{}                               & \multicolumn{1}{l|}{}                  & \multicolumn{1}{l|}{ }                                                                                                                  & \multicolumn{1}{l|}{A}                                                                                       & \multicolumn{1}{l|}{ } \\ \hline
\multicolumn{1}{|l|}{\textit{Probabilistic Attack Tree}\cite{schneier1999attack}}                                                                                & \multicolumn{1}{l|}{\OK}                               & \multicolumn{1}{l|}{}                  & \multicolumn{1}{l|}{\OK}                                                                                                                  & \multicolumn{1}{l|}{A}                                                                                       & \multicolumn{1}{l|}{ } \\ \hline
\multicolumn{1}{|l|}{\textit{Attack Tree with Costs }\cite{schneier1999attack}}                                                                                  & \multicolumn{1}{l|}{}                               & \multicolumn{1}{l|}{ }                  & \multicolumn{1}{l|}{\OK}                                                                                                                  & \multicolumn{1}{l|}{A}                                                                                       & \multicolumn{1}{l|}{ } \\ \hline
\multicolumn{1}{|l|}{\textit{Multi-Parameter Attack Tree }\cite{buldas2006rational}}                                                                             & \multicolumn{1}{l|}{\OK}                               & \multicolumn{1}{l|}{}                  & \multicolumn{1}{l|}{\OK}                                                                                                                  & \multicolumn{1}{l|}{A}                                                                                       & \multicolumn{1}{l|}{ } \\ \hline
\multicolumn{1}{|l|}{\textit{Vulnerability Attack Tree }\cite{daley2002structural}}                                                                              & \multicolumn{1}{l|}{}                               & \multicolumn{1}{l|}{ }                  & \multicolumn{1}{l|}{}                                                                                                                  & \multicolumn{1}{l|}{A}                                                                                       & \multicolumn{1}{l|}{ } \\ \hline
\multicolumn{1}{|l|}{\textit{OWA Tree }\cite{yager2006owa}}                                                                                                      & \multicolumn{1}{l|}{\OK}                               & \multicolumn{1}{l|}{}                  & \multicolumn{1}{l|}{}                                                                                                                  & \multicolumn{1}{l|}{A}                                                                                       & \multicolumn{1}{l|}{ } \\ \hline
\multicolumn{1}{|l|}{\textit{Defense Tree }\cite{bistarelli2006defense}}                                                                                         & \multicolumn{1}{l|}{}                               & \multicolumn{1}{l|}{}                  & \multicolumn{1}{l|}{\OK}                                                                                                                  & \multicolumn{1}{l|}{AD}                                                                                       & \multicolumn{1}{l|}{ } \\ \hline
\multicolumn{1}{|l|}{\textit{Protection Tree }\cite{edge2006using}}                                                                                              & \multicolumn{1}{l|}{(\OK)\footnote{extended by \cite{kordy2013adtool}}}                               & \multicolumn{1}{l|}{}                  & \multicolumn{1}{l|}{(\OK)\footnote{extended by \cite{kordy2013adtool}}}                    & \multicolumn{1}{l|}{D}                                                                                       & \multicolumn{1}{l|}{ } \\ \hline
\multicolumn{1}{|l|}{\textit{Attack-Response-Tree }\cite{zonouz2013rre}}                                                                                         & \multicolumn{1}{l|}{\OK}                  & \multicolumn{1}{l|}{\OK}                                                                                                                  & \multicolumn{1}{l|}{\OK}    & \multicolumn{1}{l|}{AD}                                                                                       & \multicolumn{1}{l|}{ } \\ \hline
\multicolumn{1}{|l|}{\textit{Attack-Countermeasure Tree }\cite{roy2010cyber}}                                                                                    & \multicolumn{1}{l|}{\OK}                               & \multicolumn{1}{l|}{\OK}                  & \multicolumn{1}{l|}{\OK}                                                                                                                  & \multicolumn{1}{l|}{AD}                                                                                       & \multicolumn{1}{l|}{ } \\ \hline
\multicolumn{1}{|l|}{\textit{improved Attack-Defense Tree }\cite{wang2014threat}}                                                                                & \multicolumn{1}{l|}{\OK}                               & \multicolumn{1}{l|}{\OK}                  & \multicolumn{1}{l|}{\OK}                                                                                                                  & \multicolumn{1}{l|}{AD}                                                                                       & \multicolumn{1}{l|}{ } \\ \hline
\multicolumn{1}{|l|}{\textit{Attack-Defense Tree }\cite{kordy2014attackdef}}                                                                                     & \multicolumn{1}{l|}{\OK}                               & \multicolumn{1}{l|}{\OK}                  & \multicolumn{1}{l|}{\OK}                                                                                                                  & \multicolumn{1}{l|}{AD}                                                                                      & \multicolumn{1}{l|}{ } \\ \hline     
\multicolumn{1}{|l|}{\textit{Cyber Threat Tree }\cite{ongsakorn2010cyber}}                                                                                       & \multicolumn{1}{l|}{ }                               & \multicolumn{1}{l|}{}                  & \multicolumn{1}{l|}{\OK}                                                                                                                  & \multicolumn{1}{l|}{A}                                                                                       & \multicolumn{1}{l|}{ } \\ \hline
\multicolumn{1}{|l|}{\textit{Enhanced Attack Tree }\cite{camtepe2007modeling}}                                                                                   & \multicolumn{1}{l|}{}                               & \multicolumn{1}{l|}{}                  & \multicolumn{1}{l|}{}                                                                                                                  & \multicolumn{1}{l|}{A}                                                                                       & \multicolumn{1}{l|}{ } \\ \hline                                  
\multicolumn{1}{|l|}{\textit{Improved- }\cite{lv2011space}, \textit{SAND Attack Tree }\cite{jhawar2015attack}}                                                   & \multicolumn{1}{l|}{}                               & \multicolumn{1}{l|}{}                  & \multicolumn{1}{l|}{}                                                                                                                  & \multicolumn{1}{l|}{A}                                                                                       & \multicolumn{1}{l|}{ } \\ \hline
\multicolumn{1}{|l|}{\textit{Attack Tree with temporal order} \cite{jurgenson2009serial}}                                                                        & \multicolumn{1}{l|}{}                               & \multicolumn{1}{l|}{}                  & \multicolumn{1}{l|}{}                                                                                                                  & \multicolumn{1}{l|}{A}                                                                                       & \multicolumn{1}{l|}{ } \\ \hline
\multicolumn{1}{|l|}{\textit{Timed Probabilistic Semantics of AT} \cite{arnold2014time}}                                                                         & \multicolumn{1}{l|}{\OK}                               & \multicolumn{1}{l|}{}                  & \multicolumn{1}{l|}{}                                                                                                                  & \multicolumn{1}{l|}{A}                                                                                       & \multicolumn{1}{l|}{} \\ \hline
\multicolumn{1}{|l|}{\textit{Decorated Attack Tree }\cite{buldas2020attribute}}                                                                                  & \multicolumn{1}{l|}{}                               & \multicolumn{1}{l|}{}                  & \multicolumn{1}{l|}{\OK}                                                                                                                  & \multicolumn{1}{l|}{A}                                                                                       & \multicolumn{1}{l|}{ } \\ \hline
\multicolumn{1}{|l|}{\textit{Hyper-alert Correlation Graphs} \cite{ning2003learning, ning2004building}}                                                 & \multicolumn{1}{l|}{}                               & \multicolumn{1}{l|}{\OK}                  & \multicolumn{1}{l|}{}                                                                                                                  & \multicolumn{1}{l|}{A}                                                                                       & \multicolumn{1}{l|}{\OK} \\ \hline
\multicolumn{1}{|l|}{\textit{Hyper-Alert Graphs}\cite{zhu2006alert}}                                                                                             & \multicolumn{1}{l|}{ }                               & \multicolumn{1}{l|}{\OK}                  & \multicolumn{1}{l|}{}                                                                                                                  & \multicolumn{1}{l|}{A}                                                                                       & \multicolumn{1}{l|}{\OK} \\ \hline
\multicolumn{1}{|l|}{\textit{Alert Dependency Graph}\cite{roschke2011new}}                                                                                       & \multicolumn{1}{l|}{}                               & \multicolumn{1}{l|}{\OK}                  & \multicolumn{1}{l|}{}                                                                                                                  & \multicolumn{1}{l|}{A}                                                                                       & \multicolumn{1}{l|}{\OK} \\ \hline
\multicolumn{1}{|l|}{\textit{Attribute Attack Graph}\cite{chen2008automating}}                                                                                  & \multicolumn{1}{l|}{}                               & \multicolumn{1}{l|}{}                  & \multicolumn{1}{l|}{\OK}                                                                                                                  & \multicolumn{1}{l|}{A}                                                                                       & \multicolumn{1}{l|}{ } \\ \hline
\multicolumn{1}{|l|}{\textit{Stage Attribute Attack Graph}\cite{li2016optimized}}                                                                               & \multicolumn{1}{l|}{}                               & \multicolumn{1}{l|}{}                  & \multicolumn{1}{l|}{\OK}                                                                                                                  & \multicolumn{1}{l|}{A}                                                                                       & \multicolumn{1}{l|}{ } \\ \hline
\multicolumn{1}{|l|}{\textit{Activity Graph}\cite{staniford1996grids}}                                                                                          & \multicolumn{1}{l|}{}                               & \multicolumn{1}{l|}{(\OK\footnote{Early attack graph model for computer worm detection})}                  & \multicolumn{1}{l|}{}                                                                             & \multicolumn{1}{l|}{A}                                                                                       & \multicolumn{1}{l|}{\OK} \\ \hline
\multicolumn{1}{|l|}{\textit{Bayesian Defense Graph}\cite{sommestad2008combining}}                                                                              & \multicolumn{1}{l|}{\OK}                               & \multicolumn{1}{l|}{}                  & \multicolumn{1}{l|}{\OK}                                                                                                                  & \multicolumn{1}{l|}{AD}                                                                                       & \multicolumn{1}{l|}{ } \\ \hline
\multicolumn{1}{|l|}{\textit{Bayesian Attack Graph}\cite{poolsappasit2011dynamic}}                                                                              & \multicolumn{1}{l|}{\OK}                               & \multicolumn{1}{l|}{}                  & \multicolumn{1}{l|}{\OK}                                                                                                                  & \multicolumn{1}{l|}{AD}                                                                                       & \multicolumn{1}{l|}{ } \\ \hline
\end{tabular}
\caption{\textbf{Agents \& Attack Analysis Features}. Summary table part 1: taxonomy facilitating users in finding the right attack graph formalism and
representation - agent domains as well as analysis features} \label{tabbyyy:summary}
\end{table}

\begin{table}
\centering
\begin{tabular}{l|c|c|c|c|c|}
\hline
\multicolumn{1}{|l|}{\textbf{Attack Graph Type}}                                                                                                                 & \multicolumn{1}{l|}{Simul.}                               & \multicolumn{1}{l|}{IDS}                      &  \multicolumn{1}{l|}{Quant.}    &  \multicolumn{1}{l|}{Agent}  &  \multicolumn{1}{l|}{Gener.} \\ \hline  \hline
\multicolumn{1}{|l|}{\textit{Cloud-level-attack Graph}\cite{sun2015inferring}}                                                                                  & \multicolumn{1}{l|}{\OK}                               & \multicolumn{1}{l|}{}                  & \multicolumn{1}{l|}{\OK}                                                                                                                  & \multicolumn{1}{l|}{A}                                                                                       & \multicolumn{1}{l|}{\OK} \\ \hline
\multicolumn{1}{|l|}{\textit{Compromize Graph}\cite{mcqueen2006quantitative}}                                                                                   & \multicolumn{1}{l|}{\OK}                               & \multicolumn{1}{l|}{}                  & \multicolumn{1}{l|}{\OK}                                                                                                                  & \multicolumn{1}{l|}{A}                                                                                       & \multicolumn{1}{l|}{ } \\ \hline   
    \multicolumn{1}{|l|}{\textit{Coordinated Attack Graphs}\cite{braynov2003representation}}                                                                    & \multicolumn{1}{l|}{}                               & \multicolumn{1}{l|}{}                  & \multicolumn{1}{l|}{}                                                                                                                  & \multicolumn{1}{l|}{A}                                                                                       & \multicolumn{1}{l|}{ } \\ \hline
 \multicolumn{1}{|l|}{\textit{Countermeasure Graph}\cite{buldas2020attribute}}                                                                                  & \multicolumn{1}{l|}{ }                               & \multicolumn{1}{l|}{}                  & \multicolumn{1}{l|}{\OK}                                                                                                                  & \multicolumn{1}{l|}{AD}                                                                                       & \multicolumn{1}{l|}{ } \\ \hline
 \multicolumn{1}{|l|}{\textit{Exploit dependency graph by Amman}\cite{ammann2002scalable}}                                                                      & \multicolumn{1}{l|}{ }                               & \multicolumn{1}{l|}{}                  & \multicolumn{1}{l|}{ }                                                                                                                  & \multicolumn{1}{l|}{A}                                                                                       & \multicolumn{1}{l|}{\OK} \\ \hline
 \multicolumn{1}{|l|}{\textit{(Exploit) Dependency Graph by Noel}\cite{noel2004managing}}                                                                       & \multicolumn{1}{l|}{\OK}                               & \multicolumn{1}{l|}{}                  & \multicolumn{1}{l|}{\OK}                                                                                                                  & \multicolumn{1}{l|}{A}                                                                                       & \multicolumn{1}{l|}{\OK} \\ \hline
 \multicolumn{1}{|l|}{\textit{Vulnerability Dependency Graph}\cite{serra2015pareto}}                                                                            & \multicolumn{1}{l|}{}                               & \multicolumn{1}{l|}{}                  & \multicolumn{1}{l|}{\OK}                                                                                                                  & \multicolumn{1}{l|}{A}                                                                                       & \multicolumn{1}{l|}{ } \\ \hline
  \multicolumn{1}{|l|}{\textit{Generalized Dependency Graph }\cite{10.5555/2041225.2041255}}                                                                     & \multicolumn{1}{l|}{}                               & \multicolumn{1}{l|}{}                  & \multicolumn{1}{l|}{\OK}                                                                                                                  & \multicolumn{1}{l|}{A}                                                                                       & \multicolumn{1}{l|}{ } \\ \hline
 \multicolumn{1}{|l|}{\textit{Probabilistic Temporal Attack Graphs }\cite{10.5555/2041225.2041255}}                                                             & \multicolumn{1}{l|}{\OK}                               & \multicolumn{1}{l|}{}                  & \multicolumn{1}{l|}{\OK}                                                                                                                  & \multicolumn{1}{l|}{A}                                                                                       & \multicolumn{1}{l|}{ } \\ \hline
 \multicolumn{1}{|l|}{\textit{Attack Scenario Graph }\cite{10.5555/2041225.2041255}}                                                                            & \multicolumn{1}{l|}{ }                               & \multicolumn{1}{l|}{\OK}                  & \multicolumn{1}{l|}{\OK}                                                                                                                  & \multicolumn{1}{l|}{A}                                                                                       & \multicolumn{1}{l|}{ } \\ \hline
 \multicolumn{1}{|l|}{\textit{Evidence Graph }\cite{wang2008graph}}                                                                                             & \multicolumn{1}{l|}{}                               & \multicolumn{1}{l|}{\OK}                  & \multicolumn{1}{l|}{ }                                                                                                                  & \multicolumn{1}{l|}{A}                                                                                       & \multicolumn{1}{l|}{ } \\ \hline
 \multicolumn{1}{|l|}{\textit{Exploitation Graph }\cite{li2006approach}}                                                                                        & \multicolumn{1}{l|}{ }                               & \multicolumn{1}{l|}{}                  & \multicolumn{1}{l|}{ }                                                                                                                  & \multicolumn{1}{l|}{A}                                                                                       & \multicolumn{1}{l|}{\OK} \\ \hline
 \multicolumn{1}{|l|}{\textit{Full attack Graph }\cite{ingols2006practical}}                                                                                    & \multicolumn{1}{l|}{ }                               & \multicolumn{1}{l|}{}                  & \multicolumn{1}{l|}{}                                                                                                                  & \multicolumn{1}{l|}{A}                                                                                       & \multicolumn{1}{l|}{\OK} \\ \hline
  \multicolumn{1}{|l|}{\textit{Predictive Attack Graph }\cite{ingols2006practical}}                                                                             & \multicolumn{1}{l|}{ }                               & \multicolumn{1}{l|}{}                  & \multicolumn{1}{l|}{}                                                                                                                  & \multicolumn{1}{l|}{A}                                                                                       & \multicolumn{1}{l|}{\OK} \\ \hline
  \multicolumn{1}{|l|}{\textit{Node-Predictive Attack Graph }\cite{ingols2006practical}}                                                                        & \multicolumn{1}{l|}{}                               & \multicolumn{1}{l|}{}                  & \multicolumn{1}{l|}{ }                                                                                                                  & \multicolumn{1}{l|}{A}                                                                                       & \multicolumn{1}{l|}{\OK} \\ \hline
  \multicolumn{1}{|l|}{\textit{Host-compromized attack Graph }\cite{lippmann2005evaluating}}                                                                    & \multicolumn{1}{l|}{ }                               & \multicolumn{1}{l|}{}                  & \multicolumn{1}{l|}{}                                                                                                                  & \multicolumn{1}{l|}{A}                                                                                       & \multicolumn{1}{l|}{\OK} \\ \hline
  \multicolumn{1}{|l|}{\textit{Host-access Graph }\cite{ammann2005host}}                                                                       & \multicolumn{1}{l|}{ }                               & \multicolumn{1}{l|}{}                  & \multicolumn{1}{l|}{}                                                                                                                  & \multicolumn{1}{l|}{A}                                                                                       & \multicolumn{1}{l|}{\OK} \\ \hline
  \multicolumn{1}{|l|}{\textit{Host-based attack Graph }\cite{malthotra2008host}}                                                                               & \multicolumn{1}{l|}{\OK}                               & \multicolumn{1}{l|}{}                  & \multicolumn{1}{l|}{\OK}                                                                                                                  & \multicolumn{1}{l|}{A}                                                                                       & \multicolumn{1}{l|}{ } \\ \hline 
  \multicolumn{1}{|l|}{\textit{Host-centric Attack Graph }\cite{hewett2008host}}                                                                               & \multicolumn{1}{l|}{ }                               & \multicolumn{1}{l|}{}                  & \multicolumn{1}{l|}{\OK}                                                                                                                  & \multicolumn{1}{l|}{A}                                                                                       & \multicolumn{1}{l|}{\OK} \\ \hline 
  \multicolumn{1}{|l|}{\textit{Multiple Prerequisites graph }\cite{ingols2006practical}}                                                                      & \multicolumn{1}{l|}{ }                               & \multicolumn{1}{l|}{}                  & \multicolumn{1}{l|}{ }                                                                                                                  & \multicolumn{1}{l|}{A}                                                                                       & \multicolumn{1}{l|}{\OK} \\ \hline
  \multicolumn{1}{|l|}{\textit{Hybrid attack Graph }\cite{louthan2014hybrid}}                                                                                 & \multicolumn{1}{l|}{\OK}                               & \multicolumn{1}{l|}{}                  & \multicolumn{1}{l|}{ }                                                                                                                  & \multicolumn{1}{l|}{A}                                                                                       & \multicolumn{1}{l|}{ } \\ \hline
  \multicolumn{1}{|l|}{\textit{Impact Dependency Graph }\cite{jakobson2011mission}}                                                                             & \multicolumn{1}{l|}{\OK}                               & \multicolumn{1}{l|}{}                  & \multicolumn{1}{l|}{\OK}                                                                                                                  & \multicolumn{1}{l|}{A}                                                                                       & \multicolumn{1}{l|}{} \\ \hline
  \multicolumn{1}{|l|}{\textit{Mission Dependency Graph } \cite{musman2009}}                                                                            & \multicolumn{1}{l|}{\OK}                               & \multicolumn{1}{l|}{}                  & \multicolumn{1}{l|}{\OK}                                                                                                                  & \multicolumn{1}{l|}{A}                                                                                       & \multicolumn{1}{l|}{} \\ \hline
  \multicolumn{1}{|l|}{\textit{Mission Impact Graph} \cite{sun2017towards}}                                                                                & \multicolumn{1}{l|}{\OK}                               & \multicolumn{1}{l|}{}                  & \multicolumn{1}{l|}{\OK}                                                                                                                  & \multicolumn{1}{l|}{A}                                                                                       & \multicolumn{1}{l|}{\OK} \\ \hline
  \multicolumn{1}{|l|}{\textit{Insecurity Flow} \cite{moskowitz1998insecurity}}                                                                                      & \multicolumn{1}{l|}{\OK}                               & \multicolumn{1}{l|}{}                  & \multicolumn{1}{l|}{\OK}                                                                                                                  & \multicolumn{1}{l|}{A}                                                                                       & \multicolumn{1}{l|}{} \\ \hline 
  \multicolumn{1}{|l|}{\textit{Risk Flow Attack Graph }\cite{dai2015exploring}}                                                                               & \multicolumn{1}{l|}{}                               & \multicolumn{1}{l|}{}                  & \multicolumn{1}{l|}{\OK}                                                                                                                  & \multicolumn{1}{l|}{A}                                                                                       & \multicolumn{1}{l|}{} \\ \hline
  \multicolumn{1}{|l|}{\textit{Intrusion DAG}  \cite{wu2003adepts}}                                                                                       & \multicolumn{1}{l|}{}                               & \multicolumn{1}{l|}{\OK}                  & \multicolumn{1}{l|}{}                                                                                                                  & \multicolumn{1}{l|}{A}                                                                                       & \multicolumn{1}{l|}{} \\ \hline
  \multicolumn{1}{|l|}{\textit{Intrusion Graph} \cite{foo2005adepts} }                                                                                     & \multicolumn{1}{l|}{\OK}                               & \multicolumn{1}{l|}{\OK}                  & \multicolumn{1}{l|}{\OK}                                                                                                                  & \multicolumn{1}{l|}{A}                                                                                       & \multicolumn{1}{l|}{} \\ \hline
  \multicolumn{1}{|l|}{\textit{Knowledge Graph } \cite{althebyan2007knowledge}}                                                                                     & \multicolumn{1}{l|}{ }                               & \multicolumn{1}{l|}{}                  & \multicolumn{1}{l|}{ }                                                                                                                  & \multicolumn{1}{l|}{A}                                                                                       & \multicolumn{1}{l|}{ } \\ \hline
  \multicolumn{1}{|l|}{\textit{Knowledge Bayesian Attack Graph} \cite{althebyan2008knowledge}}                                                                      & \multicolumn{1}{l|}{\OK}                               & \multicolumn{1}{l|}{}                  & \multicolumn{1}{l|}{\OK}                                                                                                                  & \multicolumn{1}{l|}{A}                                                                                       & \multicolumn{1}{l|}{ } \\ \hline
  \multicolumn{1}{|l|}{\textit{Logical Attack Graphs }\cite{ou2005mulval, ou2006scalable}}                                                                               & \multicolumn{1}{l|}{ }                               & \multicolumn{1}{l|}{}                  & \multicolumn{1}{l|}{  }                                                                                                                  & \multicolumn{1}{l|}{A}                                                                                       & \multicolumn{1}{l|}{\OK} \\ \hline
  \multicolumn{1}{|l|}{\textit{Personalized Attack Graphs} \cite{roberts2011personalized, urbanska2013accepting}}                                                                          & \multicolumn{1}{l|}{}                               & \multicolumn{1}{l|}{}                  & \multicolumn{1}{l|}{\OK}                                                                                                                  & \multicolumn{1}{l|}{A}                                                                                       & \multicolumn{1}{l|}{(\OK)\footnote{The attack graph is automatically personalized/customized from a raw, large attack graph, by considering the system/user properties}} \\ \hline
   \multicolumn{1}{|l|}{\textit{Privilege Graph}  \cite{dacier1994privilege} \cite{dacier1996models}}                                                                                    & \multicolumn{1}{l|}{\OK}                               & \multicolumn{1}{l|}{}                  & \multicolumn{1}{l|}{\OK}                                                                                                                  & \multicolumn{1}{l|}{A}                                                                                       & \multicolumn{1}{l|}{ } \\ \hline
   \multicolumn{1}{|l|}{\textit{Scenario Attacks; Requires-Provides Model} \cite{templeton2001requires}}                                                          & \multicolumn{1}{l|}{}                               & \multicolumn{1}{l|}{\OK}                  & \multicolumn{1}{l|}{}                                                                                                                  & \multicolumn{1}{l|}{A}                                                                                       & \multicolumn{1}{l|}{\OK} \\ \hline
   \multicolumn{1}{|l|}{\textit{Scenario Graph} \cite{sheyner2002automated, sheyner2004scenario}}                                                                                    & \multicolumn{1}{l|}{ }                               & \multicolumn{1}{l|}{}                  & \multicolumn{1}{l|}{ }                                                                                                                  & \multicolumn{1}{l|}{A}                                                                                       & \multicolumn{1}{l|}{\OK} \\ \hline
   \multicolumn{1}{|l|}{\textit{Probabilistic Attack Graph} \cite{sheyner2004scenario, jha2002mini}}                                                                         & \multicolumn{1}{l|}{\OK}                               & \multicolumn{1}{l|}{}                  & \multicolumn{1}{l|}{ }                                                                                                                  & \multicolumn{1}{l|}{A}                                                                                       & \multicolumn{1}{l|}{\OK\footnote{In a nutshell probabilistic attack graphs are a scenario graph with probability labels and are thus automatically generated in the same way as scenario graphs}} \\ \hline
   \multicolumn{1}{|l|}{\textit{State Transition Diagram} \cite{ilgun1995state}}                                                                         & \multicolumn{1}{l|}{ }                               & \multicolumn{1}{l|}{\OK}                  & \multicolumn{1}{l|}{ }                                                                                                                  & \multicolumn{1}{l|}{A}                                                                                       & \multicolumn{1}{l|}{} \\ \hline  
   \multicolumn{1}{|l|}{\textit{Vulnerability Graph }\cite{birkholz2010efficient}}                                                                              & \multicolumn{1}{l|}{ }                               & \multicolumn{1}{l|}{}                  & \multicolumn{1}{l|}{\OK}                                                                                                                  & \multicolumn{1}{l|}{A}                                                                                       & \multicolumn{1}{l|}{ } \\ \hline
   \multicolumn{1}{|l|}{\textit{Vulnerability Cause Graph}\cite{ardi2006towards}}                                                                        & \multicolumn{1}{l|}{ }                               & \multicolumn{1}{l|}{}                  & \multicolumn{1}{l|}{ }                                                                                                                  & \multicolumn{1}{l|}{A}                                                                                       & \multicolumn{1}{l|}{} \\ \hline
   \multicolumn{1}{|l|}{\textit{Security Activity Graph }\cite{ardi2006towards}\cite{byers2008cause}}                                                                           & \multicolumn{1}{l|}{}                               & \multicolumn{1}{l|}{}                  & \multicolumn{1}{l|}{ }                                                                                                                  & \multicolumn{1}{l|}{AD}                                                                                       & \multicolumn{1}{l|}{} \\ \hline
   \multicolumn{1}{|l|}{\textit{Security Goal Model  } \cite{byers2010unified}}                                                                             & \multicolumn{1}{l|}{}                               & \multicolumn{1}{l|}{}                  & \multicolumn{1}{l|}{ }                                                                                                                  & \multicolumn{1}{l|}{AD}                                                                                       & \multicolumn{1}{l|}{} \\ \hline
\end{tabular}
\caption{\textbf{Agents \& Attack Analysis Features}. Summary table part 2: taxonomy facilitating users in finding the right attack graph formalism - agent domains as well as analysis features.} \label{tabbyy:summary}
\end{table}

\section{Conclusion and open research questions} \label{sec:conclusion}
\subsection{Summary}
We present a taxonomy of attack graph research based on the knowledge representation in the attack graph model and analyse existing literature within these categories. We describe, compare and contrast the representations, form categories and put them into context with existing research. The resulting taxonomy facilitates the user in finding the right attack graph formalism and representation based on attack graph representation, agents involved as well as analysis features and automatic generation. 

We performed a systematic review of literature, involving overall 70 attack graph representations with respect to knowledge representations. We included attack graph types with a designated name in our research; models with the generic name called "attack graph" were not considered to avoid ambiguity. We describe each of the formalisms in detail in \ref{sec:knowledge_rep} and depict its use in a common use case (\emph{the malicious insider}) whenever applicable. 
We categorized the literature items with respect to these aspects and summarized the resulting categories in our taxonomy.

\subsection{Open Directions for Future Research}
Apart from the usual attack graph research questions, which mainly address scalability of attack graph generation and analysis, as well as usability and practical applications, our analysis has also revealed some less obvious research gaps and directions for future research. 

First of all, it must be mentioned that among all studied attack graph formalisms, no defender model supports automatic generation. Thus, an extension of the automatic graph generation formalism from matching vulnerabilities to exploits, towards automatic matching vulnerabilities/exploits and patches poses an open research question. While attack graphs are used for network hardening, the proposal of concrete defensive measures falls short. This especially includes the use of deception resources for defense purposes, which, unfortunately, is neglected in the majority of network hardening contexts in attack graphs. 

 Second; currently, automatic attack graph generation usually involves data from vulnerability scans only, which excludes many other data sources. This results in non-included attacks, including mainly soft attacks (e.g. social engineering, phishing). Natural language processing might become handy to identify and classify attacks from text sources and integrate them into attack graph models. The use of generative AI models this purpose might also be studied in this context to propose additional vulnerabilities present in similar systems. 

Third, it has to be mentioned that the majority of graph formalisms are design for IT-systems only. IoT and integrated systems need different models covering the virtual as well as the physical attack and effect surface. The design of graph models and analysis methods for safety and security analysis of cyber physical systems pose another open question for future research.

\section*{Acknowledgements}
This work was funded by the Karl-Popper-Kolleg “Responsible Safe and Secure Robotic Systems Engineering (SEEROSE)”, at the University of Klagenfurt (AAU). The author would like to thank her supervisor W. Faber and collegue P. Schartner for his support. Part of this research was conducted at a fellowship at Northwestern University, Northwestern Security and Artificial Intelligence Laboratory. Special thanks to V.S. Subrahmanian for this opportunity, guidance and the helpful discussions. 

\bibliographystyle{unsrt} 
\bibliography{literatur}

\newpage 

\section{Supplementary Material: Attack Graph Models} \label{sec:knowledge_rep}
In this section, we discuss the types of attack graphs on a higher level and highlight their particularities. 
After a short introduction on the early development of attack graphs in Section, the analyzed graph models are presented in clustered alphabetical order, i.e. they are presended in alphabetical order, unless they are strongly connected to other attack graph types (e.g. they were presented in the same paper). As mentioned before, we restrict our analysis towards attack graph formalisms with a special name; formalisms simply called "attack graph" are not considered in our anaylsis. 

Throughout the whole section, a running example facilitates in depicting the individual knowledge representations for the graphical cyber-attack models under scrutiny. 

\subsection{Historic Development of Graph Models for Cyber-Security}
\label{sec:evolution_of_ag}
\subsubsection{Towards Model-based Security}
\color{black}
The development of attack graphs is linked to advances in intrusion detection systems (IDS)
in the 1990s - we refer the interested reader to \cite{mukherjee1994network}
for an overview of early IDS systems. Until then, intrusion detection approaches only existed in two rudimentary forms \cite{ilgun1995state}: threshold detection and anomality detection. \emph{Threshold detection} based methods used summary statistics to identify unnaturally high occurrences of events, as in MIDAS\cite{arlowe1990mobile}, whereas \emph{anomaly detection tools}, as introduced in \cite{anderson1980computer} build usage patterns from past user activity collected over time. Anomality detection approaches were either implemented using rule-based logic systems (as in W\&S \cite{redle1992wisdom}, TIM \cite{chen1988inductive, chen1990adaptive}), often in combination statistical inference (IDES \cite{lunt1992real}, NIDES \cite{javitz1993nides}). \cite{lunt1992real} even proposed the use of neural network approaches in 1992. 

While statistical methods rather served as a complement, rule-based systems back then were developed to recognize suspicious user activity. They have been successfully employed to recognize singe auditable events and even sequences of events, representing  whole penetration scenarios \cite{ilgun1995state}. Despite these preliminary successes, rule-based approaches so far suffered from their dependence on past audit records represented in the rule base. As these approaches utilize and analyze specific user behaviour patterns, slight variations of the intrusion scenario often produce different audit record sequences that go unnoticed by such systems. 

This situation motivated the development higher-level, audit-record-independent representations of intrusion scenarios, which led to the development of model-based intrusion detection systems \cite{garvey1991model, lunt1992real}. In these works, the authors propose building abstract models for normal and abnormal user behaviour based on evidential reasoning (Dempster-Shafer Theory): Lunt et al. \cite{lunt1992real} suggest constructing frames of discernment for each unknown and connecting them using compatibility relations. Frames and compatibility relations form a network structure that expresses, which phenomena are likely to co-occur. Inserting data into the model, their system decides, whether the observed pattern matches normal or abnormal user behaviour. The authors stress that model-based reasoning allows for more intuitive explanations of alerts, as the observations can be related to the defined intrusion scenarios. The first model-based IDS emerged - and the first graph based formalisms and analysis tools followed soon after: \textit{Privilege graphs} by Dacier and Deswarte \cite{dacier1994privilege} mark the origin of attack-graph literature. The \emph{privilege graph} is a directed graph, whose nodes represent sets of privilege inheritance relations on sets of objects. 
Dacier and Deswarte \cite{dacier1994privilege} call for an incorporation of their scheme in IDS as well as an extension including a quantification of the difficulty of privilege escalation by incorporating weights in the privilege graph for a "quantitative evaluation of operational security". Unlike their wish, the first attack-graph analysis tool to be incorporated in an IDS, however, was called STAT \emph{state transition analysis}, developed by Ilgun et al. \cite{ilgun1995state}. 
s 
which was later further developed into NSTAT\cite{kemmerer1997nstat} and NetSTAT \cite{vigna1998netstat}. 

Nontheless, Dacier et al. \cite{dacier1996models} extended their privilege graph: now each arc in the graph refers to a specific exploit, while nodes still correspond to privileges. For quantitative security analysis, their privilege graph is transformed into a Stochastic Petri Net. The authors suggest to analyse the corresponding Markov process corresponding to the reachability graph to assess the risks induced by the residual flaws 
and Ortalo \& Deswarte \cite{ortalo1998quantitative} provide the first real-world graph-based quantitative security analysis use cases based on this formalism (a bank holding 30 employees in a rural area). Later, in \cite{ortalo1999experimenting}, they analysed a privilege graph model constructed from 13 major UNIX vulnerabilities. 
\begin{figure*}[ht]
    \centering
    \includegraphics[width=0.95\textwidth]{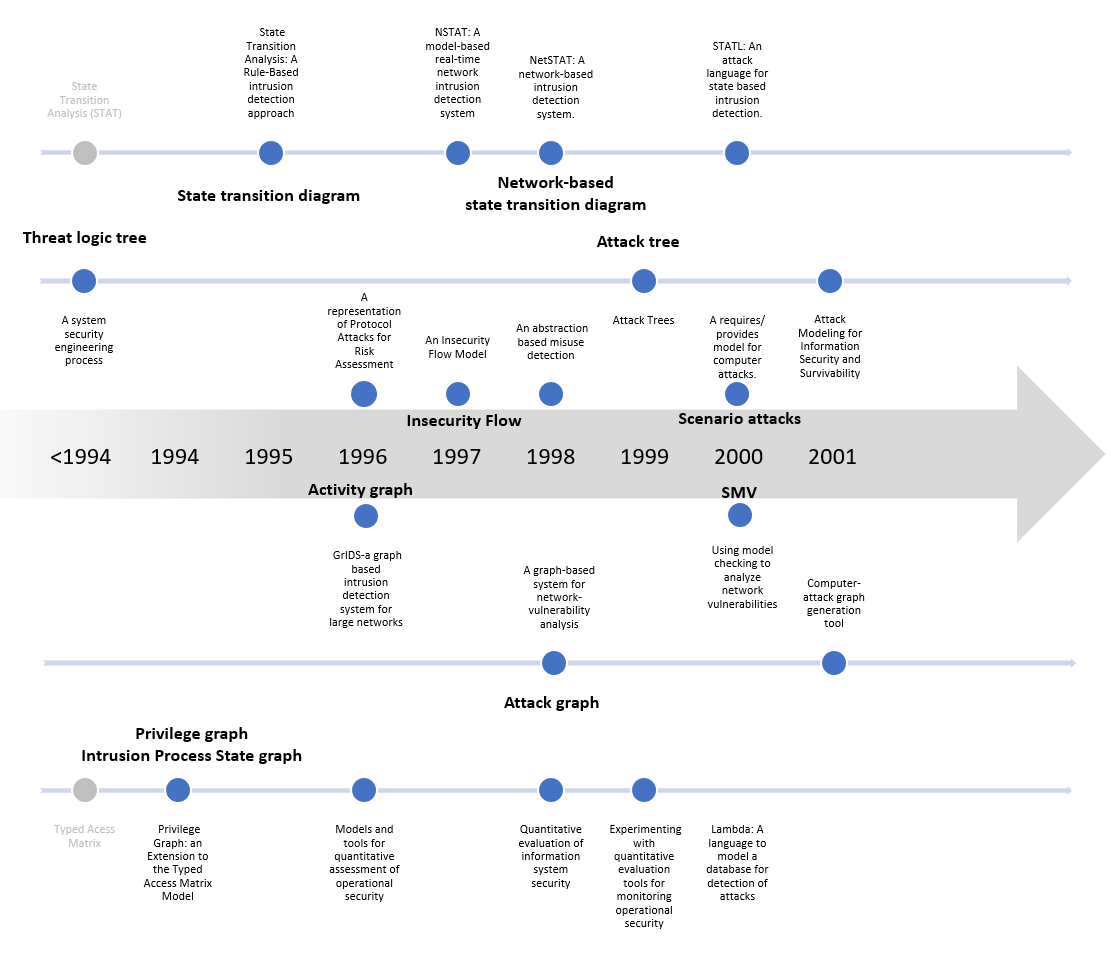}
    \caption{The Evolution of attack graphs: an overview of the first attack graph models}
    \label{fig:firstAG}
\end{figure*}

Other graph based formalisms for security that developed at about the same time, were GrIDS \cite{staniford1996grids} - the graph based intrusion detection system, Moskowitz et al.'s \cite{moskowitz1998insecurity} \emph{insecurity flows} to assess network hardening measures, 
or Meadows' representation of protocol attacks for risk assessment \cite{Meadows1996ARO} - a high level graphical model to represent attacks on cryptographic protocols.  

At the same time, Philips and Swiler \cite{phillips1998graph} developed model, for which they chose the well-known term \emph{attack graph}. In their model, nodes identify a stage of attack or its postcondition and arcs represent changes of states either caused by attacks or other actions, which makes this model somehow more general than \cite{dacier1996models}. Instead of using a Markow-Chain model, the authors directly assign probabilities of success as arc weights and propose the use of graph algorithms (e.g., shortest-path or technique of \cite{naor1993suboptimal}, computing a compact representation of all paths that are within $\delta$ optimal) to determine attack paths highest chances of success. 
Philips and Swiler also developed a tool \cite{phillips1998graph} for semi-automatic attack policy generation: "Though attack templates represent pieces of known attacks or hypothesized methods of moving from one state to another, their combinations can lead to descriptions of new attacks." 
These templates were later formalized in attack description languages, such as \emph{STATL}\cite{eckmann2002statl}, a domain-independent attack description language for the STAT family, MuSigs \cite{lin1998abstraction} and its successor \cite{ning2001abstraction}, further facilitated the \textit{automatisation} (cf. Section \ref{sec:aspect4}).


\subsubsection{Attack Trees and Schneier - Establishing and Popularizing Graph based Security Models} \label{sec:attack_trees}
The popularization of graph models for cyber-attacks, however, did not happened until 1998, when Salter et al. \cite{salter1998toward} and Schneier \cite{schneier1999attack} published articles on \emph{attack trees}. 
Note that the exact same formalism was initially proposed by Weiss \cite{weiss1991system} in 1991, though the formalism remained largely unnoticed until Schneier's paper. 
A rigorous treatment of the attack-tree formalism is presented in \cite{moore2001attack} as well as \cite{mauw2006foundations}. \cite{kordy2014dag} provides a sound overview of attack-tree types in a designated section.  
\color{black}

\subsubsection{Attack Graph Development after 2000 - Diversification}
With the popularization of graph models for cyber-attacks since the early 2000s, the number of designated attack graph model formalisms and related research items exploded. Like a tree, this field of research developed many individual branches became more diversified, making it hard for us to proceed with elaborating on attack graph development in a chronological manner: while one significant part of the research community shifted their focus towards \textit{automatic attack graph generation}, others focus on \textit{usability aspects}, \textit{analysis for security} or creating suitable \textit{representations} for various kinds of applications. We refer the reader to Section \ref{sec:related} to related work at this point. 

In this paper, we decided to present the reviewed literature in a clustered alphabetical order, cf. Section \ref{graphmodelsforcs}.  Throughout the whole section, a running example facilitates in depicting the individual knowledge representations for the graphical cyber-attack models under scrutiny. 

\subsection{Running example - the malicious insider}
The following example was adapted from \cite{LALLIE2020100219}. The description of this simple cyber-attack scenario facilitates in depicting the individual graph representations. Note that the vulnerabilities included in our example are dated and are rarely exploited nowadays. This is for ethical reasons: we do not want to provide a valid "recipe" how to compromise a system in 50+ formalisms, but solely depict the use of the formalism for defensive and network hardening purposes. 

Consider a simple network consisting of three computers, all equipped with some ssh client and daemon. Assume that an insider, who is also an authorized user from host one, aims to gain user privilege on host three. The malicious attacker is able to perform sshd buffer overflow attacks in order to acquire user privileges on reachable hosts. In this context, the attacker has two options to capture host three: either, he can attack directly, or he chooses to attack host two first, and then operates from host two to host three (pivoting). Whenever necessary, we assume the specific vulnerability is in the challenge-response authentication, as present in older OpenSSH versions, such as 2.3.1. 

Furthermore the attacker may seek to get physical access to the machine. The malicious individual considers the use of a hardware-keylogger to acquire the user credentials. The device is plugged between the keyboard and the computer, and removed at some later future point in time.
Each attack vector has its preconditions: the daemons serving incoming SSH connections of the targeted machines have to be running and accessible from the attackers ssh client in order for the attack to be successful. Depending on the connectivity, either path may lead to success or failure of the attack. For the hardware-keylogger attack, the machine needs to be physically accessible.

\begin{figure}[h!]
    \centering
    \includegraphics[scale=0.3]{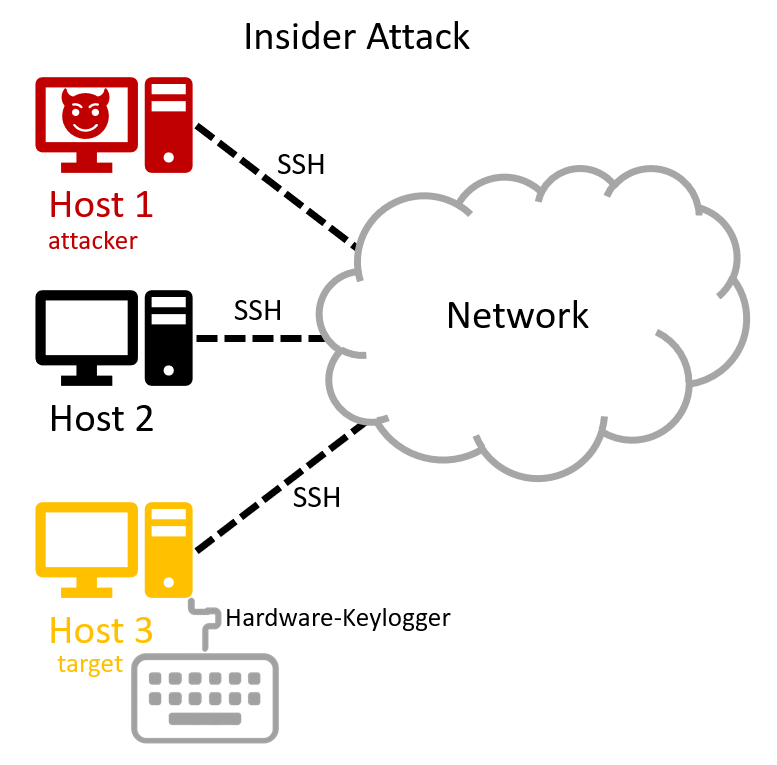}
    \caption{A running example cyber-attack scenario }
\label{fig:running_ex}
\end{figure}

Figure \ref{fig:running_ex} depicts the cyber-attack scenario. 
Whenever applicable, we apply the use case to the presented formalism: i.e. the majority of the graphics included are not just copies of the original attack graph figures from the sources, but applications thereof, tailored to the use case "the malicious insider". Whenever necessary, we will add additional information to the attack graph when needed.

In a few cases, however, only the \textit{schema} of the attack graph formalism is depicted. In these cases, we use the original images. This is the case, e.g., when the use case cannot be modelled in terms of the attack graph formalism, because it does not include the features under study (e.g. cloud) or whenever we believe the schematic figure captures the ideas of the formalism in a clear and comprehensive manner. Some historic attack graph formalism were also from the original sources. We label these graphics as "original graphics" in the figure heading. All other graphics were manually created or edited by the author to fit the use case. 
 
\newpage

\section{Graph Models for Cybersecurity} \label{graphmodelsforcs}
\begin{center}\textbf{Index}
\end{center}
\startcontents[sections]
\printcontents[sections]{}{2}[3]{}

\subsection{Anti-Goal Graph, intention-centric and goal oriented attack graphs}
According to \cite{peine2008security}, there exist two approaches to software security in software engineering: in the vulnerability driven approach, security manifests itself through the absence of predefined vulnerabilities. In this bottom-up manner, vulnerabilities are identified and avoided; security is achieved, in a bottom-up manner. In the goal-driven approach to security, the overall goal is defined first and the measures to define it are refined in a top-down manner. In this Section, the attack graph structures designated as goal-orientated are discussed.

\subsubsection{Intention-centric approach}
Depite not being designates as "goal-centric", Gorodetski and Kotenko's \cite{gorodetski2002attacks} \emph{intention-centric approach} falls into this category. Gorodetski and Kotenko \cite{gorodetski2002attacks} developed the "Computer Network Attacks" ontology to describe the domain of the adversary, as well as an attack simulator to emulate the cat-and-mouse dynamic of an attacker and network security system. Resembling the goal-oriented approach, their model is based on the observation that "any attack is target and intention centered", and the term intention is used to describe as a (sub-) goal, an attacker seeks to reach, such as reconnaissance, file access, penetrating the network etc. This graph model is especially useful for all practitioners that look for a quantitative simulation environment that also offers a rich semantic to relate attacker sub-goals (i.e. intentions) to each other - a feature most quantitative frameworks lack.

Attack intentions by Kotenko et al. are described as partially ordered sets of lower-level intentions and associated, randomized attacker actions to represent a variation attackers. Their tree-based attack strategy representation is based on stochastic context free grammars interconnected via the "grammar substitution" operator. 
The developed ontology realizes a hierarchy of attacker activities on different levels of detail, ranging from high-level intentions to macro- and micro-level description:
\begin{itemize}
\item high-level intentions can be chosen from numerous predefined intentions, either for
\begin{itemize}
    \item R - Reconnaissance (such as IH - identifications of hosts, IO - identification of host operating system) or 
    \item I - Implantation and threat realization, such as EP- escalating Privilege and CBD (creating back doors)
\end{itemize}
    \item Upper levels and lower levels are connected though "Part of", "Kind of" or "Seq of" relationships, representing decomposition, specialisation or temporal (for sequences of actions) relations 
    \item on the micro level, the "example of" relationships offers to incorporate specific exploits
    \item lower level intentions are matched to attack task specifications, realizing the lower-level intention. They are defined as quadruples $<Network~ (host)~addres$,$~attacker~intention$, $~known data,~attack~object>$, the latter being either unspecified (if there is none, such as for reconnaissance intentions, "All", "Anyone" or a specific attack
\end{itemize}
\begin{figure*}[htb]
\centering
    \subfloat[Micro-level of "Computer networks" ontology by \cite{gorodetski2002attacks}: the "Example of" relationships incorporates specific exploits; "Kind of" represents a subcategory of an attacker intention. Here, the attacker performs a TCP port scan with nmap, checking for connectivity \texttt{sshd(1,3)} of the target machine on the default port 22]{%
    \includegraphics[width=.35\textwidth]{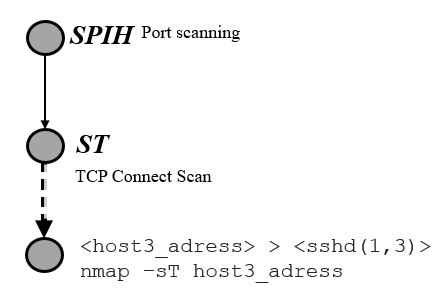}}\qquad
     \subfloat[Domain ontology model by \cite{gorodetski2002attacks} on a macro-level, depicting the top- and intermediate-level attacker intentions. For our use case, the attacker wants to check the SSH connectivity between host 3 and host 1 as before mounting his attack by scanning the default port 22 of the target host. The according path is highlighted in orange in the ontology model.]{%
    \includegraphics[width=.8\textwidth]{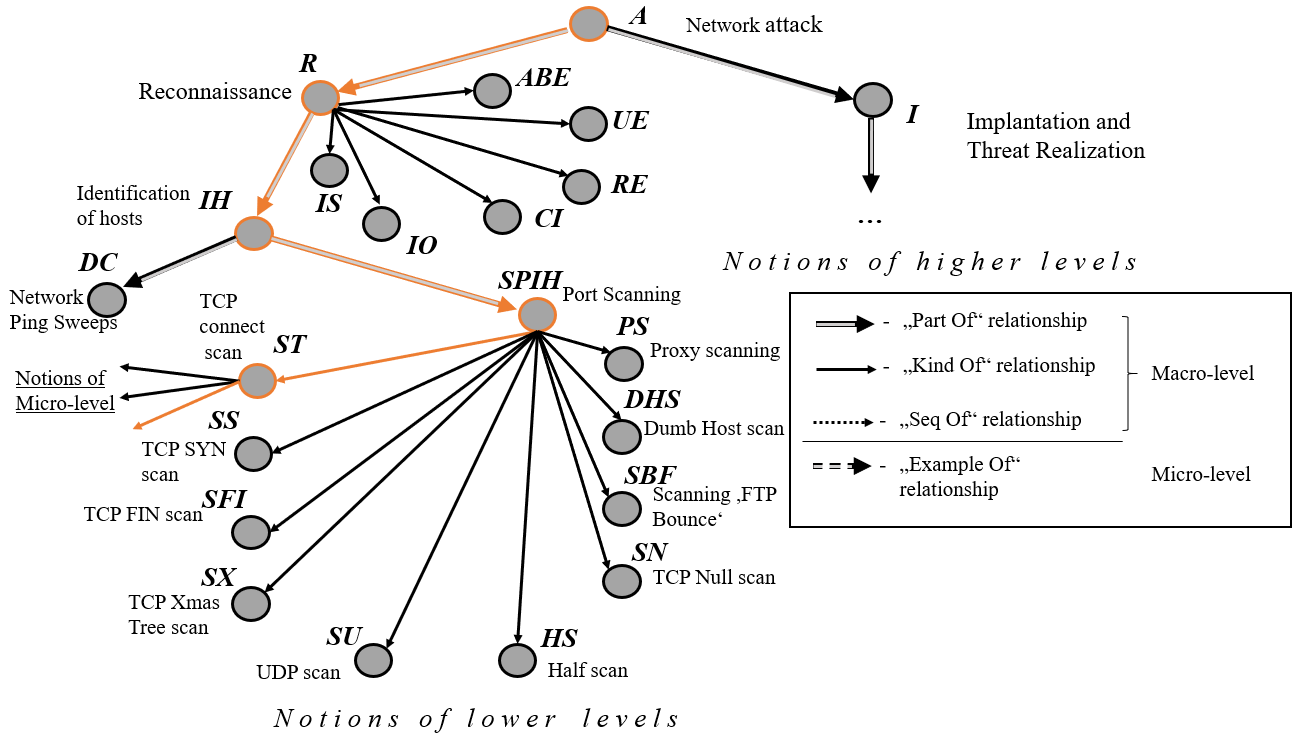}}\hfill
  \caption{Intention-oriented attack model by \cite{gorodetski2002attacks} adapted to current scenario }\label{fig:attackgraph_goaloriented}
\end{figure*}
Additionally, a model of the attacked computer network, consisting of computer network structure, host resources, attack success probabilities, model of host reaction is provided and its specification is needed for attack simulation. 
Based on the provided grammar and the computer network model, attack simulations is performed by randomizing from the possible attack intentions $X_1, X_2, ... $ and targets $Y_1, Y_2,..$ in the hierarchy $<attacker~intention~X; attack~target~Y> \to <lower-level~intention> \to actions$. As only ontology nodes of the lowest level (i.e. terminal actions) are associated with the probability of success, the network returns "success" or "failure" after each step of the simulation. This poses a potential limitation, since the aim of the tool is solely to simulate attacks, but not to explicitly quantify the overall success/severity of an attack.

\subsubsection{Anti-Goal Graphs}The goal-oriented approach for attack description stems from requirements engineering methods and was first introduced in \cite{van2003system}: "a goal is a prescriptive statement of intent about some system [...] whose satisfaction in general requires the cooperation of some of the agents forming that system." Anti-goals, on the other hand, are intentional obstacles to security goals, and can be obtained by negation of confidentiality, privacy, availability or integrity goals. Just as in attack trees, in anti-goal analysis, the anti-goals are organised into AND/OR refinement abstraction structures. The difference of anti-goal graphs and attack graphs is the process of creating them: \cite{van2003system} explicitly formalize the process of refining the way an anti-goal can be achieved in a step by step manner: the top level anti-goals are predefined. Next, refinements addressing who and why the attacker wants to achieve the anti-goal are incorporated to the graph. In the last steps, vulnerability identification, environmental effects and capabilities of the attacker are refined. The structure of the resulting structure is the \textit{anti-goal graph}. 
The explicit process of creating anti-goal graphs, as well as the fact that the high-level anti-goals are standardised elevate the usability and makes this process especially useful for process oriented practitioners.

\subsubsection{Security Goal Indicator Trees}Similarly, \emph{Security Goal Indicator Trees} (SGIT) by Peine, et al. \cite{peine2008security} follow a goal-oriented approach. The authors propose a software inspection method linking the overall security goals to a set of goal-indicators by mapping the mostly negatively-formulated security goals (e.g. "absence of buffer overflows") to positive features whenever possible\footnote{The formalism, however, also supports negative formulations of security goals whenever a positive formulation is not possible.}. These features may be detected in the software-inspection process: the presence of security feature - or their absence - is thus made explicit through the indicators and can be checked upon using the SGIT. "While  the anti-model approach focuses on security requirement elaboration, the SGIT approach is about accounting for the realization of security goals throughout the development lifecycle"\cite{peine2008security}. 

Let $A,B$ represent properties of an object $O$ to be inspected, and let $A(O)$ be true if and only if $O$ has property $A$. In this setting, Peine et al. define an indicator as a pair $I=(A, P(A))$ of a system property $A$ and its "polarity" $P(A)$. The polarity $P(A)$ is to indicate whether, in the ideal case, the designated system property is to yield true or false. I.e. $P(A)$ is to yield true (false), whenever the $A$ is beneficial (detrimental) to system security. An indicator $I$ is called detected, whenever $A(O) = P(A)$, i.e. when the desired system property $A$ is found in object $O$ or when the existence of an undesired property $A$ is negated for $O$.
\begin{figure}[htb]
\centering
    \includegraphics[width=.42\textwidth]{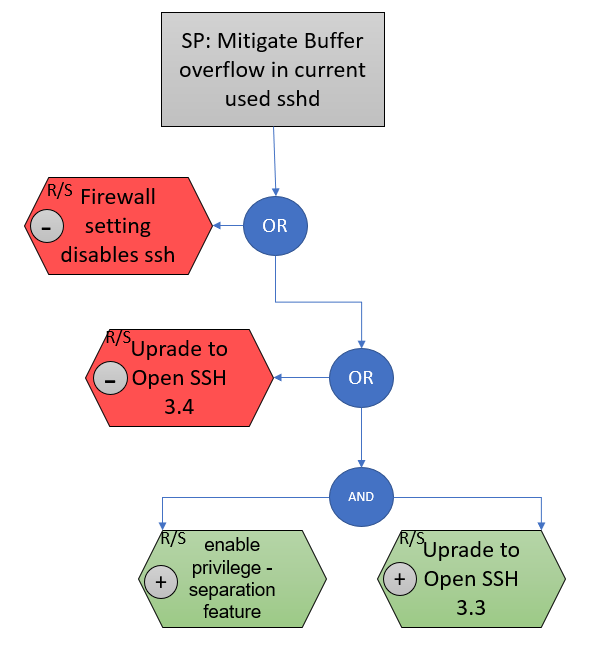}\hfill
  \caption{Security Goal Indicator Tree \cite{peine2008security} adapted to current scenario. The aim is to avoid the exploit of buffer overflows currently present in the ssh-daemon of the network.}\label{fig:attackgraph_goaloriented2}
\end{figure}
Security indicators may be connected via various operators: a directed edge indicates a conditional dependency between two indicators. $I_1 \to I_2$ is placed, whenever checking $I_2$ depends on $I_1$, i.e. checking $I_2$ can be 
skipped, whenever $I_1$ was not detected, not detecting $I_1$ corresponds to a detrimental security flaw. Other operators linking security indicators are the logical AND and OR operators. 

Furthermore, the formalism supports the property specialization operator $I_1 \implies I_2$, which helps describe situations, whenever the property in $I_2$ is a concrete version of the property in $I_1$, "usually by restriction to a certain programming platform or 
document type"\cite{peine2008security}. [...] When specializing indicators, it soon becomes apparent that this process can be recursively applied to any of the concrete indicators, resulting in a tree of indicators connected by the specialization relation. We call such a tree a \emph{specialization tree}". The authors, however, provide no formal semantics for the latter. 

Given a SGIT, the inspection process commences at the root node of the SGIT and traverses the graph in a depth-first manner. "SGITs are implemented in a prototype tool mentioned in \cite{peine2008security} They are used to formalize security inspection processes for a distributed repository of digital cultural data in an e-tourism application in \cite{jung2010practical}. The formalism is extended to dependability inspection in \cite{kloos2010systematic}." \cite{kordy2014dag}

\subsubsection{Goal-oriented attack graph by Nanda et al.}
Nanda et al. \cite{nanda2007highly} also follow a goal-oriented approach to attack graph modelling. They define the \emph{goal-oriented attack graph} as a goal-driven entity, which is decoupled from the networks or systems it is applied to. In their formalism, just as in generic attack graphs, the directed graph consists of a nodeset and a set of directed edges. Nodes represents vulnerabilities, whereas outgoing edges describe an exploit or attacker action on this vulnerability, leading to the next vulnerability to be exploited. Additionally, each attack graph possesses a unique sink node $v_f$ that represents the final vulnerability, whose exploitation realizes the final goal of the attack. I.e. the postcondition of exploiting  $v_f$ is the goal state and the ultimate idea of goal-oriented attack graphs is to model all exploit chains that lead to this exact vulnerability.

Nanda et al. propose to use this formalism in \cite{nanda2007highly} to correlate intrusion alert by matching the attack sequence to paths of vulnerabilities in the goal-oriented attack graph and to "determine the systems that are likely to be targeted by an attack in the future". The prediction algorithm matches an attack scenario to the remaining path length in the direction of the goal. Their methods, however, rely on the IDS to detect all attacks on vulnerabilities in the attack chain and it is vulnerable to false alarms. Furthermore, this method only allows for one arc between hosts which poses limitations. Furthermore, note that the success probability values are a property of the network components (i.e. the nodes), and not the exploits. This way of modelling might be problematic, as longer paths to the same target will have a lower probability of success in any case, which is not the case in many scenarios. The idea of matching the individual exploits with their possible targets might be interesting for many practitioners and adapted for their frameworks. 

\subsubsection{Goal-oriented attack graph by Liu et al. (2010)}
Liu et al. \cite{liu2010goal} model a network as an undirected graph $\mathcal{N} = \{C, E\}$, with $C$ being the components, such as hosts and servers, and the edges $E$ representing their connectivity. In their goal-oriented approach, the attack graph is described as directed multi-graph $G = (C, E, V )$ with weights on the edges. The graph is induced from the network $\mathcal{N}$, i.e. $C, E$ are inherited from $\mathcal{N}$ and the set of vulnerabilities $V$ is associated with the host nodes. The vulnerabilities $v\in V$ consist are triples $(cons, prob, host)$, where $cons \in \{g, u, r \}$ denotes the consequence of exploiting the vulnerability, with $g<u<r$ representing the ordered privilege levels $g$ - guest, $u$ - user and $r$ - root. $prob$ quantifies the probability of exploiting the vulnerability successfully and $host$ refers to the affected network component. 

Let $h$ denote a host which is connected to a node having vulnerability $v$. The exploit $e$ associated with this vulnerability and host is denoted as a tuple $e = (h, v)$. Furthermore, let $T \subseteq C$ denote the set of potential target network and define targets $T\times privilege(g, u, r) \times (0, 1)$, as triples, consisting of a target asset, a privilege level and the probability of reaching this privilege level on the target host. 

Then, using the attack graph $G = (C, E, V )$, the first aim of the goal-oriented approach is to assign an attacker goal (set) to each exploit. The authors also propose a method to compute the "weakest node" among all selected targets based on the probability of being successfully exploited. "In addition to finding out the weakest node in the network, based on the analysis of the targets of each exploit, we can predict the attacker’s target in mind on the basis of the exploits we observed so far. [...] The prediction is done by analyzing the common targets of all exploits." \cite{liu2010goal}

\begin{figure*}[htb]
\centering
  \subfloat[Goal-oriented attack graph by Nanda et al.\cite{nanda2007highly} modelling the hardware-keylogger vulnerabilits. It is not possible to model both the keylogger and buffer overflow vulnerability on host 3 in one model, as the formalism only supports one sink node $v_f$]{%
    \includegraphics[width=.38\textwidth]{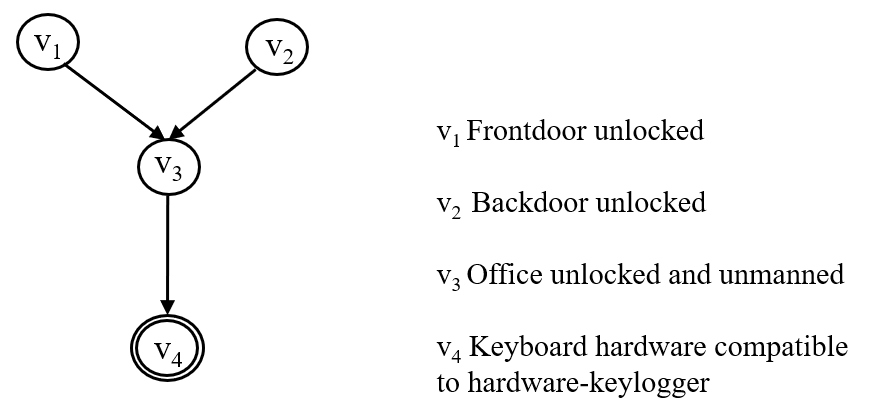}}\hfill
  \subfloat[Goal-oriented attack graph by Liu et al. (2010) \cite{liu2010goal}. As there exists only one edge between two nodes, a defender can a) only model one exploit/vulnerability between two hosts or b) he can model a generic exploit, combining the probabilities for all vulnerabilities. Note that the probability of reaching target asset is a node property, not exploit property.]{%
    \includegraphics[width=.4\textwidth]{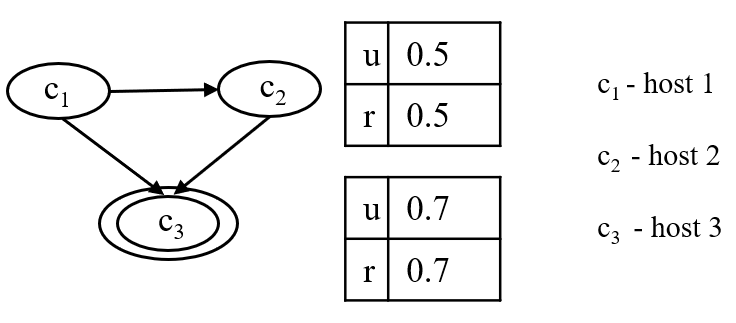}}\\
  \subfloat[Goal-oriented attack graph by Liu et al. (2015)\cite{liu2015approach}, including the target threat distance and the initial threat distance in parantheses next to the nodes ]{%
    \includegraphics[width=.76\textwidth]{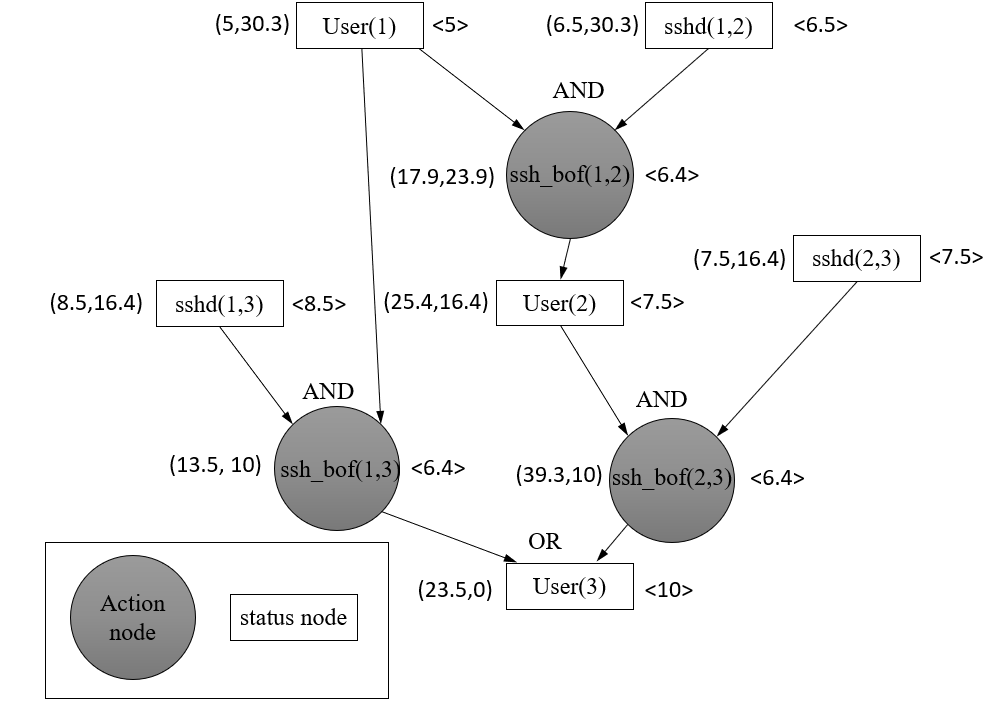}}\\
  \caption{Goal-oriented attack graph formalisms applied to the running example scenario}\label{fig:securitygoalindicator}
\end{figure*}
\subsubsection{Goal-oriented attack graph by Liu et al. (2015)}
Liu et al. \cite{liu2015approach} also use an goal-oriented attack graph for threat evaluation. Their goal-oriented attack graphs are defined as quadruples $G = \{S,A,O,E\}$, where $S$ is a set of states, $S = \{s_i(p, v, h) \mid i \in (1,N), v \in (0,10)\}$. Here, $p$ is the status information, $v$ is the significance score of the host, a number between $1$ and $10$, and $h$ is the host on which the vulnerability is located. $A$ is the set of action nodes and each action represents an exploit, which must include information on the source and destination hosts, as well as associated vulnerabilities and a value for the attack difficulty $\in [0,10]$. $O$ is the set of attacker targets $0\subseteq S$ and $E$ is the set of edges. Each edge connects either a network state from $S$ with an action in $A$ or vice versa. Edges, which are outgoing from state nodes to actions, are connected via logical AND and OR connectors; edges from action nodes to status nodes cannot be connected in this way.

Based on this information, the authors perform an attack graph analysis in two directions using two metrics: the initial threat distance and the target threat distance. The initial threat distance represent the effort taken from the target node; it sums the significance and effort scores until the target; in case of or nodes, the minimum of all values is taken. The target threat distance quantifies the effort yet to be taken to reach the target node. These two metrics quantify how hard it is for an attacker to mount an attack given a specific attack path. These metrics assume additivity of attack hardness and significance score of hosts as a heuristic, which needs to be considered when evaluating the scores. 

\subsection{Attack Trees (AT)}
\begin{figure}[htb]
\centering
  \subfloat[First ever published image of a threat logic tree by \cite{weiss1991system}]{%
    \includegraphics[width=.5\textwidth]{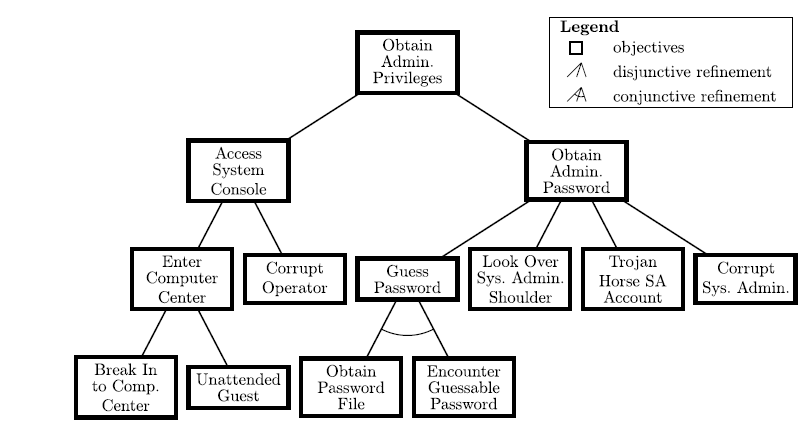 }}\hfill
  \subfloat[Simple attack tree model after \cite{schneier1999attack} for the running example]{%
    \includegraphics[width=.45\textwidth]{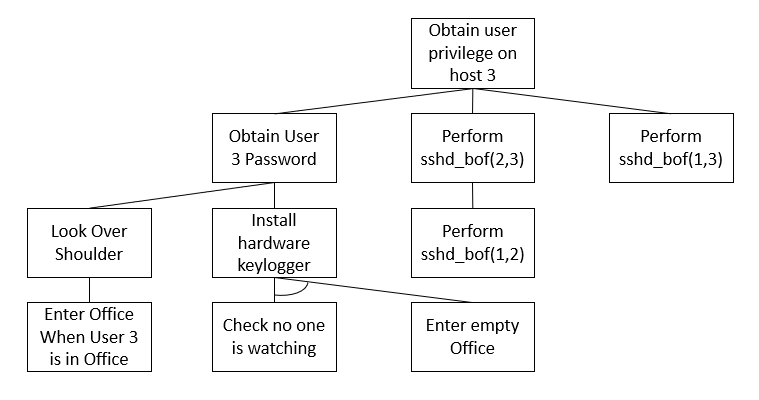}}\\
  \caption{Weiss's Threat logic tree and Schneier's Attack tree}\label{fig:attacktreeformalisms}
\end{figure}

\subsubsection{Attack trees, Probabilistic Attack trees, Attack trees with Costs}\emph{Attack trees} (AT) represent cyber-attacks in a tree structure. Rather than modelling attack sequences, attack trees are typically employed to model attack refinement. Thus, in general, attack trees are a qualitative method of cause-analysis modelling for cyber-attacks; though many sub-types of AT support quantitative analysis as well.

The root node represents the ultimate attacker goal and its children represent different means of reaching this goal, which are again interpreted as subgoals, whose children are again ways of achieving this subgoal, and so forth. Edges may be connected via AND and OR nodes. The AND-OR structure explains why these graphs do not have circles: nodes can hold numerical or Boolean values, which can be propagated upwards to the root in order to quantify an attack, yielding \textit{probabilistic attack trees} or \textit{attack trees with costs}. Attack tree tools, such as ADTool \cite{gadyatskaya2016attack, kordy2013adtool} or SecurITree \cite{securitree} work under this premise \cite{buldas2020attribute}. Logical algorithms, as well as most other forms of value propagation do not support circular structures. 

Attack trees were originally proposed by Weiss \cite{weiss1991system} as \emph{threat logic trees} (threat trees in \cite{amoroso1994fundamentals}), popularized through Schneier, Salter and Moore \cite{schneier1999attack, salter1998toward, moore2001attack}, and formally analysed by \cite{mauw2006foundations}. 
Initial tools on ATs include the parametric attack modelling language \cite{tidwell2001modeling}, which formalizes attack trees via attack tree specification templates holding a unique identifier, and a list of system element parameters.

\subsubsection{Multi-parameter attack trees} \cite{buldas2006rational} introduce multi-parameter attack trees, which extend probabilistic attack trees, as they have several interdependent attack parameters. In their work, Buldas et al. illustrated the concept using parameters for the attackers probabilities of success, their probability of getting caught, the attackers' gains in case of successful attacks, as well as the attackers' penalty, when failing an attack. They view the attack represented by the graph as an (extensive form) game played by the attacker. The computation of the attackers' expected outcome for this setting is described in \cite{jurgenson2008computing}. 
Multi-parameter attack trees offer an economic perspective to network security. Furthermore, they model both defender and attacker actions, allowing for the game theoretic analysis of the attacker-defender interaction, which has become a large research field on its own.

\subsubsection{Vulnerability attack trees / stratified network vulnerability trees} Daley et al.
\cite{daley2002structural} discuss vulnerability attack trees/stratified network vulnerability trees, an enhancement to attack trees, separating nodes into distinct classes based on functionality using the Stratified Node Topology (SNT): "These classes represent application specific exploits (event-level nodes), abstract attack ideas (state-level nodes), and attack goals (top-level nodes)." To add expressiveness, implicit node links directly imply another node, whereas explicit links only represent the potential to access a system or steal information. 
This formalism is especially helpful for practitioners who seek an AT model with a richer semantic.
\begin{figure}[htb]
\centering
    \includegraphics[width=.62\textwidth]{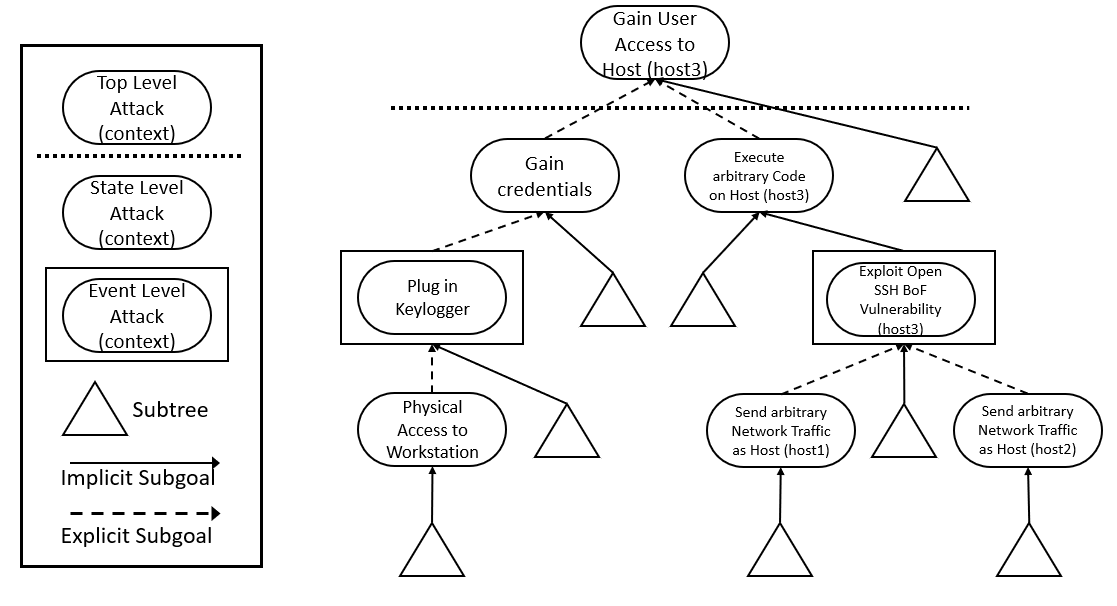}
  \caption{Vulnerability attack trees / stratified network vulnerability trees by \cite{daley2002structural}}\label{fig:stratifiedOWA}
\end{figure}
\subsubsection{OWA-trees}
Opel \cite{opel2005design} proposes extensions towards more operations (NAND, XOR, NOR), attack trees with confluent branches, leading to nodes having more than one parent.
Yager \cite{yager2006owa} also introduces new nodes to the attack graph formalism, beyond “and/or”. He generalizes attack trees to \emph{OWA trees} using Ordered Weighted Averaging (OWA) operators, a class of operators that lie between the “and” and “or” operator. "An OWA node is characterized by a vector W, called an OWA weighting vector. The vector W is of dimension equal to the number of children, $n$"\cite{yager}. The child nodes are ordered in the (temporal) order of becoming true, multiplied with the corresponding weights and summed up. "The special case of OWA node in which $w_1 = 1$ and $w_j = 0$ for $j \neq 1$ corresponds to an “or” node. [...] On the other hand the special case of OWA node in which $w_j = 0$ for $j \neq n$ and $w_n = 1$ corresponds to an “and” node" \cite{yager2006owa}. He also introduces the more general TOWA node and the SOWA node to attack graphs. A TOWA node employs the t-OWA operator, which is a mix of $t$-norms and OWA operators \cite{yager2005extending}. Likewise, a SOWA node is based on the co-t-OWA operator, a mix of $t$-co-norm \cite{dubois1980triangular} operators and OWA operators. He postulates the idea of using OWA and similar nodes for probability propagation in probabilistic attack trees, especially in case of non-independent probabilities.  Additionally, techniques for analyzing an OWA attack trees with attacker costs are provided. 

From this perspective, the formalism yields a less-complex alternative to Bayesian Networks, when stochastic independence of the involved causes is not realistic. Furthermore, this formalisms supports not only classical, probability theory based inference, but also evidence theory\cite{dempster1968generalization, shafer1976mathematical} and fuzzy logic \cite{zadeh1965fuzzy}. 

\begin{figure*}[htb]
\centering
\includegraphics[width=.6\textwidth]{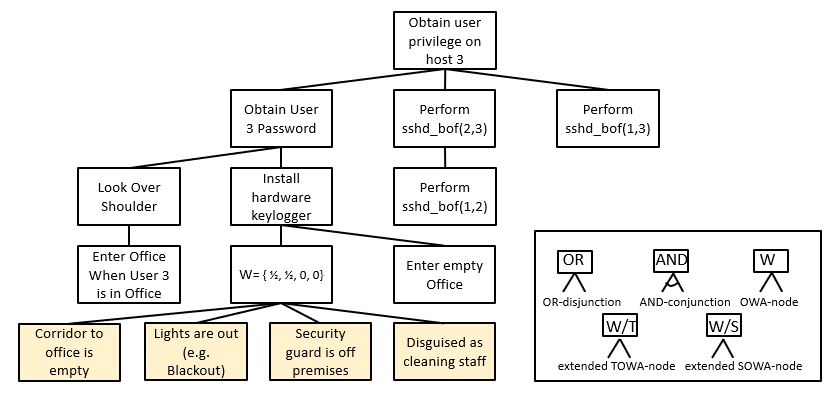}
\caption{OWA-Tree formalism by Yager \cite{yager2006owa} applied to the running use case. Assume all nodes hold Boolean values and consider the attack graph in figure \ref{fig:attacktreeformalisms} (b). Using an OWA-node, we are able to refine the „Check no one is watching node“ to indicate that an attacker continuities his attack, when whenever any two of the four conditions in the corresponding leave nodes (shaded in yellow) are met. The corresponding weights expressing this scenario are specified in the weights vector. 
}\label{fig:OWA}
\end{figure*}
\subsubsection{Defense Trees}
Opel \cite{opel2005design} calls for  additional node types by including defense nodes. This lead to \textit{defense trees}, which \cite{bistarelli2006defense} define by enclosing a set of countermeasure nodes to the leaves of an attack tree. 
Furthermore, the authors use economic indexes, such as Return on Investment/attack to quantify attack scenarios to determine optimal defence policies. Defense Trees allow for a simple and efficient analysis of defense measures. The limitation of placing defenses only to leaf nodes may not adhere to every use case and needs to be considered before application.

\subsubsection{Protection Trees}
Also seeking for protection nodes, Edge et al. \cite{edge2006using, edge2007use} define \emph{Protection trees} as AND/OR type tree structures. "The differences between the two types of trees are in what the nodes represent. A node in an attack tree represents a vulnerability. These vulnerabilities are specified but how to protect them is left out of the formal analysis."\cite{edge2006using} Protection trees are built from attack trees by specifying protective measures mitigating the vulnerabilities for every leaf node in the attack tree. Then, moving up a level in the attack tree, it is checked weather the protective measure covers the parent attack node. In case it does not, new protection nodes are added. The process is iterated until the leaf node is covered. Unlike defense trees, this formalism allows for the placement of defense nodes at all stages and defines a clear process of building Protection trees from attack trees. The formalism does not allow for quantitative analysis, however the tool ADTool by \cite{kordy2013adtool} extended the formalism for a quantitative analyisis.
\begin{figure*}[htb]
\centering
  \subfloat[Defense Tree \cite{bistarelli2006defense}]{%
    \includegraphics[width=.55\textwidth]{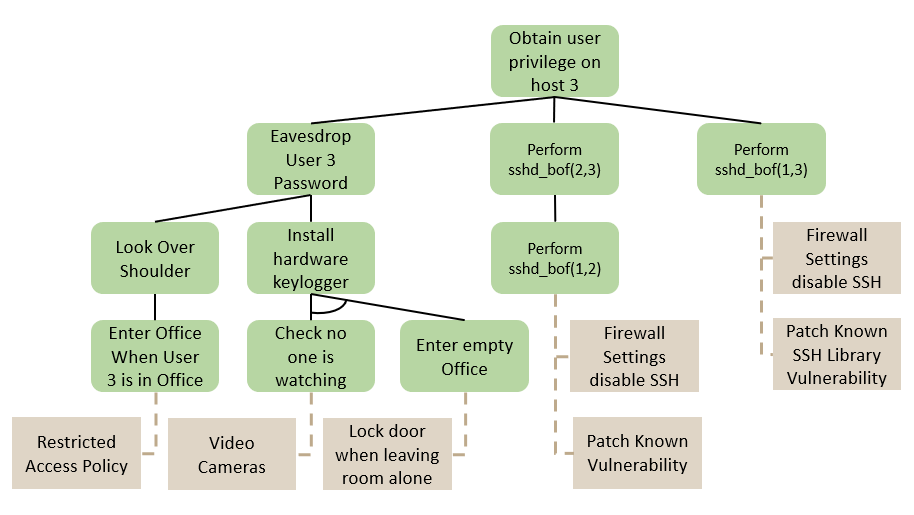}}\hfill
     \subfloat[Protection Tree \cite{edge2007use}]{%
    \includegraphics[width=.43\textwidth]{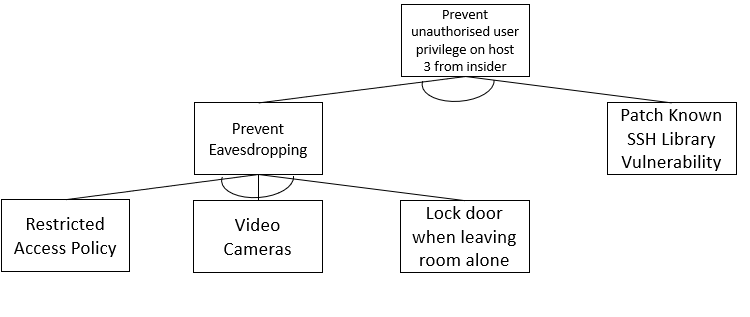}}\\
  \subfloat[Attack-Response Tree \cite{zonouz2013rre} for the running example, without host 2]{%
    \includegraphics[width=.4\textwidth]{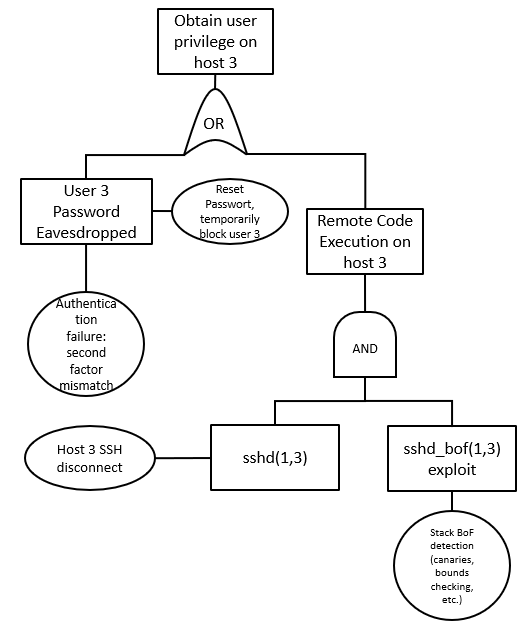}}\hfill
\subfloat[Attack–Defense Tree \cite{kordy2014attackdef} for the keylogger-attack in the running example]{%
    \includegraphics[width=.4\textwidth]{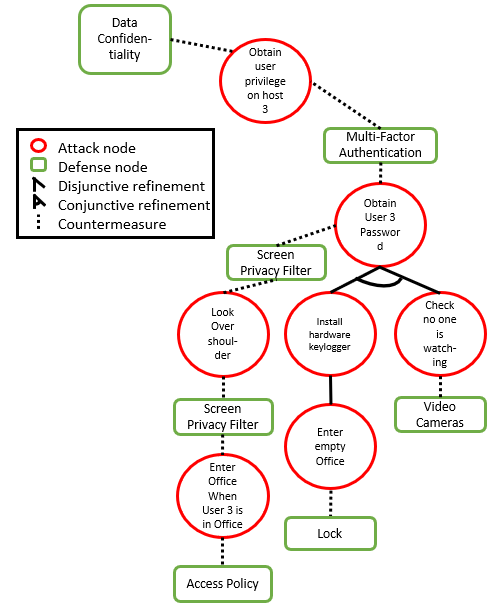}}\\
  \caption{Attack graph formalisms incorporating defense nodes}\label{fig:adtree}
\end{figure*}

\begin{figure*}[htb]
\centering
     \subfloat[Attack-Countermeasure Tree \cite{roy2010cyber} for the running example, without host 2; the improved Attack-Defence Graphs \cite{wang2014threat} also looks like this, as there are no deception nodes in our model]{%
    \includegraphics[width=.58\textwidth]{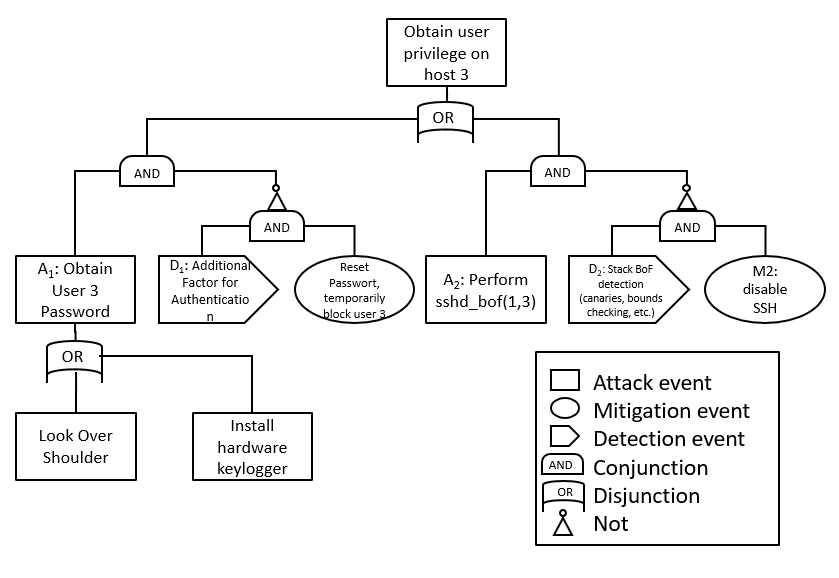}}\hfill
    \subfloat[Improved Attack–Defense Trees introduce deception nodes to Attack-Countermeasure Trees, original graphic by \cite{wang2014threat}]{%
    \includegraphics[width=.23\textwidth]{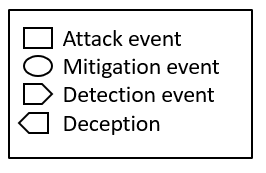}}\hfill
  \caption{More attack graph formalisms incorporating defense nodes.}\label{fig:adtree2}
\end{figure*}
\subsubsection{Attack Response Trees}
Alike, \cite{zonouz2013rre,sanders2009rre} introduced \textit{attack response trees} (ARTs) where attacks and countermeasures response may be placed at any node in the graph. Unlike classic attack trees, this formalism is based on attack consequences instead of attack scenarios; hence, not all events causing a certain consequence need to be considered, but it suffices to state the consequences (nodes) and their connections (AND/OR). Additionally, the leaf consequence nodes are connected to IDS alerts, holding a Boolean or probabilistic value, which indicates the "activity" of a leaf consequence. The values for other consequence nodes are propagated to the root in a bottom-up manner. As mentioned before, consequence nodes in the ART may be
equipped with \textit{response boxes}, which are actions a defender can take as a countermeasure. These countermeasures reset all activity values in the subtree to zero. Using game-theory, the response and recovery engine (RRE) computes an optimal response action from its action space strategy.

This model allows for modelling the connection between consequences of attacks, intrusion detection, and reaction and to compute an optimal reactive defense based on the parameters. Roy et al. \cite{roy2010cyber}, however, criticize that this formalism suffers from the curse of dimensionality, as the authors use partially observable Competitive Markov decision processes for their game-theoretic analysis.  

\subsubsection{Attack Countermeasure Trees, improved Attack-Defense Trees}
Roy et al. \cite{roy2010cyber} propose an alternative extension called \textit{attack-countermeasure trees}. In their formalism the location of defensive measures is also not limited to leaf nodes, but allowed at any vertice in the tree. In ACT, there are three classes of events, i.e. attacks, detection events and mitigation events, each represented as a different type of node. In \cite{roy2012attack} the authors show how to select optimal countermeasures via a single objective optimization approach using economic indicators (such as ROI) on ACT using reduction techniques as well as implicit enumeration (branch and bound). Wang and Liu \cite{wang2014threat} extend the formalism of attack-countermeasure trees to \textit{improved Attack–Defense Trees} by using not only attack,  detection and mitigation event nodes, but also deception nodes in their formalism. 

\subsubsection{Attack-Defense-Tree (ADTree)}
Attack–defense trees (ADTree) by Kordy et al. \cite{kordy2014attackdef} are similar to attack-countermeasure trees, as they also include attack and defense nodes, whose location is not bound to leave nodes. Child nodes of the same type represent a refinement of the parent goal into sub-goals. Such nodes may be refined using disjunctions (OR) or conjunctions (AND). Nodes without children of the same type are non-refined, representing basic actions.

Furthermore, Kordy et al. \cite{kordy2013adtool} developed the free tool ADTool for qualitative modeling and quantitative security analysis with ADTree, which also supports attack trees, protection trees as well as defense trees. "Supported measures include: attributes based on real values (e.g., time, cost, probability), attributes based on levels (e.g., required skill level, reachability of the goal in less than $k$ units of time), and Boolean properties (e.g., satisfiability of a scenario)." \cite{kordy2013adtool}. Further efficient methods for optimal countermeasure selection in ADTrees are proposed in \cite{9155095, muller2017fast}. 

\subsubsection{Cyber threat trees}
Other AT extensions include \emph{cyber threat trees}, a discrete multiple-valued version of attack trees, proposed by \cite{ongsakorn2010cyber}, where multiple-valued decision diagram (MDD) are employed to represent the cyber threat tree. For this puropose a new formal language, \texttt{CyTML}, was developed, holding 6 different types of nodes: effect, logic, threat, cause, data and state. 

\subsubsection{Enhanced -, improved -, SAND attack trees; attack trees with temporal order; timed probabilistic Semantics of Attack trees}
Further extensions of attack trees include refinements to consider temporal aspects in attack graphs: the aforementioned models consider attacks that take place in parallel. Serial attack graph models were developed to complement this static approach. 
\cite{camtepe2007modeling} incorporate temporal aspects by introducing the OAND (Ordered AND) operator
besides OR and AND, introducing \textit{enhanced attack trees} (EATs), while \textit{improved attack trees} \cite{lv2011space} and \textit{SAND attack trees} \cite{jhawar2015attack} incorporate the sequential AND (SAND) operator. \cite{jurgenson2009serial} complement attack graph by introducing a temporal order over the nodes to represent  chronological aspects of the attacker’s decision making process. 
Arnold et al.\cite{arnold2014time} add a temporal probability semantic to attack trees by introducing SEQ gates, which are AND gates with a horizontal arrow pointing in the direction of progressing time. In this formalism, each inner node is labelled with a gate and leaf nodes are labelled with a probability distribution (exponential or Erlang), representing the time it takes to carry out this attack step. The attack steps are executed sequentially: the $i+1$-st attack step commences after the $i$-th step has finished. 

\begin{figure*}[htb]
\centering
  \subfloat[Enhanced attack trees
(EATs) \cite{camtepe2007modeling} ]{%
    \includegraphics[width=.25\textwidth]{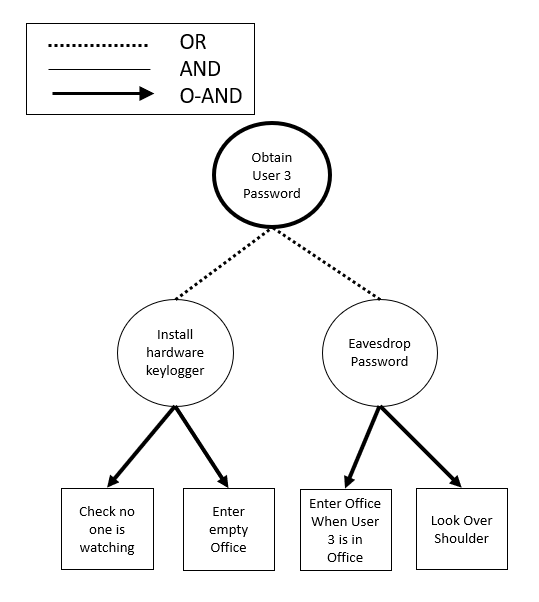}}\hfill
  \subfloat[SAND attack trees by
\cite{jhawar2015attack}]{%
    \includegraphics[width=.35\textwidth]{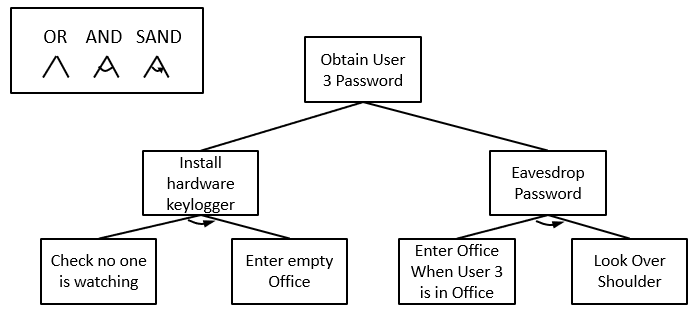}} \hfill
     \subfloat[Improved attack trees with SAND operators by \cite{lv2011space}]{%
    \includegraphics[width=.3\textwidth]{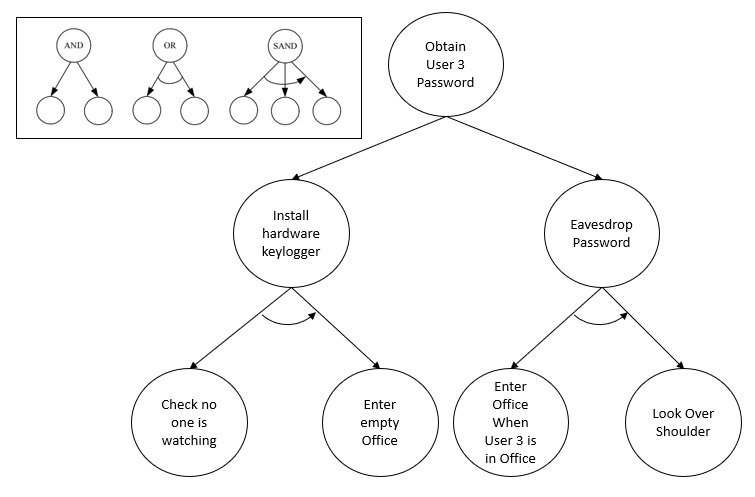}}\hfill
  \subfloat[Attack tree by \cite{jurgenson2009serial} introducing a temporal order over the nodes. The left to right ordering of the nodes $\{X_1, X_2, X_3, X_4\}$ is represented by the $id$ permutation]{
  \includegraphics[width=.4\textwidth]{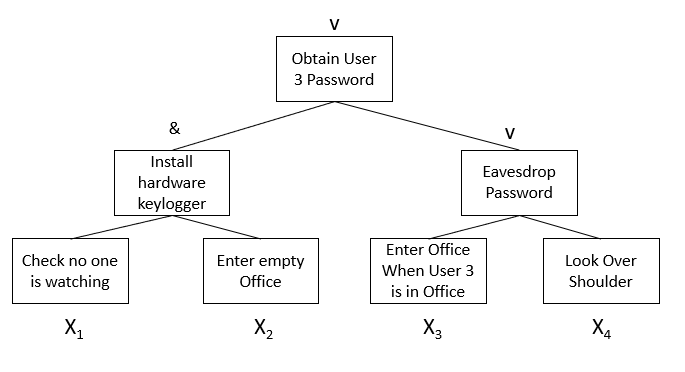}} \hfill
  \subfloat[Arnold et al.\cite{arnold2014time} introduce SEQ gates to AT]{%
    \includegraphics[width=.46\textwidth]{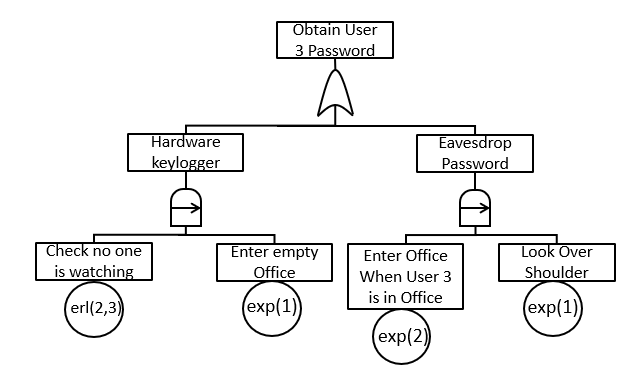}}\hfill 
  \caption{Attack tree formalisms including temporal orders}\label{fig:attacktreeformalisms2}
\end{figure*}

\subsubsection{Decorated attack trees}
\textit{Decorated attack trees} help tackle uncertainty in attack trees: Traditionally, attack trees are manually designed and may lack quantitative data necessary for the standard, quantitative bottom-up calculation methods. Therefore, in \cite{buldas2020attribute}, the authors develop a generalisation thereof, yielding \textit{decorated attack trees}. The attack-tree decoration problem corresponds to the well-known Constraint Satisfaction Problem (CSP)
, where a solution is a set of attack tree instantiations satisfying a given set of constraints. Thus, the idea of decorated attack trees is, given hard and soft facts (probabilities) on the attack tree, to finding an (approximate) quantification of the unknown values which does not conflict with the known information.

There exist many more attack tree formalisms, such as time-dependent attack trees, security goal indicator trees etc. We direct the interested reader to \cite{kordy2014dag} for a detailed overview of attack-tree types.

\subsection{(Hyper-) Alert correlation and attack strategy graphs}
\subsubsection{Hyper-alert correlation graphs}
Hyper-alert correlation graphs (HG) \cite{ning2002analyzing} were developed for IDS alert correlation, which is the process of correlating IDS alerts with specific vulnerabilitites or cyberattacks. For an overview of IDS alert correlation methods before Ning's work \cite{ning2002analyzing}, cf. \cite{gula2002correlating}.

Hyper-alert correlation graphs (HG) \cite{ning2002analyzing} describe potential attacker paths in a way similar to the requires/provides model \cite{templeton2001requires}. Hyper-alert correlation graphs, however use predicates instead of constructs and capabilities, i.e. the authors employ logical formula (combinations of predicates), to represent the prerequisites and consequences of an attack. Additionally, they transform the prerequisites and consequence relations into common conditions, represented as an equality constraint on one or more attributes, that need to be fulfilled by all instances of the strategy, in order for the attack to be successful. Thus, unlike its predecessor in \cite{templeton2001requires}, the formalization in \cite{ning2002analyzing} allows for correlating alerts from potentially unknown attack scenarios on a more fine-granular level. 

In this setup, a \emph{hyper-alert type} $T$ is a triple $(fact, prerequisite, consequence)$. A (realized) hyper-alert instantiates the prerequisite and consequence with its specific value. Given the prerequisites and consequences of different types of attacks \cite{ning2002constructing} match the consequence of earlier IDS alerts to the prerequisite of later ones, which reveal the structure of cyber-attacks, using the a hyper-alert correlation graph. A hyper-alert correlation graph (HG) $HG = (N, E)$ is a connected graph, where the nodes represent sets of hyper-alerts, connected by directed edges from $n_1$ to $n_2$ in $E$ if and only if $n_1$ is prepares for $n_2$. For brevity, the authors refer to hyper-alert correlation graphs as \emph{correlation graph} in follow up works.

In \cite{ning2002analyzing, ning2004techniques} Ning et al. present six operations on their graph structures: adjustable graph reduction reduces the number of nodes in the hyper-alert correlation graphs; focused analysis develops graphs around on individual hyper-alerts of interest; graph decomposition clusters the hyper-alerts in the graph based on their common features (such as common source and/or destination IPs) to facilitate the analysis of large sets of correlated alerts; frequency analysis helps the analyst detect patterns in a set of intrusion alerts; link analysis analyses the importance of the individual attribute values connecting the entities; association analysis analyses frequent co-ocurrences of of attribute instantiations of among different nodes and attributes. Using these methods, intrusion alerts can be aggregated and the complexity involved can be reduced.

\subsubsection{Attack strategy graphs, Hyper-alert Type graph, integrated correlation graph}
In \cite{ning2003learning} Ning et al. present techniques to automatically learn attack strategies and collect them in \emph{attack strategy graphs} from correlated intrusion alerts. Attack strategy graphs (ASG) have nodes representing exploits and edges representing the temporal order of the nodes as well as conditions that need to be fulfilled to master the exploit. Formally, $ASG = (N, E,T, C)$ is a graph over a set $\mathcal{S}$ of hyper-alert types, where $N$ is nodeset, $E$ is the edges as described above,
$T$ is a map, mapping each node to a hyper-alert type in $S$, and $C$ assigns labels to the edges, each describing the set of equality constraints for the hyper-alert types of incident nodes. The equality constraints characterize how the consequence attribute values of an earlier hyper-alert relate to the prerequisite attributes of a later alert.

Alerts are aggregated to attack strategies whenever constraints on the attack attributes in the prerequisites and conditions are fulfilled with equality, given the temporal order among these attacks. Furthermore, methods to aggregate hyper-alert correlation graphs and for automatic generalization of hyper-alert types (to hide syntactic differences between IDS alerts) are presented. A similarity measure between attacks (based on graph isomorphisms) facilitates the analysis. 

In follow up works \cite{ning2004building}techniques to construct aggregated attack scenarios, even if the IDS misses critical attacks, are presented. Two complementary types of alert correlation, namely clustering methods, which are based on the similarity between alert attributes, as well as causal correlation methods, which are based on prerequisites and consequences of attacks, are discussed. The results are employed to reason about potential attacks not detected by the IDSs based on the indirect causal relationship between subsequent intrusion alerts as well as the equality constraints they must satisfy. 

To do so, knowledge of the relationships between hyper-alert types is encoded in a \emph{(hyper-alert) type graph}, a quadruple $(N,E, T,C)$ with the same structure as ASGs. The authors derive indirect equality constraints for two hyper-alert types, whenever it is reasonable that one of them may (indirectly) prepare for the other. Indirect equality constraints are used to analyse if two hyper-alerts in two different correlation graphs are (indirectly) related. The proposed techniques can provide meaningful “guesses” of attacks possibly missed by the IDSs, and thus supply good starting points as well as supporting evidences to facilitate investigation of unknown intrusions. 

The resulting correlation graph, including reasonable guesses of undetected attack steps is called \emph{integrated correlation graph}.

 \begin{figure}[htb]
\centering
  \subfloat[two small hyper-alert correlation graphs \cite{ning2004building}]{%
    \includegraphics[width=.15\textwidth]{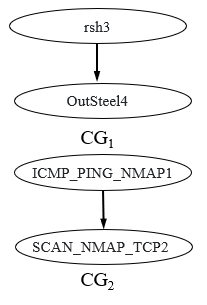}}\hfill
  \subfloat[Example of a hyper-alert type graph \cite{ning2004building}. ]{%
    \includegraphics[width=.38\textwidth]{typegr}}\hfill
    \subfloat[integrated hyper-alert correlation graph with hypotheses of missed attacks \cite{ning2004building}]{%
    \includegraphics[width=.32\textwidth]{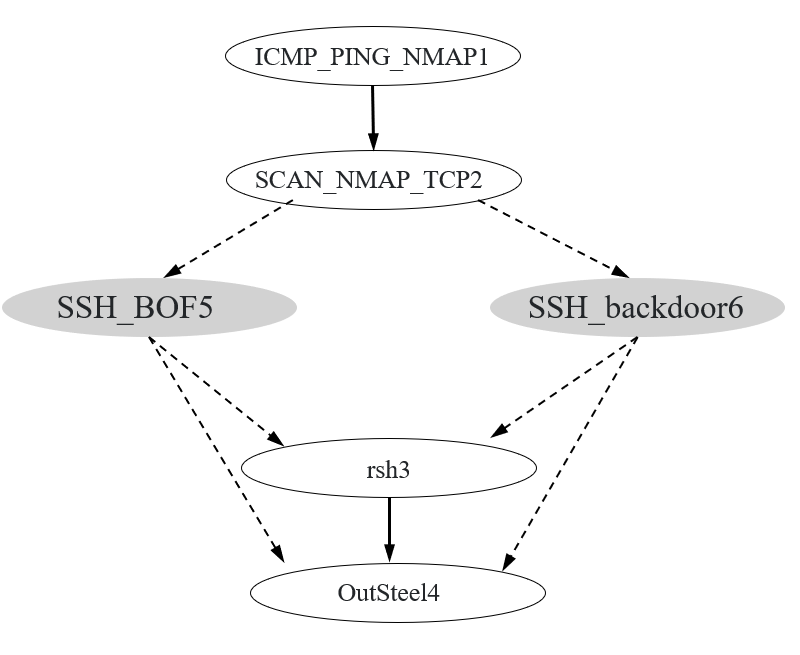}} 
  \caption{Attack graph types by Ning et. al. \cite{ning2002analyzing, ning2004building, ning2003learning} for the ssh-daemon Buffer overflow part of our example. For better demonstration, we assume the existence of an additional SSH-backdoor access and that the attackers use the malware OutSteel for data exfiltration after gaining user access. }\label{fig:ningtree}
\end{figure}

These AG Types are a suitable starting point for everyone, who seeks deeper insight into graph- or model-based IDS-alert correlation frameworks, before looking into other formalsisms, such as Generalized Dependency Graphs by Albanese et al. \cite{10.5555/2041225.2041255}. The representation has been used in, among others, \cite{anbarestani2012iterative}. 

\subsubsection{Hyper-Alert Graphs}
In 2006, Zhu and Gorbhani \cite{zhu2006alert} construct hyper-alert graphs from a large volume of raw alerts based on multi-layer perceptron learning and and Support Vector Machines. Similar to hyper-alert correlation graphs, the capability model by \cite{zhou2007modeling} is based on requires/provides formalism proposed by \cite{templeton2001requires}. 
Alike, the authors employ logical combinations of predicates to model properties of capabilities and correlate IDS alerts using several alert correlation algorithms. 

\subsubsection{Alert Dependency Graphs}
Another related formalism is the \textit{alert dependency graph} by \cite{roschke2011new}. First, consider a graph (not having any directed arcs) yet, whose vertices are triples consisting of a vulnerability, the affected host, as well as a quantification of the vulnerability impact. Additionally, a set of IDS alerts, consisting of a timestamp, a source host, a destination host, and an alert classification, is given. In this framework, similar alerts co-occurring within a close period of time are aggregated and mapped to specific nodes in the graph. The resulting graph is called alert dependency graph and "each path in the alert dependency graph DG identifies a subset of alerts that might be part of an attack scenario". The authors propose identifying the most plausible attack scenarios using the Floyd-Warshall algorithm. The authors use the use MulVAL \cite{ou2005mulval} to generate the alert dependency graph. 

\subsection{Attribute Attack Graphs, Stage Attribute Attack Graph}
\subsubsection{Attribute Attack Graph}
According to \cite{li2016optimized} the attribute attack graphs were defined by Chen et al. \cite{Chen2010two}. Unlike state-based attack graph formalisms, attribute attack graphs have two types of nodes in attribute attack graphs: exploit nodes, and condition nodes \cite{gao2021multi}. The latter represent the pre- and post conditions of exploits in form of attributes, often related to the privileges acquired. Edges are used to model the dependencies between the nodes. The formalism is similar to Ammann's exploit dependency graphs, \cite{ammann2002scalable}, without multiple edges though, which makes this representation more readable.  

Chen et. al \cite{Chen2010two} discuss the problem of analysing attack graphs with loops. First, they present method to identify all non-loop attack paths with depth smaller $n$ to some key attribute. Second, they present measures to optimise security measures in such graphs.

\subsubsection{Stage Attribute Attack Graph}
Li et al. \cite{li2016optimized} extend attribute attack graphs to stage attribute attack graph in order to model an attack capability based on Li et al.'s \cite{li2016study} four stages of advanced persistent threats: the \textit{Prepare stage}, where the adversaries collect information about their target and prepare their technologies; \textit{the Access stage} is the stage where the attackers access the network; \textit{the Resident stage}, including information collection, privilege escalation, remote control and lateral movement; and the final, \textit{Harvest stage}, where information is stolen, control is performed etc. Finally, all traces are cleaned up and evidence is removed. 

The stage attribute attack graph uses stage attacks, i.e. quadruples of the form $< name, $ $vulnerability, $ $connectivity, $ $ consequence >$ as its building blocks. The name attribute especially includes the current stage of the APT as a prefix to the exploit. 

Formally, the \emph{stage attribute attack graph} is  defined as $SG = (N_i, N_a, N_c; E)$, with $N_i$, the initial attribute nodes set, consisting of vulnerabilities and connectivities; $N_a$ is the attack nodes set, consisting of the exploits (i.e. the name of the the exploit); $N_c$ is the set of attack consequences, representing the privilege acquired by the attack. The set of edges denotes the dependencies and is defined as $E = ((N_i \cup N_c) \times N_a) \cup (N_a \times  N_c).$

This way, stage attribute attack graph explicitly label each attack with the current stage of the APT lifecycle. It is therefore suitable for qualitative attack graph modelling in the context of ATP.

\begin{figure}[htb]
\centering
  \subfloat[Attribute Attack Graph \cite{gao2021multi}]{%
    \includegraphics[width=.33\textwidth]{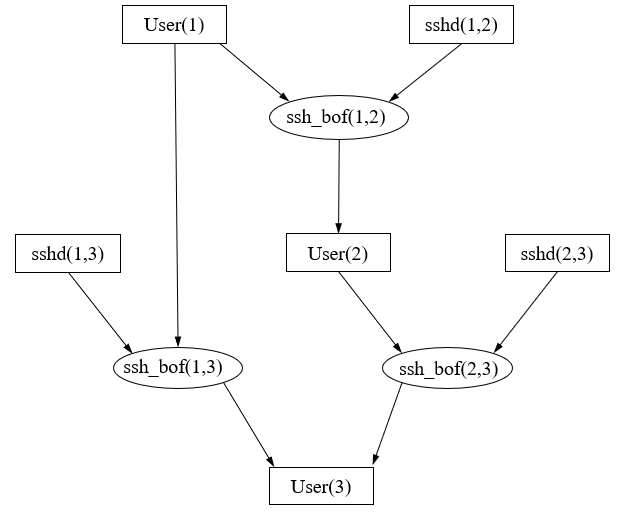}}\qquad
  \subfloat[\textcolor{black}{Stage attribute attack graph by} \cite{li2016optimized}]{%
    \includegraphics[width=.33\textwidth]{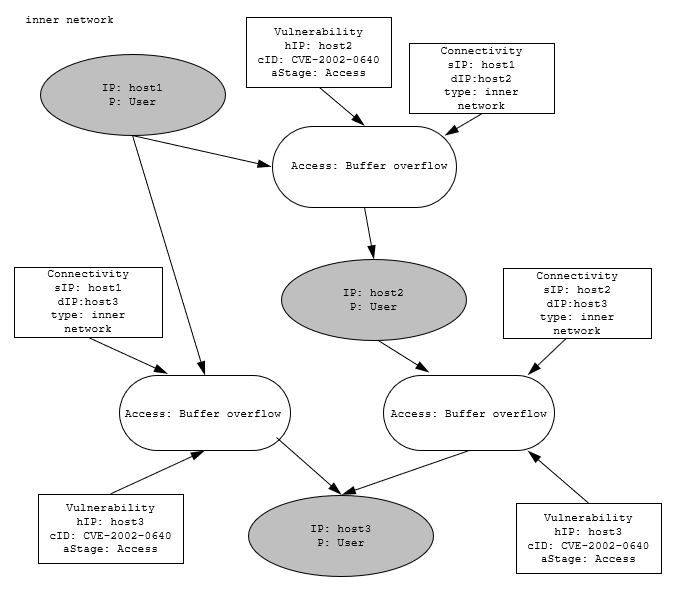}}\\
  \caption{Attribute Attack Graph}\label{fig:attrAG}
\end{figure}

\subsubsection{Activity graph}
\emph{Activity graphs} were developed in \cite{staniford1996grids} to secure large-scale networks from computer worms. In this rule-based framework, an organization is modelled as a hierarchy of departments and hosts. Each department in the hierarchy has an engine which collects data and generates activity graphs, with nodes representing hosts and arcs representing network traffic within that department. In this hierarchy, network traffic between different departments is addressed at higher levels of abstraction: As graphs propagate upward, the graph structure is aggregated to a reduced graph, where entire departments are represented as single nodes inheriting attributes of the individual subgraphs. 
Activity graphs were developed to reveal the causal structure of network activity: quickly growing, large, tree like structures are associated with worm infections. This
allows large-scale automated or coordinated attacks
to be detected in near real-time: "Recognition (detection) occurs when the counts exceed a user-specified threshold". 
\begin{figure}
    \centering
    \includegraphics[width = 0.6\textwidth]{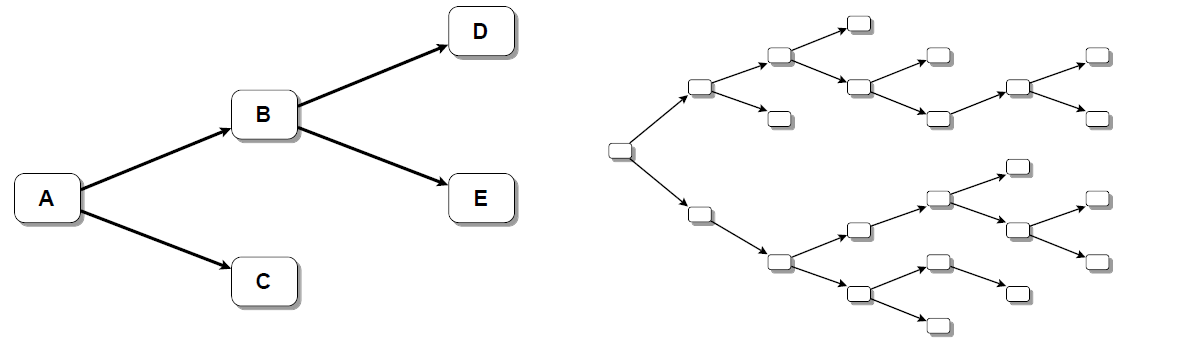}
    \caption{The beginning of a worm graph, and the activity graph after the worm has spread. Taken from the original source, as this type of attack graph does not apply to our scenario \cite{staniford1996grids}}
   \label{fig:grids}
\end{figure}

\subsection{Bayesian Attack Graphs, Bayesian Defense Graphs}
\subsubsection{Bayesian Defense Graphs}
\begin{figure}[htb]
\centering
  \subfloat[Elements of (extended) influence diagrams, acc. to \cite{sommestad2008combining}]{%
    \includegraphics[width=.33\textwidth]{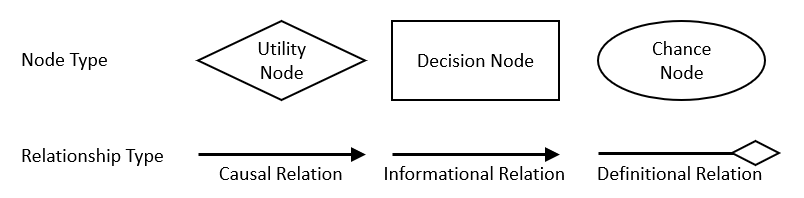}}\qquad
  \subfloat[Attack trees (cf. the AT from Figure \ref{fig:attacktreeformalisms} (b)) are expressed as an influence diagram with conditional probability tables; consequences are modelled using dependency edges and utility nodes \cite{sommestad2008combining}]{%
    \includegraphics[width=.42\textwidth]{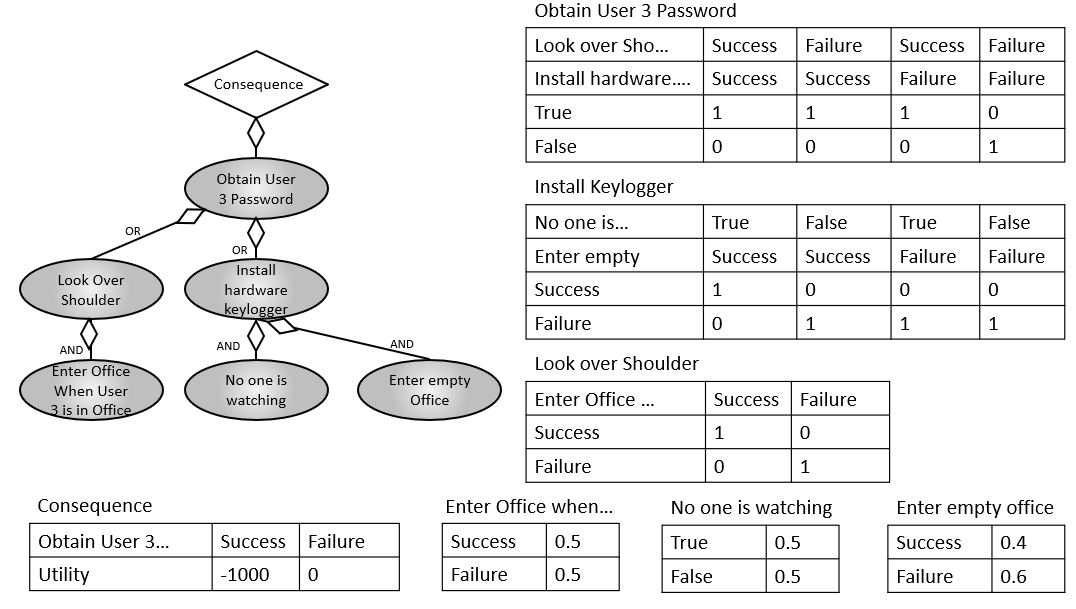}}\\
    \subfloat[Countermeasures are added using dependency edges and chance nodes yields the Bayesian Defense Graph]{%
    \includegraphics[width=.43\textwidth]{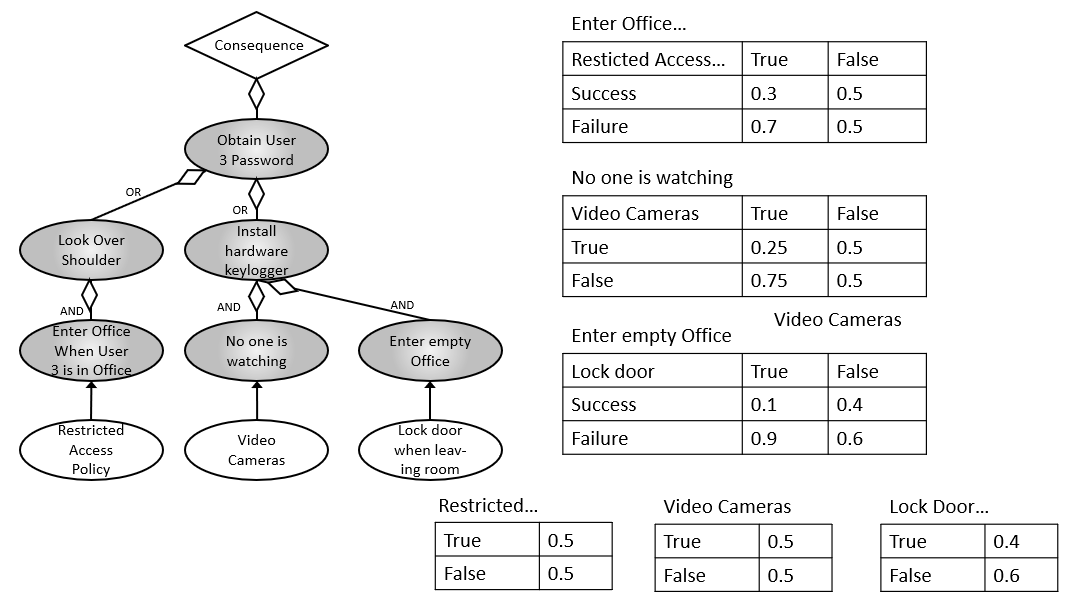}} \qquad
  \caption{Bayesian Defense Graph Formalism}\label{fig:bayes}
\end{figure}

Sommestad et al. \cite{sommestad2008combining}, present a network security assessment framework using influence diagrams (Bayesian Networks with additional decision and utility nodes) to combine attack graphs with countermeasures into defense graphs. The authors show how attack graphs, their consequences and - in the spirit of defense trees - respective countermeasures can be modeled with extended influence diagrams:

An attack graph/tree is converted into an influence diagram using the Influence Diagram elements from figure \ref{fig:bayes}, (a). Exploits are modelled as chance nodes and an attacker's success in depends on the attackers capabilities and the countermeasures in place (cf. \ref{fig:bayes} (b)). This observation is modelled using chance nodes with two states: “successful” and “failure”. AND- as well as OR relationships are expressed using definitional relations; their specification is realized in the conditional probability tables. Consequences of attacks are expressed in utility nodes. 

Sommestad et al. refer to this formalism as \emph{defense graphs} in \cite{sommestad2008combining}, or \emph{Bayesian defense graphs} in \cite{sommestad2009cyber}, see \ref{fig:bayes} (c). This formalism is not too distinct from Bayesian Networks/influence diagrams for security, and inherits all their positive - as well as negative - characteristics.

\subsubsection{Bayesian Attack Graphs}
Similar to Bayesian Defense Graphs, Poolsappasit et al. \cite{poolsappasit2011dynamic} introduce Bayesian Attack Graphs (BAGs), which adapt Bayesian networks as a means expressing different security states in cyberattacks on a network. As in ordinary attack graphs and attack trees, their AG representation features the usual cause-consequence relationships between distinct network states. Furthermore, it features the likelihoods for the individual exploits in the conditional probability tables. 

Formally, let $S$ denote a set of attributes, $A$ the set of attacks on $S$. Then a \emph{Bayesian Attack Graph} is a quadruple $BAG=(S, \tau, \varepsilon, \mathcal{P})$. Here, $S = N_{internal} \cup N_{external}\cup N_{terminal}$ is the nodeset (i.e. set of attributes), consisting of the union of internal, external and terminal nodes. External nodes have no parent nodes whereas terminal nodes have no child nodes; internal nodes have both. $\varepsilon$ is the edge set, defining a precondition-postcondition semantic on the node set. $\varepsilon$ is a set of decomposition couples, which is used to describe logical AND and OR relations between preconditions; $\mathcal{P}$ is the set of conditional probability tables.  

In their papers, the authors propose \emph{Augmented-Bayesian Attack Graphs}, an extended version the BAG with labels specifying the potential losses or gains at each node, as well as additional nodes
representing network hardening measures. The authors propose various methodologies to compute the probability of a network being compromised and employ this information for network hardening, i.e. for a security mitigation and management plan: "we propose a genetic algorithm capable of performing both single and multiobjective optimization of the administrator’s objectives. While single objective analysis uses administrator preferences to identify the optimal plan, multiobjective analysis provides a complete trade-off information before a final plan is chosen. Results are shown to
demonstrate the effectiveness of the algorithm in both static
and dynamic risk mitigation" \cite{poolsappasit2011dynamic}

Unlike Bayesian Defense Graphs, the authors quantify the expected return on investment (ROI) which is based on a user specified cost model as well as the likelihoods of system compromise. Furthermore, the authors emphasize the dynamic aspects of the Bayesian Network Formalism and supports multi-objective optimization of security goals. As in all BN-like graphs, the amount of parameters (for the conditional probabilities) is potentially large, which may take a considerable amount of work to specify.

\begin{figure}
\centering
  \subfloat[A simple Bayesian Attack Graph by Poolsappasit et al.\cite{poolsappasit2011dynamic}]{%
    \includegraphics[width=.53\textwidth]{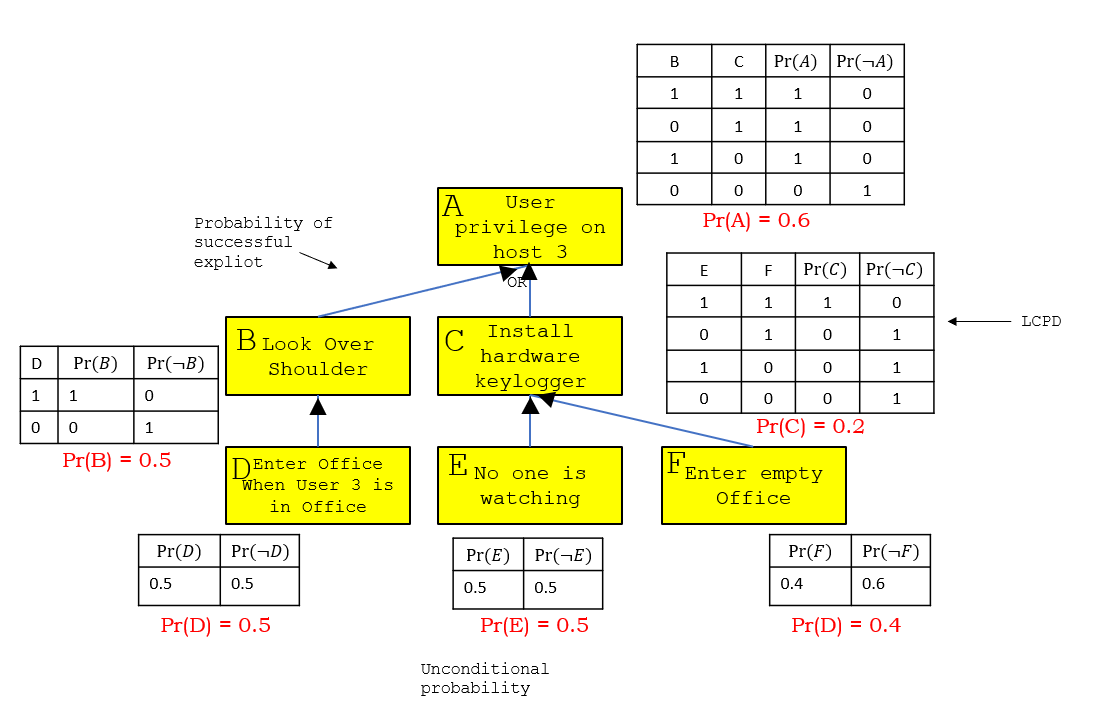}}\qquad
  \subfloat[Augmented Bayesian Attack Graph with Countermeasures and expected loss/gain at each node\cite{poolsappasit2011dynamic}]{%
    \includegraphics[width=.33\textwidth]{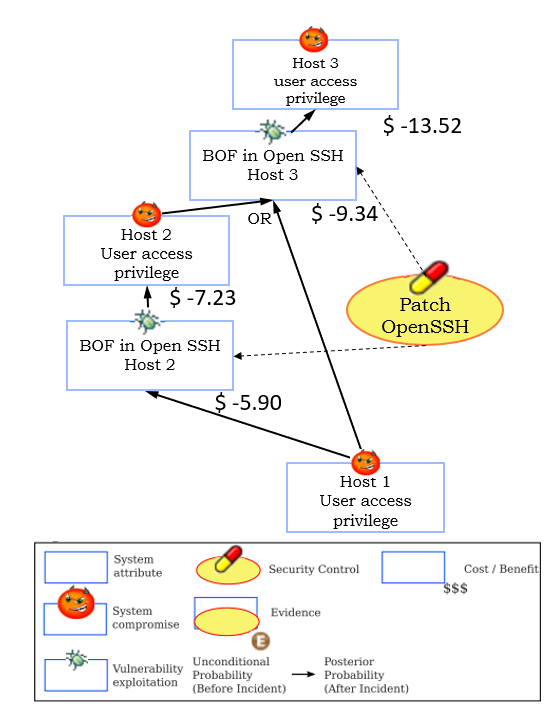}}\\
  \caption{Bayesian Attack Graphs}\label{fig:bayesatt}
\end{figure}

\subsection{Cloud-level Attack Graph}
Traditionally, an organisation or network administrator may maintain their own, individual attack graphs. "However, such individual attack graphs are confined to the enterprise networks without considering the potential threats from cloud environment."\cite{sun2015inferring}

To mitigate this issue, Sun et al. \cite{sun2015inferring} introduced cloud-level attack graphs in cloud networks. As virtualisation enables cloud computing, Sun at al.'s cloud-level attack graph aims to model the "stealthy bridges" between the virtual machine, the virtual machine image and the hosts. This way, potentially new attack paths, which are not detected by other attack graphs, may be discovered and their effects mitigated. 

The cloud-level attack graph is defined as a "cross-layer graph" with three layers: a virtual machine image (VMI) layer, a virtual machine (VMI) layer as well as a host layer.

The \textit{VMI layer} models dependencies between individual vulnerabilities/exploits inside the VMI. Next, the \textit{VM layer} models vulnerabilities in the vm instances and "stealthy bridges" between them, e.g. due to VMI sharing/inheriting vulnerabilities from the same parent VMI. Furthermore, vulnerabilities in the children instances might influence the parent VMI. Finally, the \textit{host layer} is models exploits on hosts and "stealthy bridges" between VM instances due to VM co-residency, which may lead to attacks on the other layers.

\begin{figure}[htb]
\centering
    \includegraphics[width=.68\textwidth]{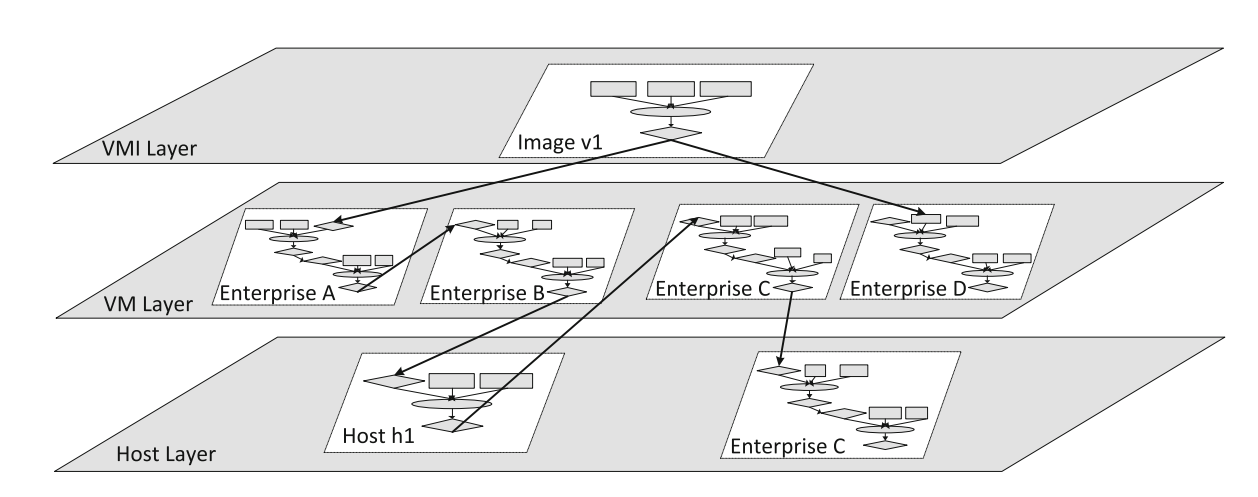}
  \caption{Concept of Cloud-level Attack Graph, original by \cite{sun2015inferring}}\label{fig:cloudy}
\end{figure}

Sun et al.\cite{sun2015inferring} use MulVAL to automatically generate their attack graphs: they introduce new primitive fact nodes as well as interaction rules on the VMI as well as host layer to capture stealthy bridges in their model. Using the attack graph, they derive Bayesian networks from it by removing the rule nodes, adding new nodes, instantiating the prior probabilities as well as the conditional probability tables.

The idea of connecting the local and inherited vulnerabilities from VMIs, as well as the stealthy bridges in the cloud network is essential to this work, which makes it an interesting reference for cloud security researchers. 

\subsection{Compromize graphs}
\emph{Compromize Graphs} were introduced by McQueen et al. in \cite{mcqueen2006quantitative} in a use case for a SCADA system. Compromize graphs were developed for network hardening and security quantification based on the time to compromise (TTC) metric \cite{mcqueen2006time}. Networks with smaller minimum TTC are easier to take over that those with higher TTC. This way, various configurations and countermeasures can be compared using compromise graphs. Furthermore, a user can use the TTC to simulate the dynamics of an attack by drawing samples from the corresponding Markov-chain process. 

Formally,\emph{Compromize graphs} are composed of nodes as well as weighted, directed edges. Nodes in the graph make up the atomic attack steps. The directed edges represent the sequential dependency between the attack steps. Furthermore, each edge holds a weight, which estimates the time required to perform the exploit. This value is derived from the difficulty of exploiting the underlying vulnerability, associated with the exploit, as well as the attacker skill level. A heuristic to find the value can be found in \cite{mcqueen2006time}.

According to \cite{kordy2014dag}, Leversage's and Byres' state-time estimation algorithm (STEA) \cite{leversage2008estimating, leversage2007comparing} is a direct spin-off of McQueen et al.'s work: "They combine a slightly modified TTC calculation approach with a decomposition of the attack according to the architectural areas of the targeted system."
The authors furthermore mention that Nzoukou et al.' work \cite{nzoukou2013unified} builds on this this approach. "The paper proposes to link the mean TTC to the CVSS metric values \cite{cvss2022} of specific vulnerabilities, which makes the employment of easily available inputs possible. To derive the overall mean TTC, the results of individual vulnerabilities are then aggregated using Bayesian networks. This allows us to lift the assumption that all attacking steps are independent."\cite{kordy2014dag} 

The link between TTC and CVSS metric values facilitates in determining the necessary parameters (except for the attacker skill level) and make this model interesting for applied researchers and practitioners, who seek to compare the tolerance of their network w.r.t. differing attacker skill levels. 

\begin{figure}
    \centering
    \includegraphics[width = 0.62\textwidth]{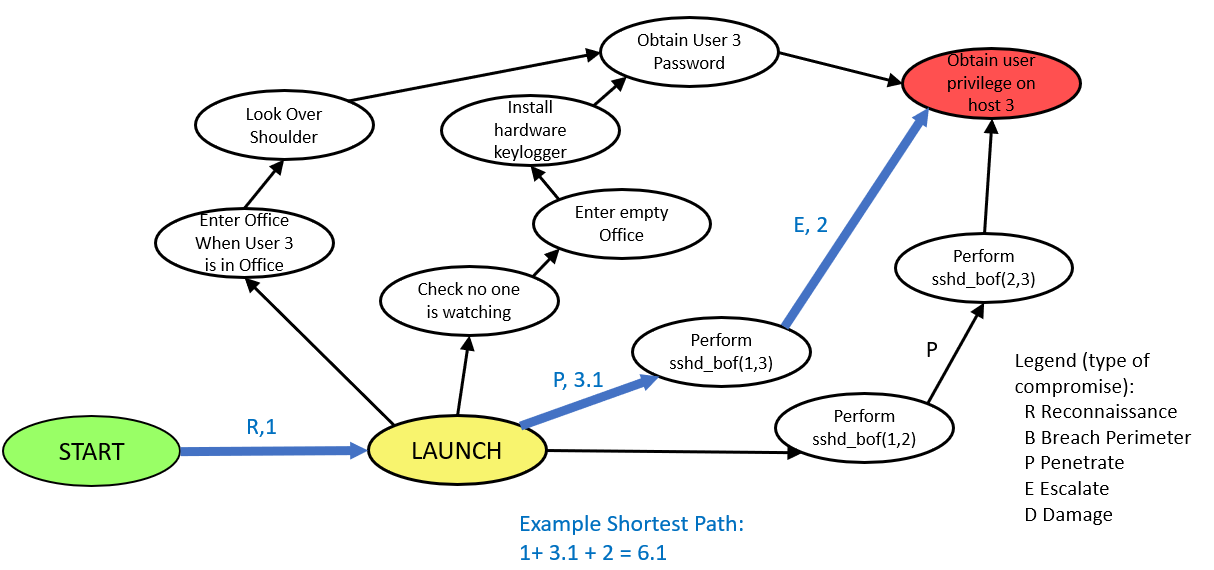}
    \caption{(Partial) Compromise Graph from \cite{mcqueen2006quantitative}. "It is a partial graph because it does not include values for each edge, it does not show all inter-node edges and does not show all target nodes" \cite{mcqueen2006quantitative}}
    \label{fig:compr}
\end{figure}

\subsection{Coordinated Attack Graphs}
Coordinated Attack Graphs (CAG) extend the notion of state enumeration attack graphs by allowing concurrent action attacks, as it is the case when multiple attackers coordinate themselves, e.g. for a DDoS attack. \cite{braynov2003representation}. CAGs are a multi-graphs of system states, which are truth assignments to some atomic formulae. Transitions are triggered by
joint actions which are concurrently executed by several attackers. The (total) order of the (joint) actions is represented as a chronological time line. 
To the best of pur knowledge, no other AG formalism is capable of representing concurrent and coordinated attacks by multiple attackers. 
\begin{figure}
    \centering
    \includegraphics[width = 0.55\textwidth]{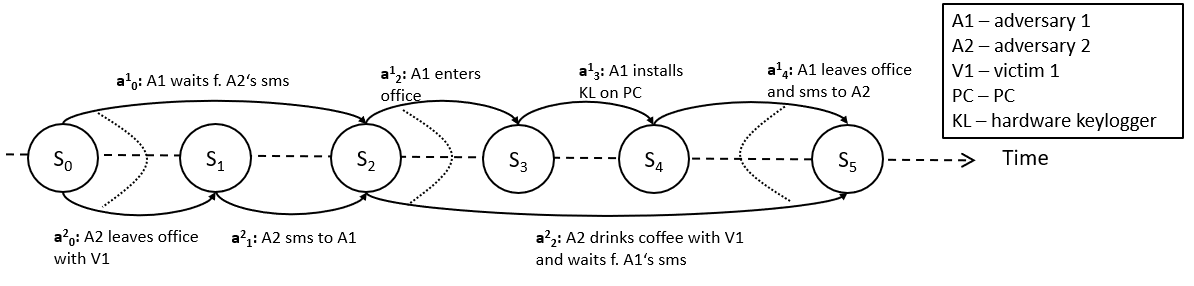}
    \caption{Coordinated Attack Graph\cite{braynov2003representation} for the use case: assume the attacker wants to install the hardware-keylogger, while a second malicious individual is having a coffee with User 3, in order to keep him out of the office. The second adversary sends his ally a text-message once they are out of sight.}
    \label{fig:coordinated}
\end{figure}

\subsection{Countermeasure Graphs}
In \cite{baca2010prioritizing}, Baca et al. introduce the \emph{Countermeasure Method for Software Security (CM-Sec)} as an extension to attack trees, including not only atomic attack steps and their dependencies, but also nodes for goals, actors and countermeasures. "The purpose of the countermeasure graph is to determine what security countermeasures would be useful to include in the product and at the same time identify what countermeasures already exist." \cite{baca2010prioritizing}. 

The nodeset in the countermeasure graph model comprises of goals (octagon), actors (hexagon), attacks (rectangles) as well as countermeasures (oval). Furthermore, a label indicating an integer value for the order of priorities is attacked to every nodes. Higher numbers indicate higher priorities and higher threats to the system in place. 

Edges connect actors to goals, to indicate which agent seeks to reach which goal; they may also connect attacks with actors whenever it it very likely that a specific agent has the capabilities to carry out the exploit. Last but not least, edges link countermeasures to the attack they prevent.

Using countermeasure graphs, hierarchical cumulative voting by Berander et al. \cite{berander2009evaluating}, a form of value propagation, can be performed to identify the most effective countermeasures using the node labels, which makes this formalism a usable and easy to comprehend model for countermeasure selection without using cost- or probability models. 

 \begin{figure}[htb]
\centering
    \includegraphics[width=.35\textwidth]{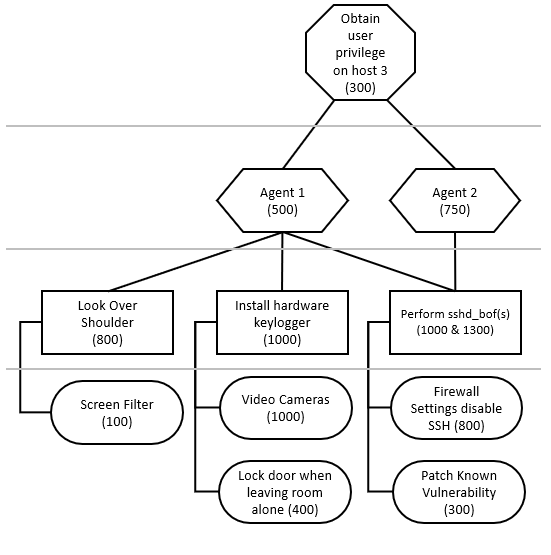}
  \caption{Countermeasure Graph by \cite{baca2010prioritizing} for the given scenario. Additionally we assume a second, more capable attacker from remote threatens the network.}\label{fig:countermeasuregraph}
\end{figure}

\subsection{Exploit dependency graphs} \label{sec:exploitdep}
\subsubsection{Exploit dependency graph by Ammann et. al.}
Until Ammann et. al work \cite{ammann2002scalable}, attack graph representations modelled all possible combinations of exploits in the attack graph. In their work, the authors incorporate monotonous logic to model attacks as dependencies among exploits, which yields the exploit dependency graph representation: the network states are described via a set of attributes, which can either be true or false, forming the nodes in the graph. Exploits are expressed as sets of edges in the graph, as one exploit can have multiple pre- and postconditions. 

The scalability of the representation improved from exponential to polynomial, by partitioning the set of attributes (nodes) into those that are satisfied in the initial state, and those that are initially unsatisfied. The idea is to divide the unsatisfied nodes into layers, according to their reachability: attributes that can be satisfied by applying a single exploit (i.e. reached via one edge) are placed in layer 1; those that require at least two exploits are placed in layer 2, etc. Thus, given the attack goal (it is modeled as a single attribute), forward reachability analysis is employed to assign each node/attribute the attack-step at which they are satisfied for the first time, yielding the attack graph. Notably, the authors present an algorithm to generate minimal attacks (in terms of levels) as well as a procedure to determine the set of exploits (edges) or attributes (nodes) that must be removed to disconnect the goal state from the initial state. This way, \cite{ammann2002scalable} represents the transition from explicit state enumeration frameworks and state-enumeration type AG models, towards dependency-type attack models.

 \begin{figure}[htb]
\centering
  \subfloat[Example of an Exploit (here: the buffer overfloe vulnerability from our running example) with three preconditions and one post-condition\cite{ammann2002scalable}]{%
    \includegraphics[width=.33\textwidth]{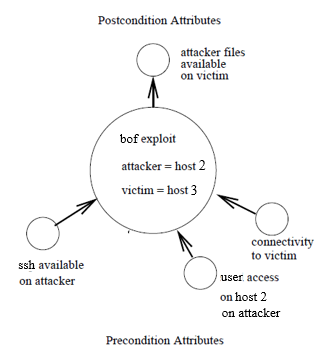}}\qquad
  \subfloat[Schema of an exploit dependency graph as a multi graph, acc. to \cite{ammann2002scalable}. Scalability is achieved by by partitioning the set of attributes in layers]{%
    \includegraphics[width=.33\textwidth]{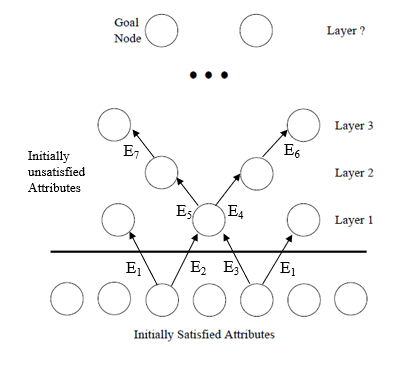}}
  \caption{Exploit dependency graph formalims by Ammann et al. \cite{ammann2002scalable}}\label{fig:ammann}
\end{figure}

\subsubsection{Exploit dependency graph by Noel \& Jajodia}
 Noel and Jajodia \cite{noel2003efficient} inherit the notions of \emph{exploit dependency graphs}.  Given a set of exploits, they automatically generate a directed graph of the 
dependencies among exploits and conditions, called \emph{exploit dependency attack graph}. This graph consists of three disjoint sets of vertices $(N_p, N_a, N_e)$ representing initial conditions, attribute nodes and exploit nodes; a set of edges $E$ mapping attributes/initial conditions to exploits or exploits to attributes; and a map $\zeta: E \mapsto E$ representing the logical relationship between adjacent edges. The latter represents whether all parent edges have to be true ($\land$) or, whether it is sufficient to have one parent edge set true ($\lor$) to set the child edge true. „In the figure, exploits appear as ovals, and conditions appear as plain text (except the goal condition, which is marked with a triple octagon). Numbers in parenthesis identify associated machines.“ \cite{noel2003efficient}. They then built the graph by foreward-chaining initial conditions to exploit preconditions, whose exploit postconditions are again mapped to another exploit pre-conditions and so forth. 
They resolve cycles by considering the graph distances of the given conditions from 
the initial conditions. 

Noel et al. further compress this representation to compute network hardening measures via \emph{resolved exploit dependency graphs}: Given an exploit dependency graph, the authors resolve the dependency terms by logically equivalent terms in a backward direction. Starting from the goal-condition exploit, the respective symbolic expression is replaced with the conjunction of its preconditions. Then the preconditions are replaced with the exploit that yields it as a postcondition in a recursive fashion; if there are several exploits having this postcondition,the disjunction of all such exploits is formed. The corresponding conditions are fond by traversing the dependency graph in a breadth first fashion. „Once the dependency graph has been fully traversed, the result is an expression for the safety of the
attack goal in terms of the initial conditions. The resulting expression for the attack goal allows us to decide if the goal is safe for a particular assignment of initial conditions.“ In consecutive works of the same authors, such as \cite{jajodia2005topological}, the formalism is simply denoted as \emph{dependency graph} (not to be confused with dependency-type attack graphs, cf. Section \ref{sec:AG-meta}).

Using this representation, minimum-cost network hardening is performed by computing a cost-minimal set of initial conditions that guarantee some safety condition: The expression for the attack goal in terms of initial conditions is then converted to the canonical conjunctive normal form, expressing the attack goal as a conjunction of maxterms\footnote{A maxterm function of Boolean variables that contains all variables in negated or non-negated form. They must all be linked by logical ORs.\cite{lee1995representations}}, each representing a safe assignment of initial conditions. "For network hardening, each non-negated initial condition in a maxterm must be assigned false to provide safety. [...] In choosing among hardening options, our approach is for the network administrator to assign a cost to each individual hardening measure (initial condition). We then select the maxterm with the lowest total cost, where we sum only the costs for non-negated [...] initial conditions."

Another application thereof is Topological Vulnerability Analysis (TVA) \cite{jajodia2005topological, jajodia2010topological}
information on network vulnerabilities (from network scans) and correlating them with existing exploits on them to build a dependency attack graph, comprised of all paths leading from the initial network state to a given network target. The intention of TVA is to complement manually performed pen-testing, identify the most suitable network configuration in the network design phase and determine the most critical network assets with respect to overall impact on network security.

\subsubsection{Vulnerability dependency graphs} 
\emph{Vulnerability dependency graphs} (VDG) \cite{10.1145/2699907} are closely related to dependency graphs. According to \cite{10.1145/2699907}, various other AG formalisms from security literature \cite{dacier1994privilege, ritchey2000using, sheyner2002automated, jha2002two, jajodia2005topological} can be translated into VDGs. 

To define VDGs, let $PR$ denote a set of software
products, $V$ a set of vulnerabilities and $PA$ a set of patches available. Each patch is able to mitigate at least one vulnerability, but there might exist vulnerabilities with no patch at all - or numerous patches. Then, vulnerability dependency graphs are directed acyclic graphs where the node set is the set of vulnerabilities and the edges connect vulnerabilities, such that an outgoing edge from vulnerability $v_1$ to vulnerability $v_2$ is added, if and only if $v_2$ can only be exploited, if $v_1$ is exploited. Root nodes may be directly exploited. Additionally, each vulnerability $v$ is labelled with an impact score $Impact(v)$, measuring the impact of leaving the vulnerability unpatched; the values of $PR(v)$, which are the software products affected by this vulnerability; as well as $PA(v)$, a list of patches mitigating vulnerability $v$. Visually, $Impact(v), PA(v)$ and $PR(v)$ are enclosed by dotted squares and linked to the corresponding node, i.e. vulnerability $v$. 

If each patch is assigned associated patching costs $CostD(pa)$ and each product labelled with an associated productivity value $Prod(pr)$, the authors propose an algorithm that finds Pareto-optimal solution defending the system while simultaneously maximizing productivity and minimizing the patching costs. Thus, VDF are designed for product developers who seek to find the most efficient way of patching vulnerabilities in their products.
 \begin{figure}[htb]
\centering
    \includegraphics[width=.57\textwidth]{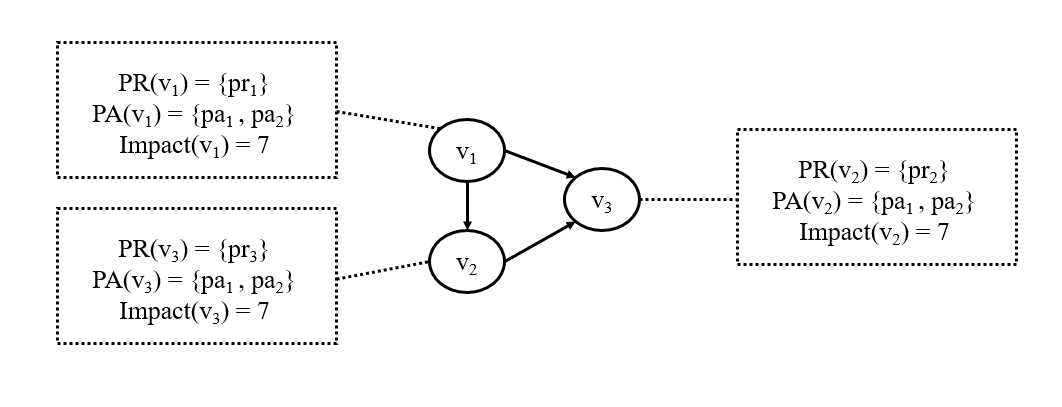}
  \caption{Vulnerability Dependency Graph \cite{10.1145/2699907} for the three buffer overflow vulnerabilities from our example. Two patches are available that would fix all three vulnerabilities.}\label{fig:vulndepgraph}
\end{figure}

\subsection{Generalized dependency graph, probabilistic temporal attack graphs, attack scenario graphs}
In \cite{10.5555/2041225.2041255}, Albanese et al. introduce \emph{Generalized Dependency Graphs (GDP)}, \emph{probabilistic temporal attack graphs} (PTAG) and \emph{attack scenario graphs} in their work. The GDP model incorporates the service-dependencies with respect to the underlying network infrastructure, while probabilistic temporal attack graphs extend compromise graphs and the mean time to compromize \cite{leversage2008estimating} by introducing the more fine-granular \textit{time stamp distributions}. \emph{Attack scenario graphs} combine both GPD and PTAG in a single representation.

To define the GDP, the definition of an \textit{attack graph} by Wang et. al. \cite{WANG20062917} is needed,
Wang et al. define an attack graph as a directed graph having having either exploits, i.e. triple $(vul, src, dest)$ or security conditions (e.g. privilege levels, trusts) as nodes. In \cite{10.5555/2041225.2041255}, exploits are described as host-bound vulnerabilities $V$, the conditions are designated as $C$. Edges are used to represent the dependencies between security conditions (ovals) and exploits (rectangles), i.e. an edge may point from an exploit to a security condition, which yields the \textit{implies} relation $R_r \subseteq C \times V$.V ice versa, edges from security conditions to exploits yields the \textit{requires} relation $R_r \subseteq C \times V$.
Thus, the attack graph is defined as $G = (V \cup C, R_r\cup R_i)$. 

\subsubsection{Generalized Dependency Graph}
Albanese et al. assume that the status of each network entity (given the status of the other entities) is represented as a dependency function $f: [0,1]^n \to [0,1]$ with $f(0,..., 0) = 0$ as well as $f(1,..., 1) = 1$ to model the service-dependencies with respect to the underlying network infrastructure. Let $\mathcal{F}$ denote the set of dependency functions. Then the \emph{generalized dependency graph} $D=(H, Q, \Phi)$ is defined as a directed acyclic graph with labels, where $H$ is the nodeset, representing the network entites, such as hosts; $Q \subseteq H \times H$ is the set of edges; and $\Phi: H \to \mathcal{F}$ is a map that links each node $h$ to a dependency function $f\in \mathcal{F}$ s.t. the number of arguments of $f$ is equal to the outdegree of $h$.

\subsubsection{Probabilistic Temporal Attack Graphs}
Albanese et al. also extend the plain attack graph model by incorporating a \textit{timespan
distribution} $\omega \in \Omega$, which is a pair $(I, \rho)$ of disjoint time intervasl $I$ and an incomplete probability
distribution $\rho$ over $I$ (it needs not be a complete distribution as the attackers might not execute the exploit at all). The timespan distribution serves as a means encoding probabilistic knowledge about the adversaries attack patters and temporal constraints on the attack. "Intuitively, a timespan distribution $(I, \rho)$ specifies a set $I$ of disjoint time intervals when a given exploit might be executed [...] $\rho(x, y)$ is the probability that the exploit will in fact be executed during the time interval $[x, y]$." \cite{10.5555/2041225.2041255} 

Then, given a timespan distribution and a DG $G = (V \cup C, R_r \cup R_i)$, \emph{(Probabilistic) Temporal Attack graphs} \cite{10.5555/2041225.2041255} are labelled DAGs, defined as quadruples $A = (V, E, \delta, \gamma)$ where $V$ are the vulnerabilities or, likewise, their associated exploits, $E = R_i \circ R_r$ the set of edges connecting them; $V^s$ and $V^e$ are the start node(s) and the unique end node respectively; $\delta$ is a mapping that labelling each edge in the graph with a timespan distribution and $\gamma$ is a function mapping each exploit to its the clause of preconditions that need to be fulfilled in order to mount the exploit. 

\subsubsection{Attack Scenario Graphs}
"Intuitively, attack scenario graphs merge attack and dependency graphs introducing edges between vulnerability exploits and network entities encoding how
the latter are affected by the former." \cite{10.5555/2041225.2041255}. Thus, attack scenario graphs fuse a probabilistic temporal attack graph $A = (V, E, \delta, \gamma)$ and a corresponding generalized dependency graph $D=(H, O, \Phi)$ to an
an \emph{attack scenario graph} $(A, D, F, \nu)$ where $F \subseteq V \times H$ links the hosts to their vulnerabilities or exploits and  $\nu: F \to [0,1]$ labels each vulnerability-host pair $(v, h) \in F$ with the performance reduction factor $h$ encounters when compromised due to vulnerability $v$. Thus, attack scenario graphs can be used to simulate or model the stages of a cyber-attack, monitor the network status while at the same time assessing the potential losses in service performance. The authors also present a data structure for processing incoming IDS alert and a method for a real-time alert correlation, which indexes incoming intrusion alerts, as well as an algorithm to update the index. They furthermore propose an algorithm to predict and rank potential future attack scenarios on the network based on scenarios based on the aforementioned data structures. The methods were tested using synthetic data and were operating in real time - the large amount of manual work to construct the status functions and timespan distributions unfortunately limit the applicability of the model.


\begin{figure*}[htb]
\centering
  \subfloat[Attack Graph by \cite{WANG20062917}]{%
    \includegraphics[width=.37\textwidth]{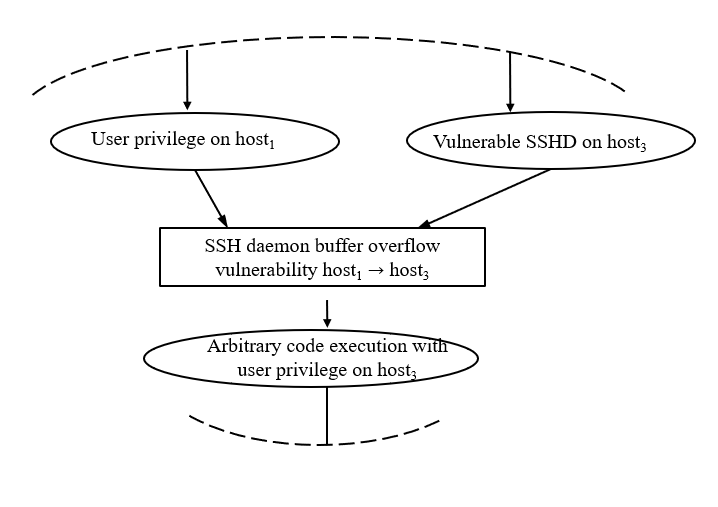}}\hfill
     \subfloat[\textcolor{black}{Generalized Dependency Graph: }\cite{10.5555/2041225.2041255}]{%
    \includegraphics[width=.42\textwidth]{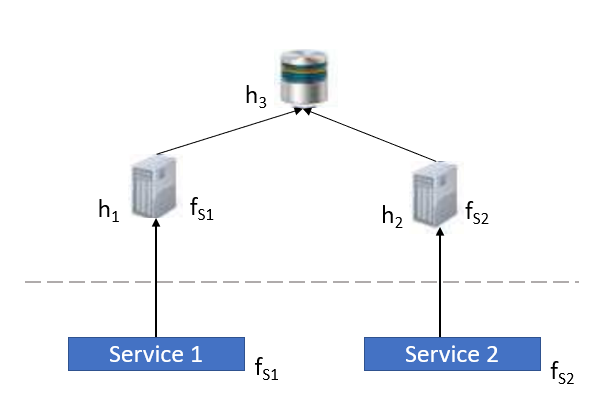}}\\
  \subfloat[Probabilistic Temporal Attack Graph\cite{10.5555/2041225.2041255} for the scenario ]{%
    \includegraphics[width=.35\textwidth]{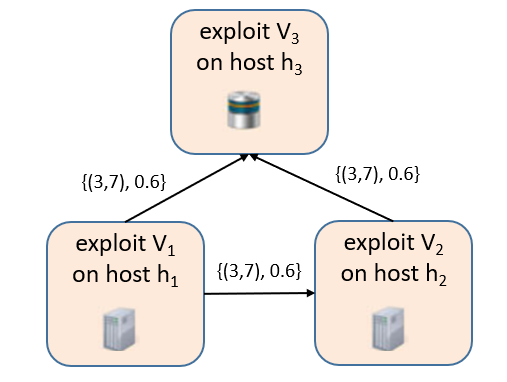}}\hfill
     \subfloat[Attack Scenario Graph \cite{10.5555/2041225.2041255} combining the GDG and the PTAG]{%
    \includegraphics[width=.42\textwidth]{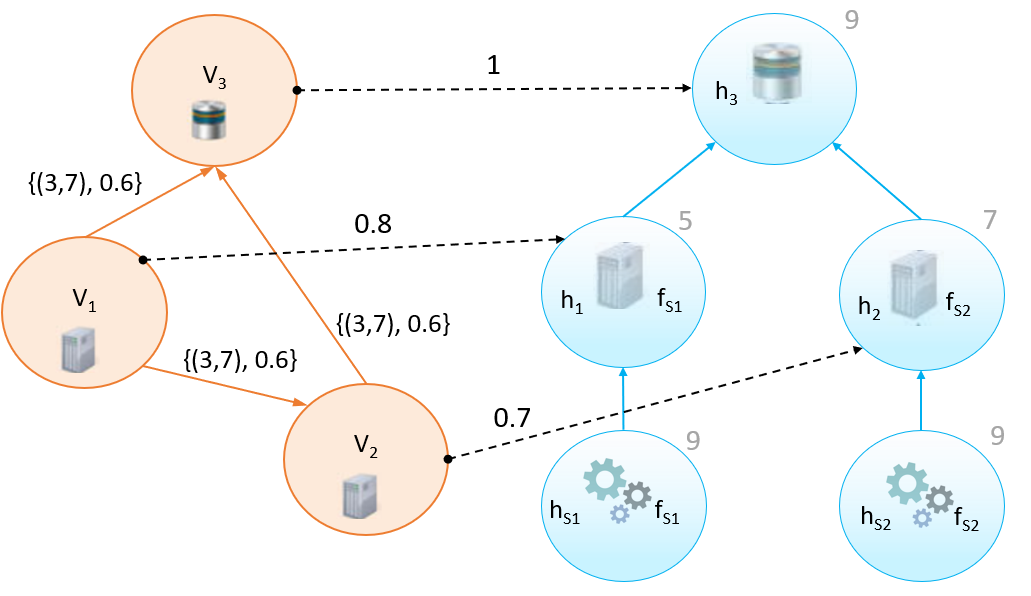}}
  \caption{GDG, PTAG and Attack Scenario Graph for the use case. Additionally, we assume that host 3 is especially critical, as it contains the databases needed for service 1 and service 2. To assess damage, each host is given a theoretical utility $u(h)$ (in grey) whenever the host $h$ is fully operational. The damage $\Delta u(h)$ for each host is then quantified by multiplying the service reduction factor with the utility of the host. The overall damage is the sum of damages.} \label{fig:generalized}
\end{figure*}
\subsection{Evidence Graph}
Wang and Daniels\cite{wang2008graph} have developed the evidence graph formalism which is build from collected forensic intrusion evidence. First, intrusion alerts are preproccessed to hyper-alerts based on the similarity of their attributes as well as context requirements. Then, the evidence graph model, a host-based graph, is generated from this intrusion data. The aim of the analysis is evidence presentation as well as automated local reasoning to infer if the network components still functioning accordingly. Furthermore, a method to extract important nodes from the network and "extract groups of densely correlated participants in the attack scenario" \cite{wang2008graph}.

Wang and Daniels define evidence graphs as quadruples $G = (N, E, L_N, L_E),$ where $N$ is the set of hosts or host-level entities, $E$ is the set of directed edges, representing preprocessed forensic evidence. Furthermore, $L_N$ is the set of node attributes (which are labels) and $L_E$ is the set of edge attributes." \cite{wang2008graph}
The set of labels for the nodes include the node IDs; a set of fuzzy functional states, indicates the belief whether the current node is an Attacker, Victim, Stepping Stone, Affiliated; time stamps; as well as the Value $\in [0,1]$ attribute, indicating the (relative) asset value of the host. Thus, evidence graphs are means to model intrusion evidence and their inter-dependencies.

The set of labels for the edges includes the Weight attribute, a fuzzy value $\in [0,1]$; a Relevancy attribute, specifying whether the evidence indicates a relevant action (true positive) or an irrelevant action (false positive, non-relevant true positive); a fuzzy parameter for Host Importance $\in [0,1]$, denoting the criticality of individual hosts. 

\subsection{Exploitation Graph}
In \cite{li2006approach}, Li et al. introduce exploitation graphs as a means of automatically generating attack graphs for the system in use. Their approach is based on a vulnerability knowledge
database. In this database, single vulnerabilities are represented using a template, according to a pre- and postcondition semantic. \cite{li2006approach}

Their modeling process is performed in two stages: first, the aforementioned vulnerability knowledge base with vulnerability graphs is created. The template to model vulnerabilities is represented in Table \ref{tab:template}.

\begin{tiny}
\begin{table}[]
\begin{tabular}{|l|ll|l|}
\hline
\multicolumn{1}{|c|}{\textbf{Condition}}                                             & \multicolumn{2}{c|}{\textbf{Category}}                                             & \multicolumn{1}{c|}{\textbf{Name}} \\ \hline
\multirow{11}{*}{\textbf{\begin{tabular}[c]{@{}l@{}}Pre-\\ conditions\end{tabular}}} & \multicolumn{1}{l|}{\multirow{4}{*}{\begin{tabular}[c]{@{}l@{}}Operating\\ System\end{tabular} }}    & Name                   & OS\_name                           \\ \cline{3-4} 
                                                                                     & \multicolumn{1}{l|}{}                                     & Version                & OS\_version                        \\ \cline{3-4} 
                                                                                     & \multicolumn{1}{l|}{}                                     & Architechture          & OS\_archi                          \\ \cline{3-4} 
                                                                                     & \multicolumn{1}{l|}{}                                     & Kernel                 & OS\_kernel                         \\ \cline{2-4} 
                                                                                     & \multicolumn{1}{l|}{\multirow{2}{*}{Application}}         & Name                   & App\_name                          \\ \cline{3-4} 
                                                                                     & \multicolumn{1}{l|}{}                                     & Version                & App\_version                       \\ \cline{2-4} 
                                                                                     & \multicolumn{1}{l|}{\multirow{2}{*}{Access}}              & Range                  & Access\_range                      \\ \cline{3-4} 
                                                                                     & \multicolumn{1}{l|}{}                                     & User Level             & Access\_level                      \\ \cline{2-4} 
                                                                                     & \multicolumn{1}{l|}{\multirow{3}{*}{Additional}}          & Open Port(s)           & Addition\_port                     \\ \cline{3-4} 
                                                                                     & \multicolumn{1}{l|}{}                                     & Running Application(s) & Addition\_runapp                   \\ \cline{3-4} 
                                                                                     & \multicolumn{1}{l|}{}                                     & Other                  & Addition\_other                    \\ \hline
\multirow{6}{*}{\textbf{\begin{tabular}[c]{@{}l@{}}Post-\\ conditions\end{tabular}}} & \multicolumn{2}{l|}{Availability}                                                  & Availability                       \\ \cline{2-4} 
                                                                                     & \multicolumn{2}{l|}{Confidentiality}                                               & Confidentiality                    \\ \cline{2-4} 
                                                                                     & \multicolumn{2}{l|}{Integrity}                                                     & Integrity                          \\ \cline{2-4} 
                                                                                     & \multicolumn{1}{l|}{\multirow{3}{*}{\begin{tabular}[c]{@{}l@{}}Security\\Protection\end{tabular}}} & Super User Access      & SecPro\_superuser                  \\ \cline{3-4} 
                                                                                     & \multicolumn{1}{l|}{}                                     & User Access            & SecPro\_user                       \\ \cline{3-4} 
                                                                                     & \multicolumn{1}{l|}{}                                     & Other Access           & SecPro\_other                      \\ \hline
\end{tabular}
\caption{Template for preconditions and postconditions \cite{li2005building} to create exploitation Graph} \label{tab:template}
\end{table}
\end{tiny}

Next, the exploitation graph is tailored to the system in use by adding system configuration information as well as data from scanners to the vulnerability database. The representation and algorithm to build exploitation graphs are similar to Amman et al.\cite{ammann2002scalable}, but their formalism uses nodes to represent exploit and edges to represent state changes instead of the multi-graph formalism. Then, as in \cite{ammann2002scalable}, the authors pose a monotonicity assumption, stating that a (successful) exploit is never repeated. "The result of this process is a layered structure of network exploitations, in which the  execution of an exploitation in a layer depends on exploitations in the lower layers. This process starts with an empty set of exploitations at the lowest layer, which means no exploitation has been performed before an 
attack. This process proceeds by checking pre-conditions of non-executed exploitations against available system attributes, Newly-available exploitations and newlysatisfied system attributes are added into new layers. This incremental process ends when there are no newly available exploitations or the goal state is reached. "\cite{li2005building}

Due to the monotonicity, there is no need to backtrack in the creation of exploitation graphs. The exploit-oriented graph structure adtioninally reduces the complexity. After the exploitation graphs are created, abstraction techniques based on exploitation similarity, and extended exploitation similarity are used to reduce the complexity of exploitation graphs- these are described in detail in \cite{li2005building}. 

\subsection{Full -, Predictive- and Node-Predictive Attack Graph; Host-centeric, host-access and host-based attack graph}
\subsubsection*{Full Attack Graphs}
According to \cite{artz2002netspa, lippmann2006validating}, \emph{full attack graphs} show all possible paths or sequences of compromised hosts and vulnerabilities an attacker may exploit to capture the network. The vertices in full attack graphs represent hosts and the edges represent vulnerabilities that may be exploited to compromise the hosts. 
"The earliest work in this line is the Kuang system \cite{baldwin1994kuang}, and it’s extension to a network environment, the NetKuang system \cite{zerkle1996netkuang}. In these systems a backwards, goal-based search scans UNIX systems for poor configurations. The output is a combination of operations that lead to compromise"\cite{ammann2002scalable}. NetSPA, in  the version of \cite{artz2002netspa} also produces full attack graphs. In the updated version of NetSPA\cite{lippmann2005evaluating}, full graph are generated in a breadth-first manner: Starting from the root node, other nodes, representing target hosts, are removed from the queue and analyzed for reachability. Whenever a host is reachable via a port, all vulnerabilities on the port are tried out. In case the vulnerability is exploitable, the exploit is added to the graph. In case the target host does not exist in the queue, the target nodes is created first. Note that if there exist more than one vulnerability to compromise a specific host, multiple edges may be added – one per vulnerability. 
\begin{figure}[htb]
\centering
  \subfloat[Full attack graph \cite{ingols2006practical} for the running example]{%
    \includegraphics[width=.48\textwidth]{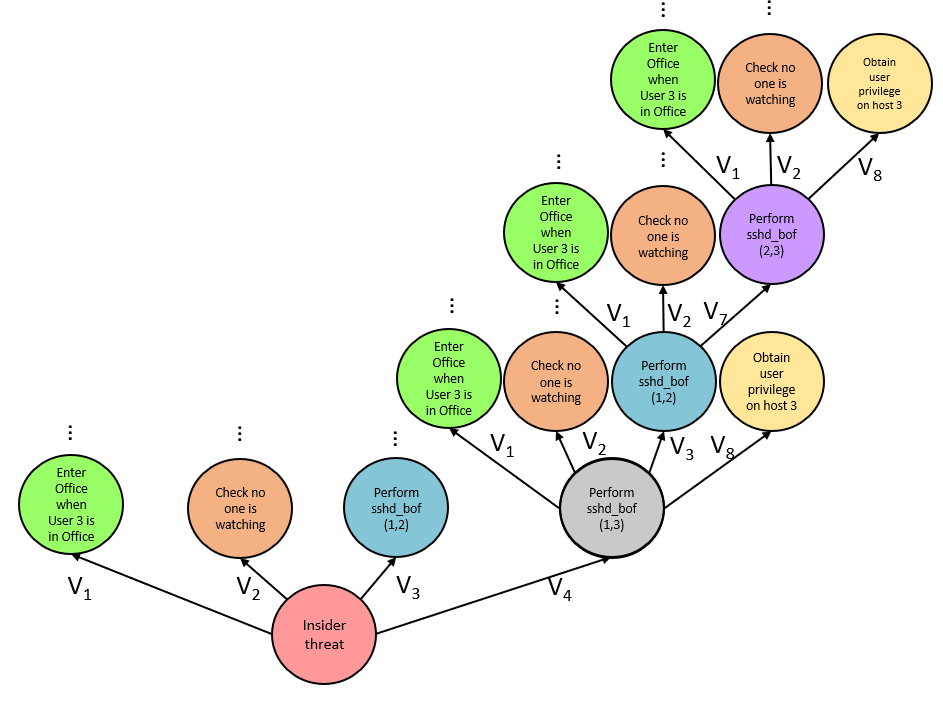}}\\
  \subfloat[Predictive attack graph corresponding to the full AG from (a) \cite{ingols2006practical}]{%
    \includegraphics[width=.33\textwidth]{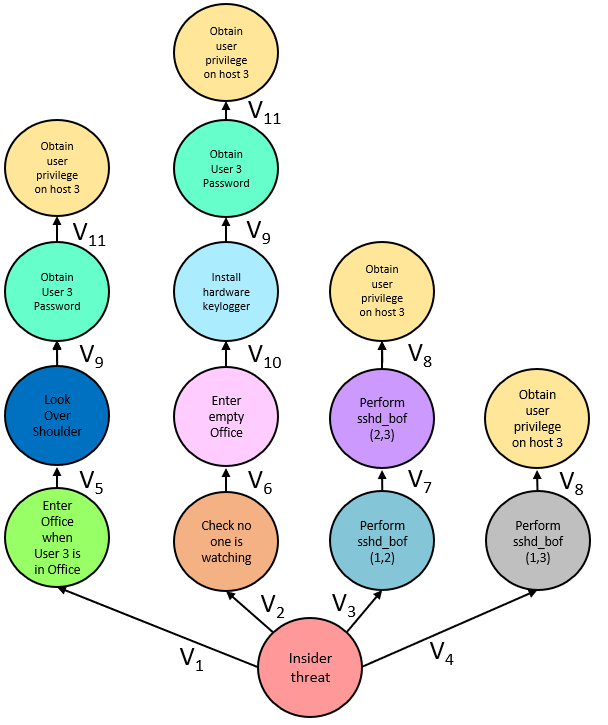}}\hfill
  \caption{Attack graph formalisms: full and predictive attack graph}\label{fig:attacktreeformalisms3}
\end{figure}

\begin{figure}[htb]
\centering
    \includegraphics[width=.57\textwidth]{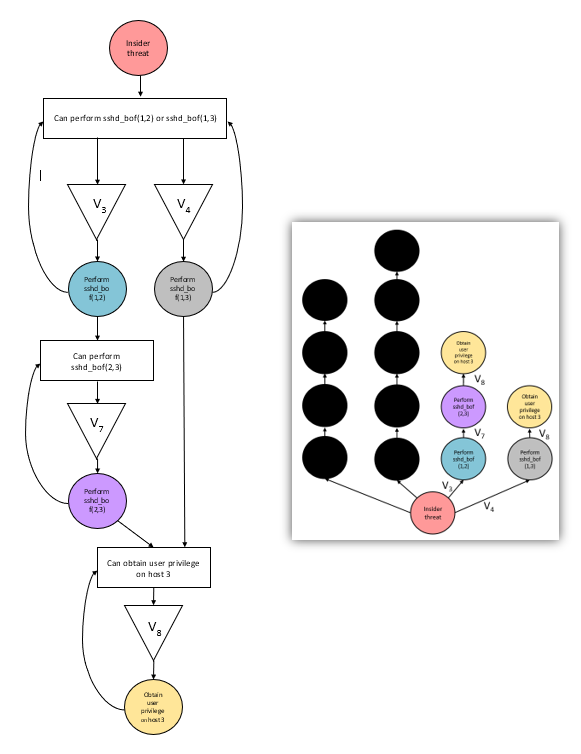}
  \caption{Multiple Prerequisites graph \cite{ingols2006practical} corresponding to the predictive AG in the gray square, which represents the right two branches of the predictive graph in \ref{fig:attacktreeformalism3} (b) }\label{fig:attacktreeformalisms3}
\end{figure}

\begin{figure}[htb]
\centering
    \subfloat[Predictive (left) attack graph: assume the same vulnerability resides on $200$ user machines and it is available via two paths. The corresponding Node-Predictive (right) attack graph \cite{ingols2006practical} represents this fact in a more compact manner. Original graphic from \cite{ingols2006practical}]{%
    \includegraphics[width=.37\textwidth]{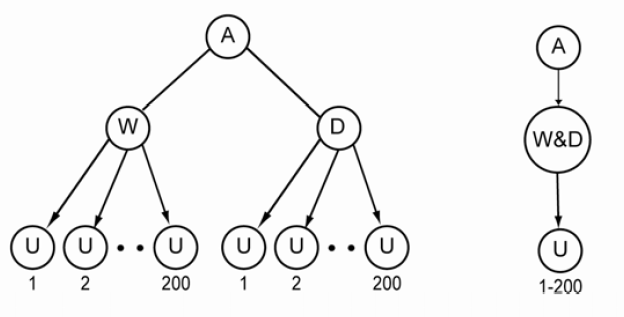}} \qquad
  \subfloat[Predictive (left) and corresponding Host-compromised attack graph (right). Original graphic by\cite{lippmann2005evaluating}]{%
    \includegraphics[width=.3\textwidth]{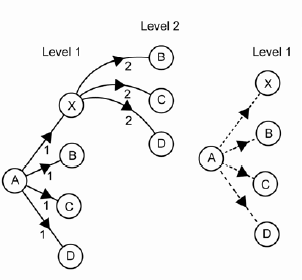}}\hfill
  \caption{Attack graph formalisms}\label{fig:attacktreeformalisms72}
\end{figure}

\begin{figure}[htb]
\centering
  \subfloat[Access-graph, host-access graph, host-based access graph \cite{ammann2005host} for the use case]{%
    \includegraphics[width=.38\textwidth]{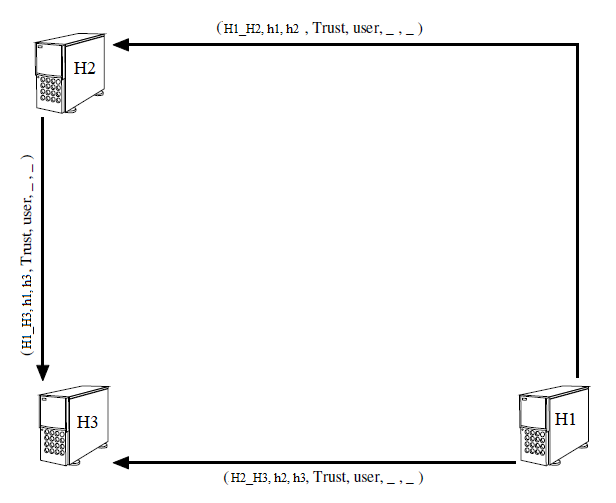}}\qquad
\subfloat[Host-based attack graph for the use case \cite{malthotra2008host}. Probabilities were assigned randomly (for illustration)]{%
    \includegraphics[width=.35\textwidth]{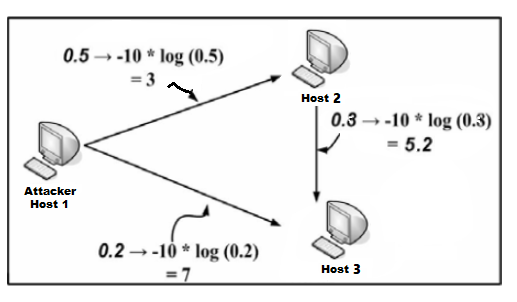}}
  \caption{Attack graph formalisms with hosts}\label{fig:hostagform}
\end{figure}
Full attack graphs, however, are intractable for almost all real networks, and provide more information than required for many application. Therefore, in it first version, NetSPA offered a function to prune the full graph attack graph to contain only paths leading to a predetermined goal state \cite{artz2002netspa}. 

\subsubsection{Host-Compromised Attack Graphs}
In case one is only interested in the fact, whether the system can be compromised or not, and in case, the highest privilege level an attacker may gain, Ingols, Lipman et al. present \emph{host-compromised attack graphs} \cite{lippmann2005evaluating} an alternative to full attack graphs. "Nodes in a host-compromised graph indicate all hosts that can be compromised by the attacker and the level of privilege gained. Edges indicate one\footnote{and only one} of possibly many sequences of vulnerabilities that can lead 
to the compromise" \cite{lippmann2005evaluating}. 
Hence, the host-compromised graph cannot be employed to check whether patching a given 
vulnerability blocks the attacker's path to the goal, since it would require building the whole attack graph, i.e. the full attack graph is more suitable to do so. 

Based on the company's patent description \cite{cohen2005system}, \cite{ingols2006practical} claim that the attack graph software Skybox View \cite{acm:skybox} uses such host-compromised graphs, including only the shortest attack paths to the compromised hosts. 

\subsubsection{Host-access Graph}
Note that there are similarities to Amman et al.'s \cite{ammann2005host} "host-centered" model that was developed at the same time. Their \emph{access-graph} (host-access graph, host-based access graph acc. to \cite{hewett2008host}) also presents hosts as nodes, rather than states or conditions. Furthermore, it only visualizes the "degrees of compromise" on the host instead of an exhaustive representation of all possible attack sequences to complement classical attack graph or attack tree approaches.

Given a set of hosts $H$, access edges $E$, exploits $X$ and vulnerabilities $V$, the model automatically outputs a graph and the maximal obtainable privilege between the hosts in the network by leveraging each host's exploits. "If sufficient connectivity on the appropriate ports exists between two hosts, an exploitable vulnerability exists on the destination host, and the source host has all of the prerequisites for the exploit, an edge can be added from source to destination. For book- keeping purposes, edges are tagged with a route ID, source host, destination host, the means by which the edge was achieved (trust relationship, exploit name, etc.)." \cite{ammann2005host}

\subsubsection{Host-based attack graph}
Malhotra et al. \cite{malthotra2008host} also describe their attack graph as \emph{host-based attack graph}. Instead of computing all possible attack paths, their method identifies the most critical attack path in terms of an attack surface measure, combining \textit{access 
levels} (specified in an access control matrix model, as described in \cite{bishop2005introduction}) with the \textit{attackability} of the host. The latter is defined as "the cost  benefit ratio of the damage potential of the resource and the effort that the attacker has to spend in order to 
acquire that resource. [...] SA is the sum total of the attackability of the system attack class." \cite{malthotra2008host}

First, they relax the access control matrix by pruning all edges, except for the ones representing the highest access level between the hosts. The access levels are then quantified in a way that preserves the total ordering of the access levels. Then, given the relaxed matrix and the individual attack surface measures $SA$ of all nodes, the host-based attack graph is computed. 

Additionally, for probabilistic analysis, the authors label the resulting edges with the probability of an attacker choosing a specific edge: each edge is assigned proportional of to the number of critical vulnerabilities reported for the host. 

\subsubsection{Host-centric attack graphs} 
Hewett et al. \cite{hewett2008host} adopt the host-centric approach and propose \emph{host-centric attack graph}, which can be generated using a modified version of NuSMV at improved scalability as follows: "The states of a state transition system in model checking traditionally represent attributes of the entire system. [...] We propose host-centric model checking, in which each state is specified by the attributes of a single host. [...] The fact that almost all exploits are concerned with attributes of only an attacking host and a victim host makes the change from network-level to host-level pre(post)-conditions transparent". 

In their model, the attributes of the individual hosts are services, software vulnerabilities, configuration vulnerabilities, connectivity, trust relationships and attacker’s access privilege. All host attributes except for the latter are considered static under the common monotonicity assumption that acquired privileges are never lost. The transitions in their state transition system may either be exploits or the attacker’s atomic actions. Transitions are specified by the following attributes; attacker pre-conditions, host pre-conditions, attacker post-conditions, host post-conditions as well as exploit mode (e.g. remote or local). 

In their host-centric attack graph, each node is a pair consisting of the host name $h$ and the access privilege $a$ of an attacker. The network-centric attack graph contains the attributes of all hosts, and the nodes are labelled by an $n$-tuple consisting of the attacker access privileges on all $n$ hosts in the network. Furthermore, the authors show how to generate the host-based access graph from \cite{ammann2005host} from their host-centric AG. 

\subsubsection{Predictive attack graph}
While offering a compact representation and good scalability, both formalism only presents the hosts that may be compromised maximum one exploit to do so, which is why these formalism "cannot be used to determine whether eliminating a vulnerability on an edge prevents compromise or whether there are other ways to compromise hosts." \cite{lippmann2005evaluating} 
For this purpose, \cite{lippmann2005evaluating} developed the notion of \emph{predictive attack graph}. "As in the full graph, nodes represent hosts in the network which the attacker has compromised, and edges represent vulnerabilities the attacker could use to achieve the compromise. However, all of the redundant paths have been removed. The remaining structure fulfills the predictive requirement."\cite{lippmann2005evaluating} 

\subsubsection{Node-Predictive attack graph}
An alternative approach to the predictive attack graphs, also introduced in \cite{lippmann2005evaluating}, is the
\emph{node-predictive attack graph}. The idea of the node-predictive attack graph is to cope with seemingly redundant structures in predictive attack graphs. To illustrate this redundancy, the authors make use of a sample network, consisting of the Attacker Host (A), a firewall, a Web Server (W), a Database Server (D) as well as 200 User host machines (U), all suffering from the same remote-to-admin vulnerability. Despite representing one and the same vulnerability, the associated predictive attack graph would have 400 leaf nodes - two for every host machine. To overcome unnecessary complexity, the node-predictive attack graph formalism is employed: first, hosts are grouped together using a technique called Dynamic Host Collapse (DHC). Then the predictive attack graph is build using the host groups instead of all individual hosts, because "if two hosts can be compromised at the same level (user, administrator, other, DoS) from every host the attacker has already compromised, then the hosts are equivalent from the point of view of the attack graph." \cite{lippmann2005evaluating}. Thus, node compromizes AG are a useful form of visualizing scenarios, where a vulnerability is present in repetitive structures in the network.

\subsubsection{Multiple Prerequisites graph}
In subsequent works, Ingols et al. \cite{ingols2006practical, williams2008interactive} developed the \emph{multiple-prerequisite graphs} formalism (MP graphs), which replaced the predictive attack graph in an updated version of the NetSPA suite \cite{lippmann2005evaluating}. MP graph consists of unlabeled edges and three node types: states, prerequisites and vulnerability instances. 
\begin{itemize}
\item State nodes are represented as circles. They depict an attacker’s privilege or access level on a particular host. Outgoing edges always point to the prerequisites they provide to an attacker. 
\item Prerequisite nodes are represented as rectangles. They may represent a reachability group or a credential data type. Outgoing edges can only point to the vulnerability instances that require the prerequisite for successful exploitation. 
\item Vulnerability instance nodes are represented as triangles. They each represent a specific vulnerability on a given port. Outgoing edges point to the state the attacker may reach when exploiting the vulnerability.  
\end{itemize}
Thus, the order of these node types in the graph is given as follows: states point to prerequisites, which point to vulnerability instances the attacker may exploit. Vulnerability instances in turn point to states they provide to the attacker. Implicitly, the attack paths are monotonic, as acquired privileges/access levels are never lost.

MP graphs are faster to generate while at the same time offering more options of representing AG. "Their visualization employs spatial grouping and color-coding to represent levels of potential compromise. Groups of machines with similar levels of exploitability can then be collapsed, reducing the overall complexity of the graph." \cite{homer2008improving}
\cite{lippmann2006validating} present a recommender based on MP graphs, stating a list of vulnerabilities that need to be patched in order to prevent attackers from reaching a given prerequisite. MP graphs are efficient. They are, however, not very suitable for visual display. 

\subsection{Hybrid attack graph}
Louthan et al. \cite{louthan2014hybrid} extend traditional attack graphs introducing \emph{hybrid attack graphs} (HAG). "The objective of the HAG is to capture blended attack vectors (comprising discrete and continuous exploit events) in CPSs and  hybrid systems. As in hybrid automata, a hallmark of hybrid systems is that their discrete behavior places them into a mode in  which the passage of time causes some kind of state evolution.  Likewise, a hybrid or blended attack takes the form of a discrete action within a mode causing the system to evolve in some malicious way. The hybrid extensions for stateful attack graphs [...] address these considerations in two ways. The first is the  purely mechanical addition of real valued continuous qualities and topologies. [...]The second contribution is a method for modeling the progression of time."\cite{louthan2014hybrid}

The nodeset in HAGs represent states, i.e. network model instances. Network model instances comprize of an asset list as well as a factbase. As a hybrid model, facts can either be discrete, i.e. \textit{token facts}, or \textit{real facts}. Token facts assigned and enforced by the $=$ operator, and may also be tested using the $!=$ operator. Real facts, on the other hand, are instantiated using $:=$. Permitted relational operators include $==, >, >=, <=$, and $<>$ for $\neq$. Note that the initial state, i.e. the network assets and the initial fact base, needs to be inputted in a specification language. 

In addition to the nodeset, the model features an exploit framework with exploits representing the edges (state transitions) in the network. "Exploit patterns resemble functions whose bodies are lists of preconditions, which are identical in form to the facts in the network model. Preconditions document those properties that must be in  place for an exploit to bind (execute) to a given state. Postconditions are simply deletions or insertions (with overwrite semantics) of facts that result in a change of state." \cite{louthan2014hybrid}

Nichols et al. \cite{nichols2017introducing} extend the hybrid attack graph to introduce a notion priorities to attack graphs in order to tackle the state explosion problem in AG. 

\begin{figure}[htb]
\centering
  \subfloat[Example of the malicious insider network model, specified according to the syntax defined in \cite{louthan2014hybrid}]{%
    \includegraphics[width=.39\textwidth]{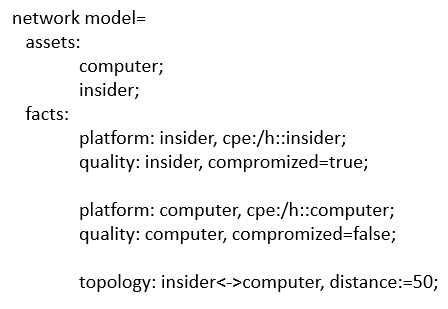}}\qquad
  \subfloat[Hybrid Attack Graph for our use case acc. to the HAG formalism \cite{louthan2014hybrid}]{%
    \includegraphics[width=.32\textwidth]{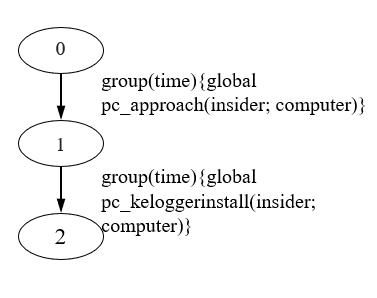}}\
  \caption{In the above example of a hybrid attack graph for the compromise of host $3$, a the intruder is 50 meters away from the computer. The computer, uncompromised at the beginning, has three states. Exploit transitions are collected in a group called “time,” which represents certain points in time as triggers for the state transition. \cite{louthan2014hybrid}}\label{fig:hybridAG}
\end{figure}

\subsection{Impact Dependency Graphs, Mission Dependency Graph}
\subsubsection{Impact Dependency Graphs}
In \cite{jakobson2011mission}, Jakobson introduces the Impact Depencency Graph (IDG) to model the impact of a cyber attack on a systems assets, services, and missions. Thus, IDG serves as a mean to assess mission dependency of a system in case of a cyberattack. The formalisms comprises of an attack model, as well as an domain model, the Impact Dependency Graph (IDG), on which the effect of individual attacks on the system can be evaluated. 

The attack model comprised of nodes for attacks, hardware platforms, assets as well as vulnerabilities (on assets). Furthermore, the model formalism offers binary relations: \textit{Target}-relations may exist between an attack and a hardware-platform; \textit{Exploit}-relations may exist between an attack and a vulnerability; \textit{Houses} represent the relationship between hardware-platforms and assets; \textit{Has-Vulnerability} is the relation between an asset and a vulnerability. Furthermore, the attack model consist of two parameters: the impact factor (IF) of an attack, a number in $[0,1]$ indicating the potential of compromize an attack may achieve, as well as the operational capacity (OC) $\in[0,1]$, indicating "to what level the asset, service, mission step or mission was compromised".

\begin{figure}[htb]
\centering
  \subfloat[Model of a Cyber-attack, original graphic by Jacobson \cite{jakobson2011mission}]{%
    \includegraphics[width=.4\textwidth]{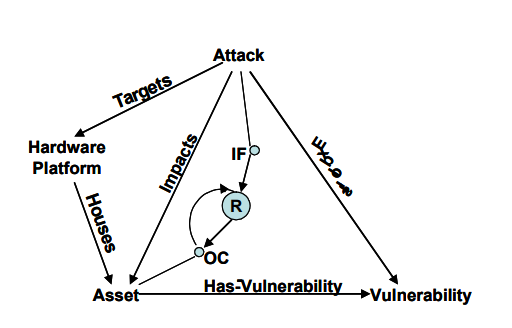}}\qquad
  \subfloat[Schema of an Impact Dependency Graph, original graphic from \cite{jakobson2011mission}]{%
    \includegraphics[width=.4\textwidth]{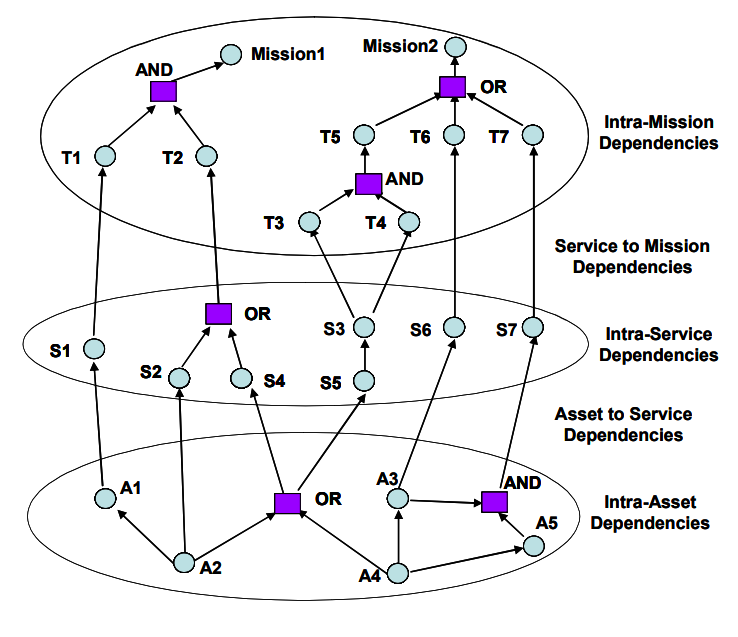}}\
  \caption{Impact Dependency Graph}\label{fig:idg}
\end{figure}

The Impact Dependency Graph is an AND/OR graph consisting of nodes and directed edges. Edges in the formalism represent the dependency relationships between the nodes. The nodes of the IDG represent assets, services, mission steps or mission. A mission hereby is defined as a "goal-directed sequential or parallel flow of mission steps" \cite{jakobson2011mission}. The individual mission step are composed of other missions or elementary actions, called tasks. Missions follow a temporal order: all mission steps need to be executed in the order prescribed by the diagram. Furthermore, each mission possesses a start-time, end-time and a duration. The mission state (completed, in progress, planned) depends on the individual states of the mission components. 

As mentioned before, an IDG may have AND as well as OR-nodes, which represent the logical dependencies among the child nodes in the IDG. "The AND/OR dependencies may themselves be "time-dependent", i.e. a  dependency holds during certain time intervals. Such time-dependency may be monitored or automatically set as a function of mission context". 

Using IDG, mission impact assessment can be performed as follows: in the first step, \textit{Attack Point Detection} the target, i.e. the primary target software asset under attack and the vulnerability exploited are determined using constraint programming. Next, in \textit{Direct Cyber Attack Impact Assessment} the operational capacity (OC) of the target asset is computed, while in \textit{Impact Propagation through the Cyber Terrain} the operational capacities of all dependent nodes are computed. In the final step, \textit{Mission Impact Assessment} is performed: the mission impact of an attack depends on the mission state (completed, in progress, planned) and the impact the attack has on the individual steps of the mission. Depending on these factors, the operational capacity of the mission is propagated through the IDG and the overall value computed. 

\subsection{Mission dependency graph, mission impact graph}
Sun et al. \cite{sun2017towards} also follow the mission-centric approach and present mission impact graphs. The \emph{mission impact graph} integrates mission dependency graphs and cloud-level attack graphs. 

According to \cite{sun2017towards} \emph{mission dependency graph} were developed by Musman et al. in \cite{musman2009} in 2009 \footnote{We could not directly verify this claim, as we could only find the 2010 version of the paper, which does not contain the definition. Therefore, we cite this definition from \cite{sun2017towards} as a secondary source}
: the nodes of the \emph{mission dependency graph} represent an entity, i.e. an asset, a service, a mission or an individual task, and the edges represent general dependencies among nodes. Furthermore, this formalism also incorporates AND and OR nodes. In this abstract description, mission dependency graphs resemble impact dependency graphs. 

\begin{figure}[htb]
\centering
    \includegraphics[width=.55\textwidth]{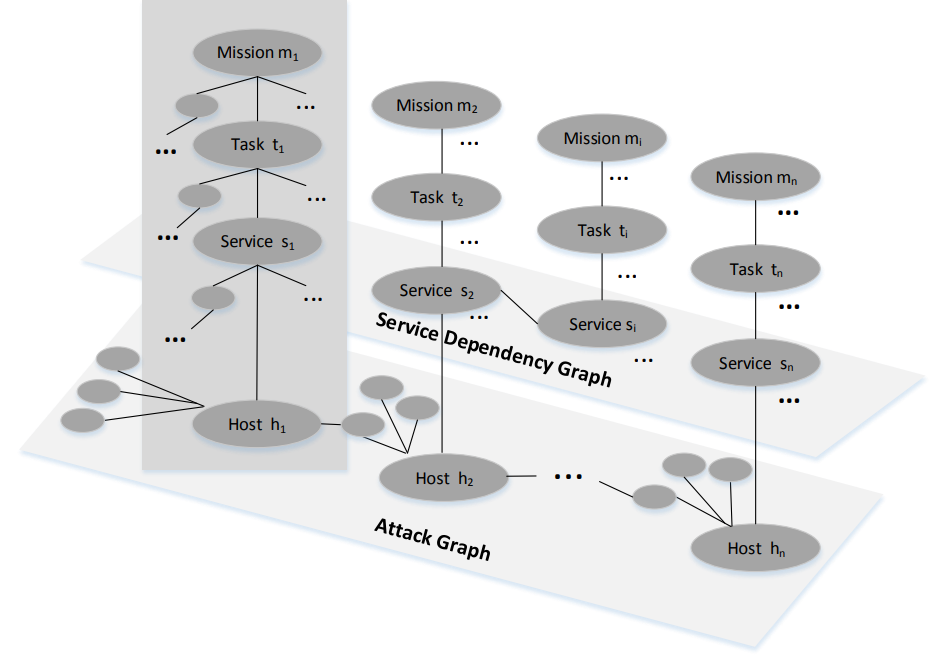}
  \caption{Schema of a Mission Dependency Graph, original graphic from \cite{sun2017towards}}\label{fig:mission}
\end{figure}

Their \emph{mission impact graph} (MIG) extends 
cloud-level attack graphs and fuses it with the mission dependency graph. 
MIGs are directed graphs with three
parts: an attack graph, service dependency part and mission-task-service-host dependency part.

This graph itself offers two node types: \textit{derivation} \textit{nodes} as well as \textit{fact nodes}. Fact nodes represent logical statements - they can be primitive (containing some information, e.g. host configuration) or derived, i.e. they were derived as a result of some interaction rule. Interaction rules are represented as derivation nodes.

The directed edges represent causalities between the nodes: derived fact nodes may depend on several derivation nodes, connected with logical ANDs; derived fact nodes depend on one or several derivation nodes, combined by logical OR relations.
Furthermore the authors extended MulVAL to automatically generate mission impact graphs, cf. \cite{sun2017towards}.

\subsection{Insecurity Flows, Risk Flow Attack Graph}
These types of attack graphs model risk in terms of flow, which is their distinct feature. 
\subsubsection{Insecurity Flow}
Moskowitz et al.'s \textit{insecurity flows} \cite{moskowitz1998insecurity} are a Bernoulli percolation type model which models the dynamics of insecurity flowing from a source (the starting point of the invader) to a sink (the domain we seek to to protect). In this early model, each node corresponds to a security mechanism, modelled by an  (independent) Bernoulli random variable whose probability $p$ denotes the success chances of the intruder at this node. A security breach is interpreted as a (probabilistic) flow through a network, whose connected nodes and links are referred to as a circuit. 

\begin{figure}[htb]
\centering
  \subfloat[Schema of a circuit, original graphic by \cite{moskowitz1998insecurity}, modelling insecurity flows as an undirected network]{%
    \includegraphics[width=.34\textwidth]{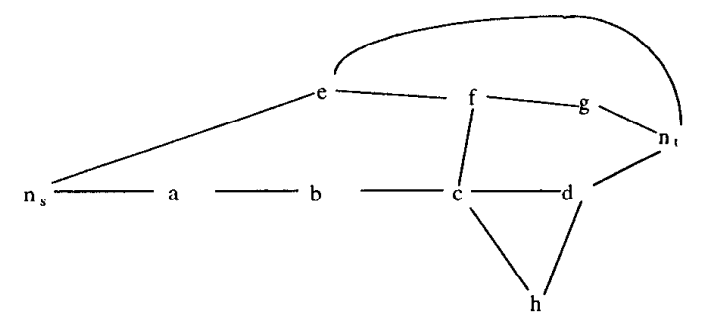}}\qquad
  \subfloat[Risk Flow Attack Graph by \cite{dai2015exploring}]{%
    \includegraphics[width=.35\textwidth]{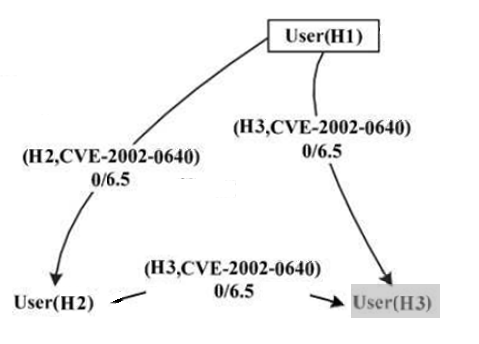}}\
  \caption{Attack graph formalisms involving flow}\label{fig:insecflow}
\end{figure}

\subsubsection{Risk Flow Attack Graph}
Alike, \emph{Risk flow attack graph} (RFAG) \cite{dai2015exploring} employ methods from network-flows (maximum flow) to identify critical attack paths. They propose a fuzzy comprehensive risk assessment to quantify risk severity. Similar to dependency AG, they consist of three disjoint sets of vertices $(N_s, N_g, N_m)$ representing initial attacker capabilities, the ultimate goal node and intermediate multi-sets of nodes, representing exploit pre- and postconditions; each node holds the value true or false. Edges map from $N_s \times N_m$ to $N_m \times N_g$ and represent exploits in network. Furthermore the RFAG defines, a mapping $V$ from an edge to its corresponding vulnerability and a map $\tau$, joining pre- and postcondition nodes, as well as risk flow values $C$ (derived from CVSS) and capacity values $F$ of the edges.

\subsection{Intrusion DAG, Intrusion Graph}
\subsubsection{Intrusion DAGs}
The intrusion DAG (I-DAG) was developed in the framework of the Adaptive Intrusion Tolerant System (ADEPTS) by Wu et al. \cite{wu2003adepts}.

In I-DAGs, the root node or several root nodes represent the ultimate goal. Each goal is refined into its sub-goals yielding the nodes in the graph. The subgoals in the chain need to be achieved in order to reach the overall goal. They may be associated with alerts from an intrusion detection system. Necessary and sufficient conditions for reaching a subgoal are modelled with AND as well as OR semantics on the edges leading to the node. To facilitate the expression of more complex Boolean aggregations, \textit{intermediate nodes} without any associated goals are introduced. 

Note that every node provides two "services": the \textit{Cause Service Set (CSS)} comprises of all services that need to be compromised to reach the goal; the \textit{Effect Service Set (ESS)} is made up of all services, which are compromised when the corresponding sub-goal is reached. Furthermore, every node holds a confidence value, expressing the probability of the IDS alert being correct.

The I-DAG models representation forms the knowledge representation in the ADEPTS intrusion response system to prevent an adversary from moving upstream by automatically deducing countermeasures or reactions of adequate strength. This process consists of three steps: 
First, in case of an intrusion alert, an event identifier, corresponding to a node in the I-DAG, as well as a confidence value $\in [0,1]$ are send to the IRS, updating the confidence value of the corresponding node in the I-DAG. Then, the Compromised Confidence Index (CCI), representing the probability that a given node is compromised, is computed for every node: It is based on the confidence values of the individual nodes and propagated to the root. Then, user defined threshold is compared to the CCI of the nodes in a top-down manner: whenever the CCI is greater than the node's threshold, the node is flagged as a Strong Candidate (SC) for being compromised. Weak Candidates (WC) are nodes below the threshold, but they have an AND-path (only including AND-edges) to an SC node. Likewise, Very Weak Candiates (VWC) are nodes below the threshold with an OR-path to an SC node. Finally, non-candidates (NC) are all remaining, non SC, WC and VWC nodes. 
Second, a response index (RI) is computed, which determines the response to the incident. Third, the response feedback algorithms evaluates the effectiveness of the incident response and decides for follow-up response actions.

\subsubsection{Intrusion Graph}
The intrusion DAG formalism in ADEPTS was later replaced by the intrusion graph (i-graph) formalism in \cite{foo2005adepts} by Foo et al. In this formalism each intrusion goal is represented by one node in the graph. Nodes reached by the outgoing edges of a node correspond to higher-order goals relative to the goal of the original node.

"In the I-GRAPH, edges are categorized into three types –
OR, AND, and Quorum edges. [...] For Quorum edges, one can
assign a Minimum Required Quorum (MRQ) on it, which
represents the minimum number of child nodes whose
goals need to be achieved in order for the node with
incoming Quorum edges to be achieved."\cite{foo2005adepts} 

I-GRAPHs are semi-automatically generated or updated from vulnerability descriptions and system services description 
(SNet) using the Portable I-GRAPH Generation (PIG) method. 
They can be used for IDS: whenever a detector flags an intrusion, the alerts are placed in the I-GRAPH nodes with the corresponding intrusion event and the Compromised Confidence Index (CCI) of a node (i.e. a measure of the likelihood that the node has been compromized) is computed. Furthermore, I-Graphs can be employed for intrusion reaction as well as quantification of effectiveness of the reaction: "For each selected response command, the Response Control Center computes 
the Response Index (RI). The RI takes into the account the 
estimated effectiveness of the response to the particular 
attack, measured by the Effectiveness Index (EI), and the 
perceived disruptiveness of the response to legitimate users 
of the system, measured by the Disruptiveness Index (DI)." \cite{foo2005adepts} 
Furthermore, adepts can be employed for simulation purposes and security quantification in terms of \textit{survivability} of 
the system under given attack scenarios. Simulation can be performed under the assumpton of unavailable or only local intrusion responses.

\begin{figure}[htb]
\centering
  \subfloat[Intrusion DAG from \cite{wu2003adepts}]{%
    \includegraphics[width=.3\textwidth]{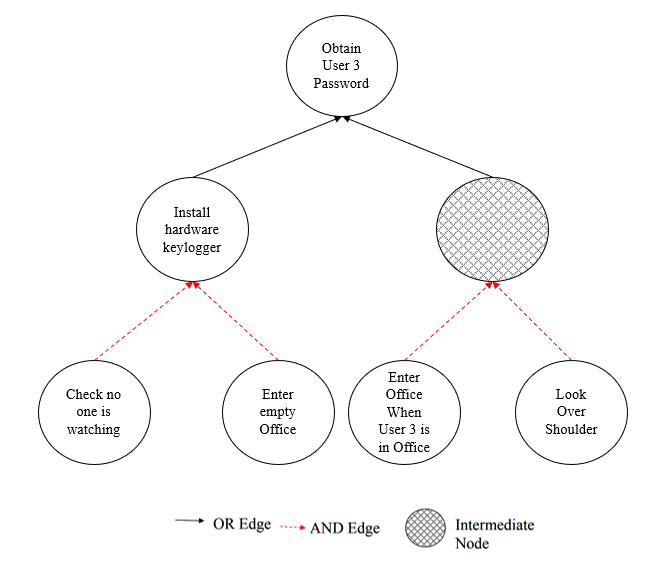}}\qquad
  \subfloat[Intrusion graph formalism from \cite{foo2005adepts}]{%
    \includegraphics[width=.35\textwidth]{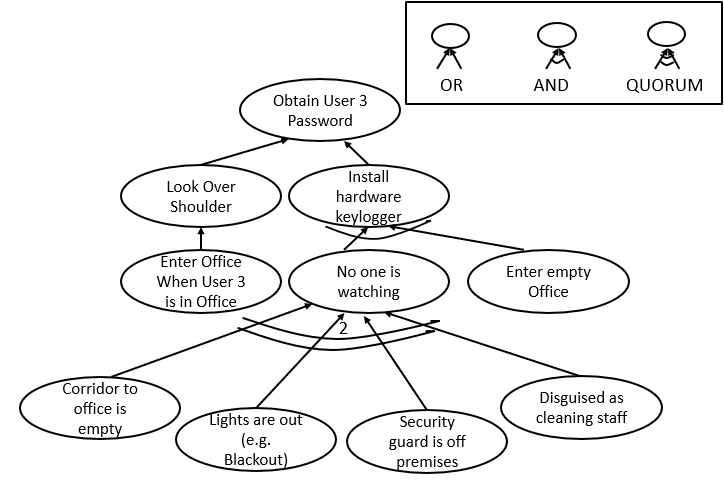}}\
  \caption{Intrusion DAG and Intrusion Graph}\label{fig:igraph}
\end{figure}

\subsection{Knowledge Graph, Knowledge Bayesian Attack Graph}
\subsubsection{Knowledge Graph}
A \emph{Knowledge Graph (KG)} \cite{althebyan2007knowledge} represents the dependency of \textit{knowledge units} $K_i$ of an insider. The term knowledge unit is used to represent any kind of information insiders may gain and save in their knowledge base: "A given insider has a specific KG that is initiated the first time he/she accessed the organizations’ resources. It is updated after each access to any object of the system."\cite{althebyan2008knowledge} Thus, knowledge graphs model the way an attacker may attack a system in order to acquire full knowledge.

Formally, knowledge graphs are directed graphs $G=(V,E)$, where the nodeset $V$ comprises of object nodes $O_i$, knowledge units $K_i$ and $CK$, the \textit{composed knowledge}, representing the union of all knowledge units in the knowledge base. A directed edge $E$ from node $V_i$ to $V_j$ is drawn whenever the knowledge unit in $V_j$ is accessible knowing $V_j$. Edges may only connect knowledge units with knowledge units, an object to a knowledge unit or a knowledge unit to the composed knowledge. \cite{althebyan2008knowledge}

\subsubsection{Knowledge Bayesian Attack Graph}
The Knowledge Bayesian Attack Graph (KBAG) is a formalism, which combines Bayesian Networks, Knowledge Graphs and Dependency Graphs for the system objects. In this context, \emph{Dependency Graphs (DG)} are defined to describe the dependency of the objects from the KB/KBAG, and must not be confused with dependency (attack) graphs. The DG is defined as a directed graph, where objects $O_i$ make up the node set. The directed edges represent the dependency direction between the objects, in a sense that $(O_i, O_j) \in E$, whenever information stored in object $O_i$ is needed to derive some piece of information in $O_j$. \cite{althebyan2008knowledge}

The \emph{Knowledge Bayesian Attack Graph (KBAG)} is a DAG indicating the probabilistic dependencies among different objects, which make up the nodeset. The conditional probabilities quantify the probability of mounting a successful attack on that node, i.e. the probability of compromise.

Assume $o_hot$ is a placeholder for \textit{hot nodes}, i.e. nodes of special importance to the system, e.g. because they contain data an adversary seeks to possess. Assume the list of hot nodes is given and an object $o_hacked$ is compromized. Then, the KBAG is constructed as follows: Initially only object nodes from the KG are considered. Next, the dependency graph is traversed and edges between the aforementioned objects are added, if and only if the objects are adjacent in the corresponding dependency graph. Next, all isolated vertices are removed from the graph. In the final step, all vertices in the newly created graph are checked for having a path in the DG to any of the hot nodes $o_hot$. These paths, including all intermediate vertices (i.e. objects) are added to the graph.

After construction, the conditional probabilities are assigned to the nodes. The authors propose a method similar to the approach in Dantu and Kolan \cite{dantu2005risk} to assign the probability values. Based on these probabilties, the computation of a node's risk value is performed by computing the posterior probabilities using belief updating. 

\cite{althebyan2008knowledge}
\begin{figure}[htb]
\centering
  \subfloat[The Knowledge Graph of our insider \cite{althebyan2008knowledge}]{%
    \includegraphics[width=.3\textwidth]{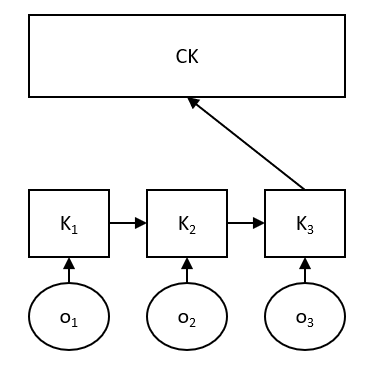}}\qquad
    \subfloat[Dependency Graph\cite{althebyan2008knowledge} for the common use case. ]{%
    \includegraphics[width=.25\textwidth]{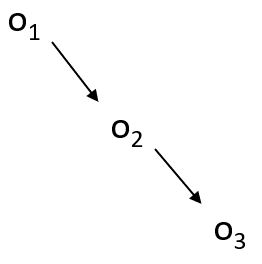}}
  \caption{Knowledge Graph and Knowledge Bayesian Attack Graph for the running example: assume the insider wishes to gain user privilege on host 3 to access a (password secured) directory. Let $o_1$ denote the admin, $o_2$ the user's machine and $o_3$ the directory. The insider may either directly try to find out the credentials of user $3$ ($K_2$), to gain user access on host, or try to  acquire the credentials using the admin's credentials ($K_1$). The adversary will still need to acquire the password ($K_3$) of the directory ($CK$) to access it.\\Note that in this case, the Dependency Graph and the Knowledge Bayesian Attack Graph from \cite{althebyan2008knowledge} (without conditional probability table) have the same structure. }\label{fig:kbgraph}
\end{figure}

\subsection{Logical Attack Graphs}
Some attack graph models are based on predicate or first order logic, including symbolic model checkers\cite{ritchey2000using, sheyner2002automated, jha2002two}. These works involve some attributes or model variables to formulated as a propositional clauses involving network configuration information. 

Still, Ou et al. \cite{ou2005mulval, ou2006scalable} are the first to define the logical attack graph explicitly as a special sub-type of attack graph: they use logical deduction to build their attack graphs using the Multihost, multistage Vulnerability Analysis Language framework, MulVAL. MulVAL is based on Datalog \cite{datalog2018Maier}, a declarative logic programming language, whose programs comprise of a fact base, containing logical predicate that are true for a particular combination of attributes, as well as a rule base that can "be considered an inference rule for deducing new facts from existing ones".\cite{datalog2018Maier}. MulVAL uses the output of an OVAL \cite{acm:oval2022} 
vulnerability scan, combines it with the vulnerability effects from the ICAT database \cite{MellICAT}
and converts the result into Datalog clauses. While vulnerabilities, host, access policies and network configurations are encoded as facts, the operating system or attacker behavior, are encoded in a collection of Datalog rules that comprise the reasoning engine. The MulVAL reasoning rules specify the general semantics of different attack methodologies and methods of privilege escalation, which are not tied to not specific vulnerabilities. 

In this framework, data from vulnerability scans and the network data are represented as facts and all possible data accesses are are derived using logical deduction. In the following policy checking phase, the data access output from attack simulation are compared with the given security policy.
In case policy violations are found, the results are collected in an attack graph, which is automatically generated by the program; "The nodes do not represent the entire state of the network (as in Sheyner’s attack graph), but each node represents one Boolean variable which is logical conclusion of the facts given. The edge relation in logical attack graph represents causality relation between system configurations and attackers potential privileges" \cite{saha2008extending}.

Graphically, "a traditional attack graph generated by MulVAL is composed of two types of nodes, fact nodes (including primitive fact nodes and derived fact nodes) and derivation nodes (also known as rule nodes). Primitive fact nodes (denoted with rectangles [...]) present objective conditions of the network, such as the network, host, and vulnerability information. Derived fact nodes (denoted with diamonds) are the facts inferred by applying the derivation rule. Each derivation node (denoted with ellipse) represents the application of a derivation rule. The derivation rules are implemented as interaction rules in MulVAL. Simply put, one or more fact nodes could be the preconditions of a derivation node, while the derived fact node is the post-condition of the derivation node." \cite{sun2017towards}

\begin{figure}[htb]
\centering
    \includegraphics[width=.6\textwidth]{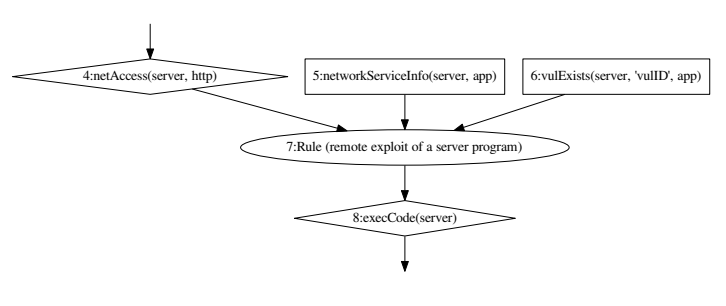}
  \caption{Part of a Simplified Logical Attack Graph generated by MulVAL, original graphic from \cite{sun2017towards}}\label{fig:logAG}
\end{figure}

According to \cite{barik2016attack} logic attack graph has the limitation, that some system properties may not be expressible in terms of propositional formulas. "Moreover, logical attack graph cannot provide proper causality reasoning when rules describing the causality relation contain negation. For instance, if a privilege of an attacker negatively depends on another privilege of the attacker, logical attack graph can represent that the first privilege is attainable by the attacker as second privilege cannot be attained. However, it cannot give any reason as to why the second privilege cannot be attained." 

In their extension to MulVAL, Saha \cite{saha2008extending} addresses this issue and presents a methodology to generate logical attack graphs which contain negation in rules. Furthermore, the author presents method to incrementally re-generate logical attack graphs (e.g. due new vulnerabilities, parameter changes or updates). Today, MulVal and its extensions are still a popular choice for attack graph generation from network scans. 

\subsubsection{Personalized Attack Graphs}
The personalized attack graph (PAG)\footnote{In the text it was also referred to as \emph{enhanced attack graph}.} was introduced in \cite{roberts2011personalized} and formalized in \cite{urbanska2013accepting}. Instead of solely focusing on technical properties, PAGs integrate user behaviour and personalization into attack trees. This way, the security of a specific user and hardware/software combination can be analysed.

Initially, in \cite{roberts2011personalized}, PAGs are defined as plans resulting from domain models in the Planner Domain Definition Language (PDDL). The PDDL was developed by Fox and Long \cite{fox2003pddl2} and proposed to be used for attack graph models by \cite{sarraute2008advances, futoransky2010simulating, obes2013attack}.

The PDDL model fpr PAGs comprizes of facts, representing information on the system, its users etc., and actions, causing changes in system states. First, a set of exploit trees (which are not personalised yet) are constructed by defining chains of actions, specified in a pre- and postcondition semantic in a PDDL model. This set of exploit trees is called \emph{monolithic personalized attack graph} \cite{urbanska2013accepting}.

These exploit trees are similar to classical attack trees: they comprise of root nodes representing the attacker goal, and interior vertices, which represent intermediate system states of the attack, and edges represent actions that cause a transition from one state into another. Subtrees may be connected via logical "AND" and "OR" connectors. But unlike ordinary attack trees, the formalism is not constrained to attacker actions, but it offers four types of actions: the Attacker actions, marking the individual steps of the exploits; the User Actions, enabling a vulnerability; the system actions, through which vulnerabilities are manifested, "e.g., a browser accepting a (compromised) certificate or a key-logger capturing and forwarding keystrokes."\cite{roberts2011personalized} as well as intervention actions, which prevent the use of exploits or mitigate vulnerabilities.

After the set of exploit trees, i.e. the monolithic PAG\cite{urbanska2013accepting}, was defined via the (potential) actions of users, attackers, system and interventions, which may lead to compromise, the specific user and system attributes are specified as facts and added to the model using the PDDL problem description. The problem description consists of a set of world objects, the initial conditions, as well as some goal definition. 

Using the exploit trees as a domain model, and the problem description including the user and system attributes, the authors use a planner to construct the sequences of actions that lead to the system compromise for the given facts on the system properties and user behaviour. This reduced graph is called \emph{instantiated PAG}.

Using the the PDDL representation, this planner based approach can be used to narrow the set "raw" exploit trees down to those vulnerabilities that are specific for the system configuration in combination with the user properties. This approach can also be used for network hardening in order to identify suitable intervention actions or simulate the effect of various system/user combinations on system security. 

\begin{figure}[htb]
\centering
    \includegraphics[width=.45\textwidth]{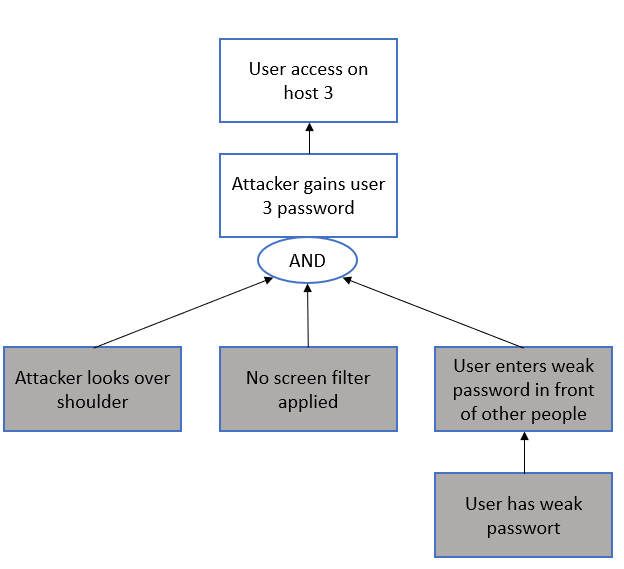}
  \caption{Personalized Attack Graph Model \cite{roberts2011personalized}}\label{fig:personall}
\end{figure}

In \cite{roberts2012using}, Roberts et al. compare the performance of various A$^\ast$-based algorithms for planning in the context of personal attack graphs. Mukherjee's PhD thesis \cite{mukherjee2017heuristic} extends the personal attack graph approach using tools from natural language processing (NLP) to extract the information needed to construct PAGs. The software and versions used, attacker and user actions as well as the post-conditions are extracted from English language vulnerability descriptions from the National Vulnerability Database (NVD) descriptions, collected in the Joshi Korpus \cite{joshi2013extracting}

Potential future applications of personalized attack graph include personalizing generic attack graph templates (e.g. from vulnerability scans) to the current patch level. 

\subsubsection{Privilege graph}
The \emph{privilege graph} is a directed graph, whose nodes represent sets of privileges on (sets of) objects and directed edges are mark how they are inherited by individual users: Every user holding the parent node's privileges can also acquire the child node's privileges. Privilege graphs are the first known attack graph type. Note that 
 in \cite{dacier1996models} the privilege graph is extended: instead of privileges, more general attack exploits are considered. In addition, it is annotated with attack rates and transformed into a Stochastic Petri Net, whose reachability graph is called \emph{intrusion process state graph}, which corresponds to a Markov Chain model. The authors suggest to analyse the corresponding Markov process with respect the risks induced by the residual flaws. To do so, the authors suggest a quantification of vulnerabilities, based on "Effort" or "Time" spent by a potential attacker to reach the security target, namely the Mean Time to Failure (MTTF) or Mean Effort to failure (METF). 
In addition, the authors define the \emph{intrusion process state graph}, a tree collecting all attack paths, which is similar to attack graphs in \cite{phillips1998graph}, but without probability labels. 
In \cite{ortalo1998quantitative}, Ortalo and Deswarte adopted the privilege graph and provide the first real-world security use-case of a small bank holding 30 employees in a rural area, while Ortalo et al. \cite{ortalo1999experimenting} later analysed a privilege graph model constructed from 13 major UNIX vulnerabilities.
\begin{figure}[htb]
\centering
    \includegraphics[width=.73\textwidth]{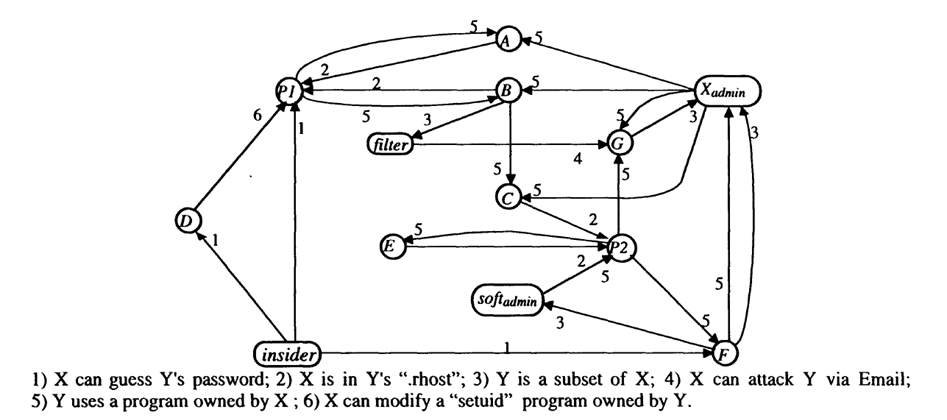}
  \caption{The Privilege Graph by \cite{dacier1996models} (original graphic) is the oldest AG formalism}\label{fig:privilege}
\end{figure}
\subsubsection{Scenario attacks}  
Scenario Attacks \footnote{For attack scenario graphs by Albanese et al. \cite{10.5555/2041225.2041255} Section \ref{sec:exploitdep}} were proposed by \cite{templeton2001requires} as part of the requires/provides model for attack representation. They define \textit{scenario attacks} as an abstract set of \textit{attack capabilities} that provide the requirements for other new capabilities. They serve as a complement to the notion of \textit{concepts}, which a set of required capabilities to perform an exploit, as well as the provided capabilities, resulting from the exploit. This model is also known as requires-provides formalism. Thus, scenario attacks provide one of the first methods for tailoring complete attacks in a precondition-postcondition format, which proved to be helpful for discovering new attack possibilities as well as mid-level intrusion detection. Scenario attacks were implemented using their own attack specification language, JIGSAW. Today, this formalism is not being used anymore, as there exist more advanced approaches. Still, this work is fundamental as it paved the way towards IDS correlation with attack graphs and gave input for future work towards scalable attack graph generation. 
 The main criticism of this (generic) method that "they do not capture the conditions that have to be met by related attacks" \cite{ning2003learning}, which led to the development of (hyper-)alert correlation graphs. Later, \cite{alserhani2015knowledge} also employed the requires-provides formalism.

\subsection{Scenario graph, Probabilisitic Attack Graphs} \subsubsection{Scenario graph} Sheyner \cite{sheyner2002automated} was the first to use model checking techniques to generate attack graphs. The formal model behind this idea is the scenario graph, elaborated by Sheyner in \cite{sheyner2004scenario}. 
Scenario graphs are formally described in terms of Büchi automata, which are automata that accept infinite words if and only if an accepting state is visited infinitely often. Büchi automata are formal models that can be checked for satisfying a given safety property $P$. The authors use this property to generate \emph{safety scenario graphs} using a routine, whose input is a Büchi model and a safety property written in CTL. Using an iterative fixpoint symbolic procedure the set of states where the desired property is not satisfied is computed and collected in the scenario graph. The individual counterexamples violating the safety property yield the individual attack paths.

In \cite{sheyner2004scenario}, a second algorithm is presented that constructs the intersection of the model automaton and another Büchi automaton representing the negation of the safety property, which yields the \emph{scenario automaton}. Here, the scenario graph is a subset of the language recognized by the scenario (Büchi) automaton, and a scenario is simply a word. Overall, Sheyner's model checking approach yields one of the cornerstones on the way to automatizing the generation of attack graphs. 


\subsubsection{Probabilistic Attack Graph}
In the same work, the term \emph{Probabilistic Attack Graph} is found. Probabilistic attack graph are described as annotated attack graphs, where probabilities are used to describe random state transitions. \cite{sheyner2004scenario, jha2002mini}  define  probabilistic attack graphs \cite{jha2002mini} as mixed attack graphs having a mix of probabilistic and nondeterministic state transitions. They convert these mixed, probabilistic attack graphs (PAG) into an alternating probabilistic attack graph (APAG) and then interpret the result as a Markov Decision Process \footnote{Each action in the MDP represents a transition from a nondeterministic to a probabilistic state. Hence, an APAG can be obtained from a PAG by adding hidden states whenever there is a transition between two nondeterministic or probabilistic states. Note that the probability of an execution fragment is exactly the product the probabilities of the probabilistic}. \cite{sheyner2002automated} propose to make use of these probabilistic attack graphs to model attacker-defender interaction in computer networks as extensive form games on such graphs, introducing  attack graph games. In Sheyner's thesis \cite{sheyner2004scenario}, probabilistic attack graphs (PAG) were called probabilistic scenario graphs (PSG) and alternating probabilistic attack graph (APAG) were called alternating probabilistic scenario graphs (APSG). 

Today, similar notions of probabilisitic attack graphs are still used in attack graph analysis, especially for game-theoretic analysis of attack graphs, cf., e.g. \cite{durkota2015optimal}. 
\subsection{State Transition Diagram}
In \emph{state transition analysis}, \cite{ilgun1995state}, employ state transition diagrams to model cyber-attacks as a series of state changes leading "from an initial secure state to a target compromised state", linked via signature actions\footnote{"those action that, if omitted from the execution of an attack scenario, would prevent the attack from completing successfully"}. This early graphical representation a cyber-attack uses a representation similar to state machine diagrams and "identifies precisely the requirements and the compromise of a penetration".
\begin{figure}[htb]
\centering
    \includegraphics[width=.58\textwidth]{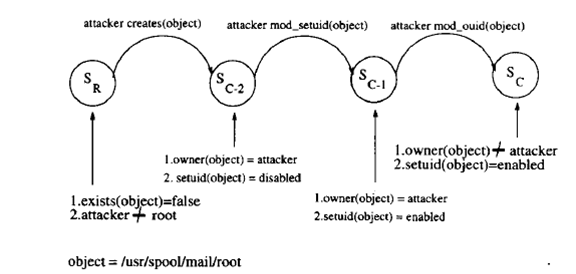}
   \caption{State Transition Diagram, an early attack graph type. Original illustration by Ilgun \cite{ilgun1995state}}
\end{figure}
Later, Kemmerer \cite{kemmerer1997nstat} the state transition approach  to allow for for modelling shared resources in a tool called NSTAT, before the tool was further developed into a real-time network based tool for intrusion detection called NetStat \cite{vigna1998netstat}. In this work, the authors combine the formalism of State Transision Diagrams with a model of the network topology. Formally, the network is represented as a hypergraph, which is a graph, where edged are allowed to connect multiple nodes. In the hypergraph, networks constituate of interfaces, modelled as nodes, as well as hosts and links, modelled as a set of interfaces (i.e. hyper-edges) holding distinct attributes. 

\subsection{Vulnerability graph}\label{sec:vuln_graph}
\begin{figure}[htb]
\centering
  \subfloat[Communication graph and vulnerability graph 
 \cite{birkholz2010efficient} for the Buffer-Overflow vulnerabilities in the running example]
    {\includegraphics[width=.65\textwidth]{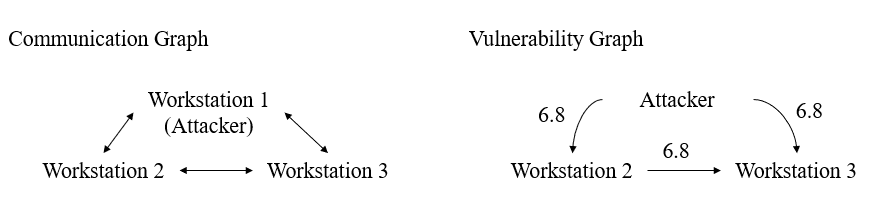}} \qquad
   \caption{Vulnerability graph formalisms by Birkholz et al. \cite{birkholz2010efficient}}\label{fig:attackgraphformalismsbirkholz}
\end{figure}

Birkholz et al. \cite{birkholz2010efficient} derive their vulnerability graphs from \emph{communication graphs}, which are undirected graphs $G_C = (V_C , E_C )$, where the set of nodes $V_C$ is a set of hosts. Two nodes are connected by edges $E_C \subseteq V_C \times V_C$, whenever the two hosts can directly communicate with each other over the network. 

Note that, in their formalism, the attacker’s initial location $s$ and target $t$ are both elements of the nodeset, and an attack path is defined as a sequence of nodes from $G_C$. 

From this data, the \emph{vulnerability graph} is defined as a directed subgraph $G_V\subseteq G_C$ of the communication graph, with the nodeset $V_V \subseteq V_C$ and edges $E_V \subseteq V_V \times V_V$.
Additionally, edges in the vulnerability graph are labelled with weights corresponding to the CVSS score of the vulnerability. I.e. for an edge $(u, v)$, the weight $w$ is defines as $w(u, v) = score(v)$. 

Using vulnerability graphs, the authors seek to compress this attack graph formalism into a more compact tree representation: by transforming the edge weights to $w'$, $w'(u,v)=10-w(u,v)$ and applying Dijkstra's shortest path algorithm on the graph as well as the inverse graph (graph with inverted nodes), and merging the results, a more compact tree representation of the network is obtained. 

\subsection{Vulnerability cause graph, Security Activity Graph, Security Goal Model}
\subsubsection{Vulnerability cause graph}
\emph{Vulnerability cause graphs} (VCG) are defined by Ardi et. al in \cite{ardi2006towards}, and its application to model software vulnerabilities are illustrated in and \cite{byers2006modeling}. 
In this directed acyclic graph formalism, all nodes except for one represent causes, while edges are used to model the relationships between causes in a precondition-consequence semantic. Unlike other forms of attack graphs, vulnerability cause graphs specifically represent how vulnerabilities are introduced into code and how they can be avoided in software engineering, rather than patched. 

First generation Vulnerability cause graphs (VCGs) from \cite{ardi2006towards} are directed acyclic graphs, having software vulnerabilities or their causes as vertices. The directed edges represents the causal connections between them. "Vertices with no successors are known as vulnerabilities [...]. Vertices with successors are known as causes, and represent conditions or
events that may lead to vulnerabilities being present in the software being developed. [...]
Intuitively, vertices in the VCG with no predecessors represent root causes of vulnerabilities, and traversing the graph from the root causes indicates progressively lower-level causes of vulnerabilities." \cite{ardi2006towards}

\begin{figure}[htb]
\centering
  \subfloat[Schema of the first generation vulnerability Cause graph. Original graphic by \cite{ardi2006towards} ]
  {%
    \includegraphics[width=.24\textwidth]{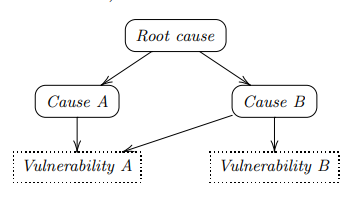}}\hfill
     \subfloat[Schema of the generation Vulnerability Cause graph. Original graphic by \cite{ardi2006towards} ]{%
    \includegraphics[width=.24\textwidth]{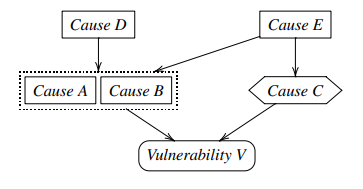}}\hfill
     \subfloat[Example of a Vulnerability Cause graph \cite{ardi2006towards} for the use case]{%
    \includegraphics[width=.34\textwidth]{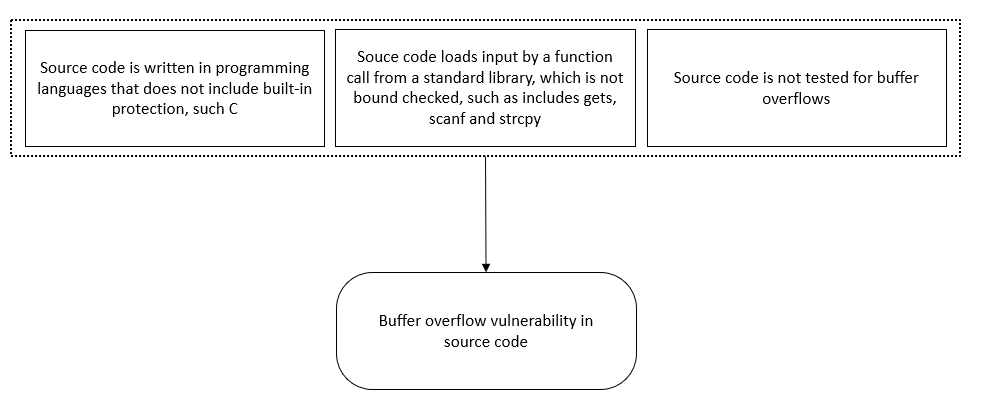}}\hfill
  \caption{Vulnerability Cause Graph}\label{fig:attackgraph_vcg}
\end{figure}
The second generation VCG formalism \cite{byers2006modeling} is more refined and offers additional nodes. The second generation VCG consists of four different node-types: simple nodes, compound nodes, conjunction nodes and exit nodes. As in the first generation graph, simple nodes represent conditions or causes, leading to vulnerabilities in the software. Visually they are represented as squares. Conjunction nodes are represented as dashed squares around other nodes. They used to represent joint preconditions. I.e. a conjunction node represents the logical conjunction (i.e. logical "and"s) of the nodes it encloses. 
Compound nodes are represented as hexagons. They have a similar purpose as functions in a programming language: they represent "entire VCGs that model reusable or complex analysis elements."\cite{byers2006modeling}
Furthermore, "every VCG has a designated exit node, which must be the only node in the graph without successors, and is the only node that does not represent a cause"\cite{byers2006modeling}. Exit nodes are represented by rounded rectangles.

Vulnerability Cause graphs are designed for security aware software engineers of software managers as a means of detecting and proactively avoiding insecure coding practise and patterns.

\subsubsection{Security Activity Graph}
In \cite{ardi2006towards}, Ardi, Byers, and Shahmehri also introduced the \emph{security activity graphs (SAGs)}, which can be generated from vulnerability cause graphs. To construct a security activity graph from a VCG, first, mitigations for each vulnerability in the VCG need to be identified and expressed as first order predicate logic formulas having atomic actions ("activities") as predicates. The predicates can be joined by AND, OR and SPLIT gates, which represent the same action being applied to different parent nodes.

Then, SAG fragments are created from these first order logic formulae. Finally, the SAG fragments are joined together: the VCG is traversed in a bottom-up manner and the SAG fragment representing the mitigation of the node is joined using an OR gate. Post-generation simplification of the logical expression helps simplify the SAG without affecting the meaning \cite{ardi2006towards}

Note that in a later work, \cite{byers2008cause}, the syntax of SAGs was altered: "Split gates no longer appear in the formalism. The functionality that simple activities can be distinguished from compound activities (complex activities that may require further breakdown) was added. Moreover cause references (possible attack points) serve as placeholders for a different SAG associated with a particular cause." \cite{kordy2014dag}. For analysis, Boolean variables are assigned to the leaf-nodes of the SAG, indicating weather the activities were performed accordingly or not. In order to determine whether the actions prevent the vulnerability during software development or not, the value of the root node is deduced using logic calculus.

\begin{figure*}[htb]
\centering
  \subfloat[Security Activity Graph Fragments. Original graphic by \cite{ardi2006towards}]{%
    \includegraphics[width=.25\textwidth]{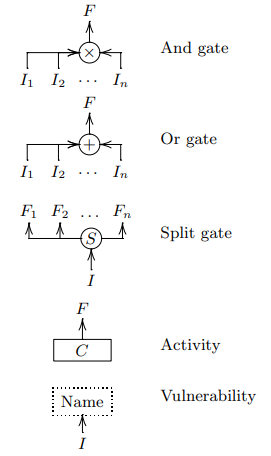}} \hfill
     \subfloat[Elements of Security Goal Models. Original graphic by from \cite{byers2010unified}]
    {\includegraphics[width=.65\textwidth]{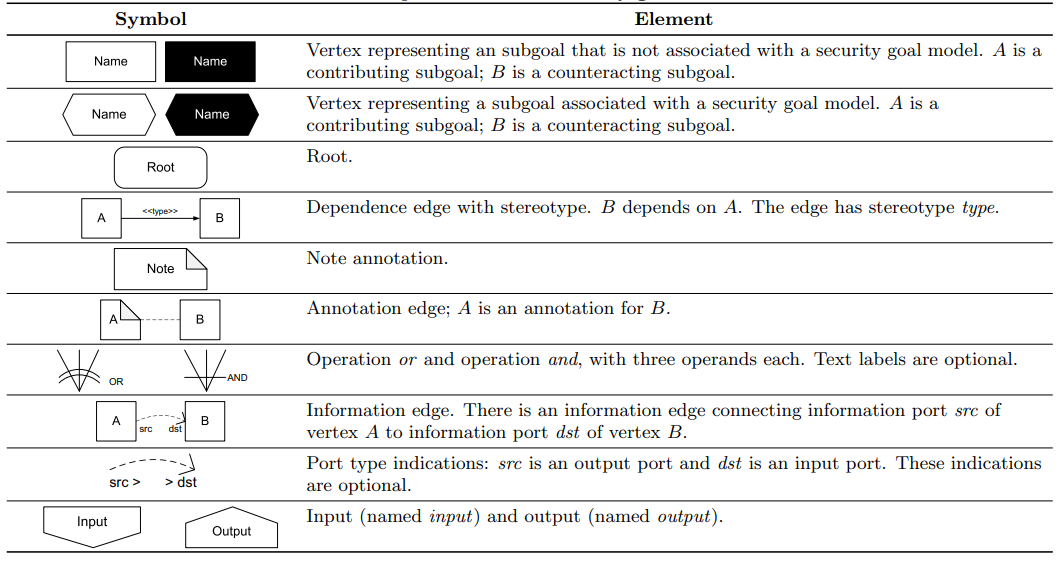}}\\
  \subfloat[Schema of a Security Activity Graph. Original graphic by\cite{ardi2006towards}]
    {\includegraphics[width=.45\textwidth]{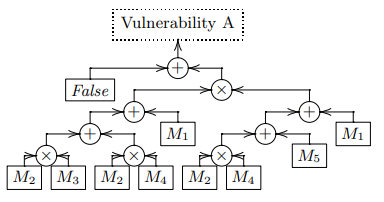}} \hfill
    \subfloat[\textcolor{black}{Security Goal Model to avoid buffer overflows, which is the vulnerability in the ssh-daemon in our running example }\cite{byers2010unified}]
    {\includegraphics[width=.5\textwidth]{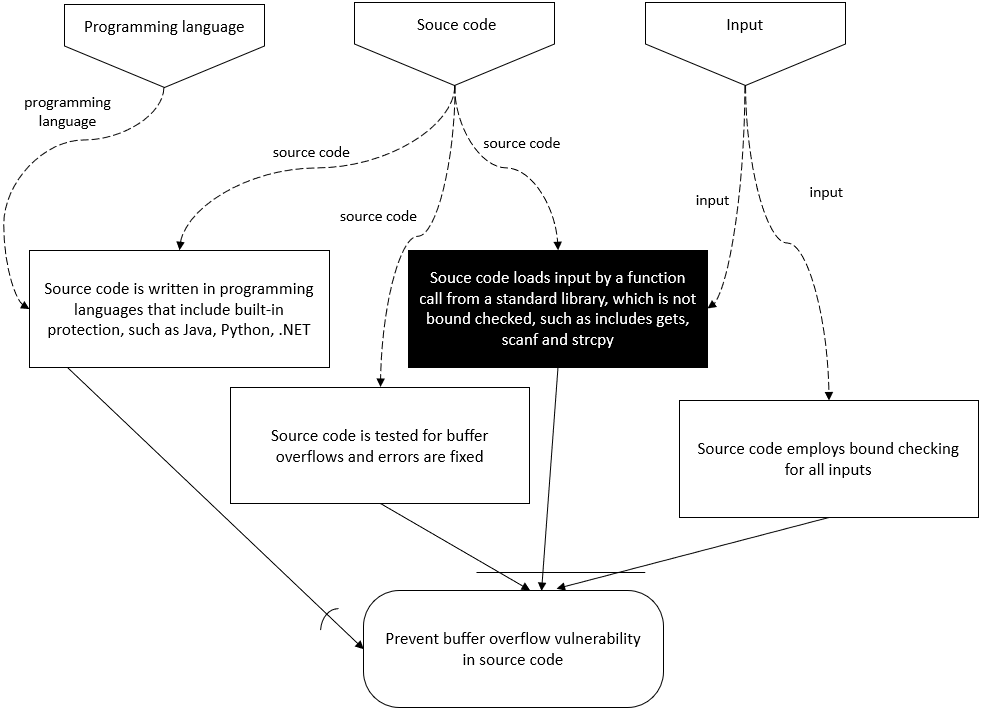}}
  \caption{Security Activity Graph and Secuity Goal Model}\label{fig:attackgraph_sagsgm}
\end{figure*}

\subsubsection{Security Goal Model}
 In later a work, Byers and Shahmehri introduced the \emph{Security Goal Model} (SGM) language "that can be used in place of attack trees, security activity graphs (SAG), vulnerability cause graphs (VCG), and security goal indicator trees (SGIT)." \cite{byers2010unified}. The authors furthermore claim that transformations to and from attack trees, VCGs, SAGs, and SGITs to SGMs were developed, but not published in the paper due to the page limit.

Similar to \textit{goal-oriented attack graphs}, security goal models represent the many ways how a predetermined goal can be realized. "In this context a goal is anything that affects security or affects some other goal. Typical examples are vulnerabilities, security functionality, security-related software development activities, and attacks." \cite{byers2010unified}

A security goal model is a DAG with nodes representing (sub)goals. The final node is the root node, which represents the overall goal, and the root node must not have any successors. Solid edges in the graph represent a temporal and causal dependency relation between the subgoals, which must be adhered to in order to reach the overall goal. Furthermore, the formalism supports the combination of dependencies via AND and OR operators. 
Dashed edges are used to model the information flow between the subgoals: An InputPort is used for information needed to achieve a subgoal, whereas an OutputPort represents the information obtained when achieving the subgoal. To facilitate understanding, the model formalism makes use of annotations: "Annotations may be associated with other model elements through AnnotationEdges." \cite{byers2010unified} Additionally, special annotations on the edges, called Stereotypes, are employed to explain why an edge exists; they may either indicate causes, prerequisites or uses. 

To analyze weather a security goal is met or not, every vertex in the SGM, representing a Boolean variable, is assigned its truth value. Using logic calculus and the instantiation of the variables, the truth value of the root node is computed.  

SGM were used by Shamehri et al. \cite{SHAHMEHRI2012997}in a case study on a buffer overflow vulnerability in the xine media player. In this work, the authors present an approach to automatically convert the SGM into \emph{vulnerability detection conditions (VDC)}- formal vulnerability models that serve as input for passive software testing tools, checking for evidence of the modelled vulnerability during when executing the program code. 

Similar to Security Goal Indicator Trees, this formalism helps identify indicators (here: actions) that prevent certain vulnerabilities in software development. It helps implement concrete measures realizing a certain security feature. 

\end{document}